\documentclass[11pt,a4paper]{article}
\usepackage{authblk}
\usepackage{epsfig}
\usepackage{cite}
\usepackage{amsmath}
\usepackage{mcite}
\usepackage{array,tabularx,epsfig,mathrsfs,graphicx,rotating}
\usepackage{ifthen}
\usepackage{fancybox}
\usepackage{amsfonts}
\usepackage{float}

\PassOptionsToPackage{hyphens}{url}
\usepackage[hyphens]{url}

\usepackage{hyperref}
\hypersetup{
  colorlinks=true,
  linkcolor=blue,
  citecolor=blue,
  urlcolor=blue
}

\usepackage{breakurl}

\newcommand{\beq}{\begin{equation}}
\newcommand{\eeq}{\end{equation}}

\newcommand{\pt}{p_{T}}

\newcommand{\ptjet}{p_{T}(\mathrm{jet})}
\newcommand{\pttop}{p_{T}(\mathrm{top})}

\newcommand{\ptfrac}{p_{T}^\mathrm{frac}}

\newcommand{\mmu}{\langle \mu \rangle}

\newcommand{\mjet}{M_{\mathrm{jet}}}
\chardef\til=126

\newcommand{\gev}{{\,\mathrm{GeV}}}

\newcommand{\tev}{{\,\mathrm{TeV}}}
\newcommand{\pythia}{{\tt PYTHIA8\,}}
\newcommand{\herwig}{{\tt HERWIG++\,}}

\newcommand{\ttbar}{ t \bar t}

\newcommand{\eplus}{e^+}
\newcommand{\eminus}{e^-}
\newcommand{\epem}{\eplus\eminus}
\def\invfb{ \mbox{fb}^{-1} }

\newcommand{\afbt}{A^t_{FB}}

\newcommand{\cthel}{\mathrm{cos} \theta_{hel}}

\newcommand{\tpq}{t}
\newcommand{\Wboson}{W}
\newcommand{\bottom}{b}
\newcommand{\quark}{q}

\newcommand{\pem} { {\cal P} }
\newcommand{\pep} {{\cal P'} }
\def\beq{\begin{equation}}
\def\eeq#1{\label{#1}\end{equation}}
\def\eeqn{\end{equation}}
\newcommand{\gsim}{\hbox{ \raise3pt\hbox to 0pt{$>$}\raise-3pt\hbox{$\sim$} }}
\newcommand{\lsim}{\hbox{ \raise3pt\hbox to 0pt{$<$}\raise-3pt\hbox{$\sim$} }}

\providecommand{\ee}   {\rm{e^+e^-}}
\graphicspath{./fig/}
\usepackage{subfigure}

\begin{document}

\clearpage
\pagestyle{empty}
\setcounter{footnote}{0}\setcounter{page}{0}%
\thispagestyle{empty}\pagestyle{plain}\pagenumbering{arabic}%

\hfill  ANL-HEP-CP-13-31

\hfill August 3,  2013

\hfill Version 2.0

\vspace{2.0cm}

\date{\today \\ version 2.0 \\ ANL-HEP-CP-13-31}

\begin{center}
{\Large \bf Reconstructing top quarks at the upgraded LHC \\ and at future accelerators}

\vspace{10 mm} 
{\large Summary of the Snowmass ``Top algorithms and detectors'' \\ High Energy Frontier Study Group \vspace{1 cm}}

\end{center}

\vspace{10 mm}

\begin{center}
{\large R.~Calkins$^{\mathrm{[a]}}$, S.~Chekanov$^{\mathrm{[b]}}$, J.~Conway$^{\mathrm{[c]}}$, J.~Dolen$^{\mathrm{[d]}}$, R.~Erbacher$^{\mathrm{[c]}}$, J.~Pilot$^{\mathrm{[c]}}$, R.~P\"oschl$^{\mathrm{[e]}}$, S.~Rappoccio$^{\mathrm{[d]}}$, Z.~Sullivan$^{\mathrm{[f]}}$,  B.~Tweedie$^{\mathrm{[g]}}$ 
}
\end{center}

\vspace{1.0cm}
\itemsep=-1mm
\normalsize
\small

\noindent $^{[a]}$ Northern Illinois University, Northern Illinois Center for Accelerator and Detector Development
(NICADD), DeKalb, IL60115, USA.   Email: Robert.Calkins@cern.ch \\ 

\noindent $^{[b]}$  HEP Division, Argonne National Laboratory,
9700 S.Cass Avenue, 
Argonne, IL60439, USA.  Email: chakanov@anl.gov \\ 

\noindent $^{[c]}$  Department of Physics,
UC Davis,
One Shields Avenue,
Davis, CA 95616, 
USA. Email: jrpilot@ucdavis.edu \\

\noindent $^{[d]}$ State University of New York at Buffalo Department of Physics, 
239 Fronczak Hall,
Buffalo, NY 14260 USA.  Email: james.dolen@gmail.com \\ 

\noindent $^{[e]}$  Laboratoire de l'Acc\'el\'erateur Lin\'eaire, Universit\'e Paris Sud, 
F-91898 Orsay CEDEX, France. Email: poeschl@lal.in2p3.fr \\

\noindent $^{[f]}$ Department of Physics, Illinois Insttitute of Technology, 3300 S Federal St, Chicago, IL 60616 
Email: Zack.Sullivan@iit.edu \\

\noindent $^{[g]}$ Physics Department, Boston University, 590 Commonwealth Avenue, Boston, MA 02215, USA. Email: btweedie@bu.edu \\

\normalsize
\vspace{1.0cm}

\begin{abstract}
This report describes the studies performed for the Snowmass ``Top algorithms and detectors'' High Energy Frontier Study Group. 
\end{abstract}

\newpage
\setcounter{page}{1}

\newpage
\tableofcontents
\newpage

\section{Introduction}
\label{sec:intro}
The long-awaited  measurements of top quarks and their properties 
by ATLAS \cite{ATLAS_detector} and  CMS \cite{1748-0221-3-08-S08004} have proved the LHC 
to be the world's top factory.
The Standard Model (SM) top-quark studies at the LHC include\footnote{We give only the most representative or recent 
references which are only needed to make our point, without attempt to collect a comprehensive list of references from ATLAS and CMS.} 
reconstruction of the total and differential
$t\bar{t}$ cross sections \cite{Aad:2010ey,Aad:2011yb,Aad:2012qf,ATLAS:2012aa,Aad:2012mza,Aad:2012hg,Chatrchyan:1359133,Chatrchyan:1472749,Chatrchyan:1502676}
top-quark mass measurements \cite{ATLAS:2012aj,Chatrchyan:1477737,Chatrchyan:1354223}, 
cross sections of invariant masses of $t\bar{t}$ pairs, 
charge asymmetry  \cite{ATLAS:2012an} and single-top measurements \cite{Chatrchyan:1479139}.
A number of searches have been performed aiming to find heavy resonances decaying to $t\bar{t}$ pairs \cite{Chatrchyan:2012cx,Aad:2012raa}.

\begin{figure}[ht]
\begin{center}
 \subfigure[ATLAS top mass combination]{
 \includegraphics[scale=0.32, angle=0]{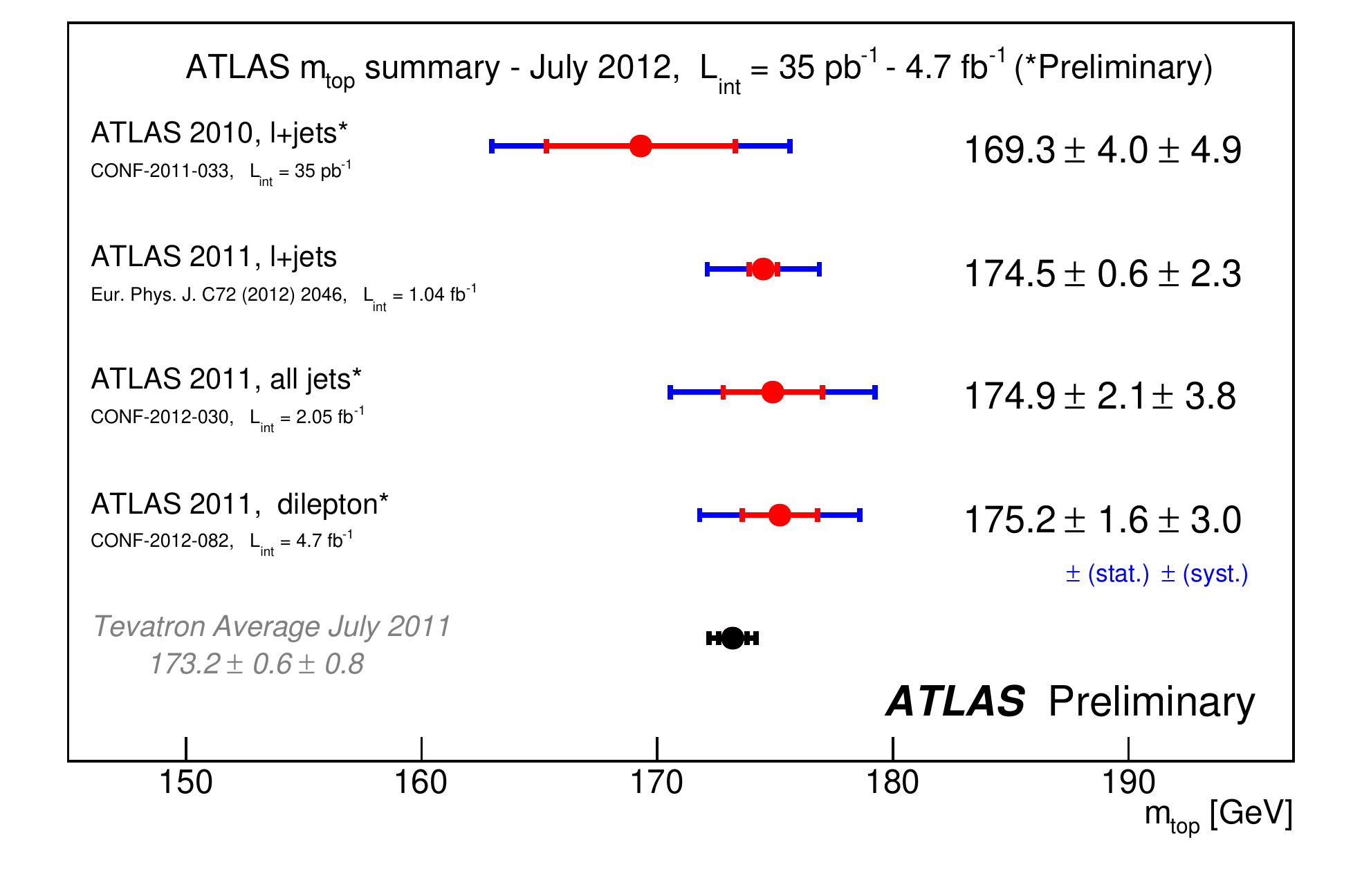}
 }
 \subfigure[CMS top mass combination]{
 \includegraphics[scale=0.32, angle=0]{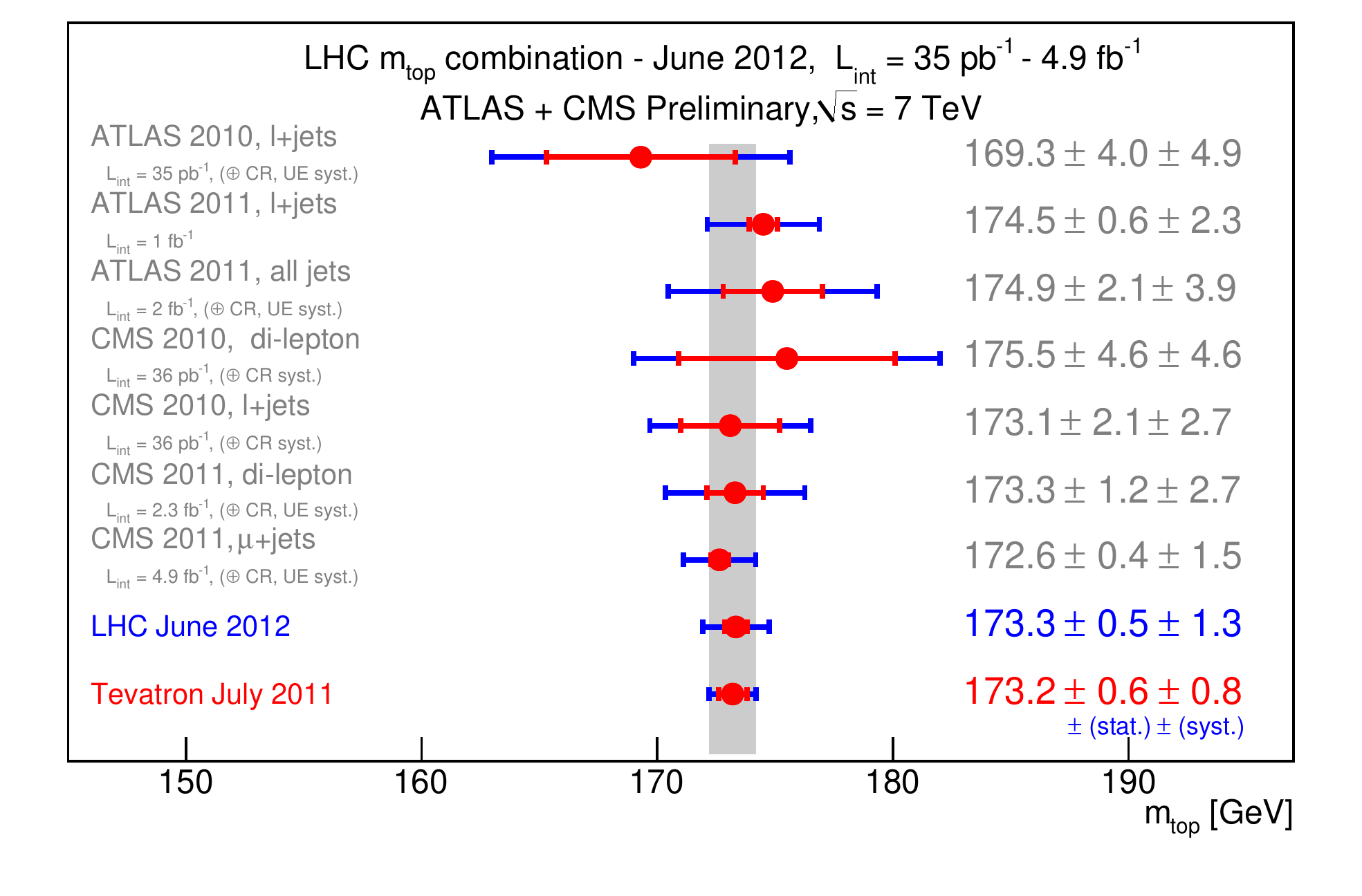}
 }
 \subfigure[ATLAS top cross section combination]{
 \includegraphics[scale=0.32, angle=0]{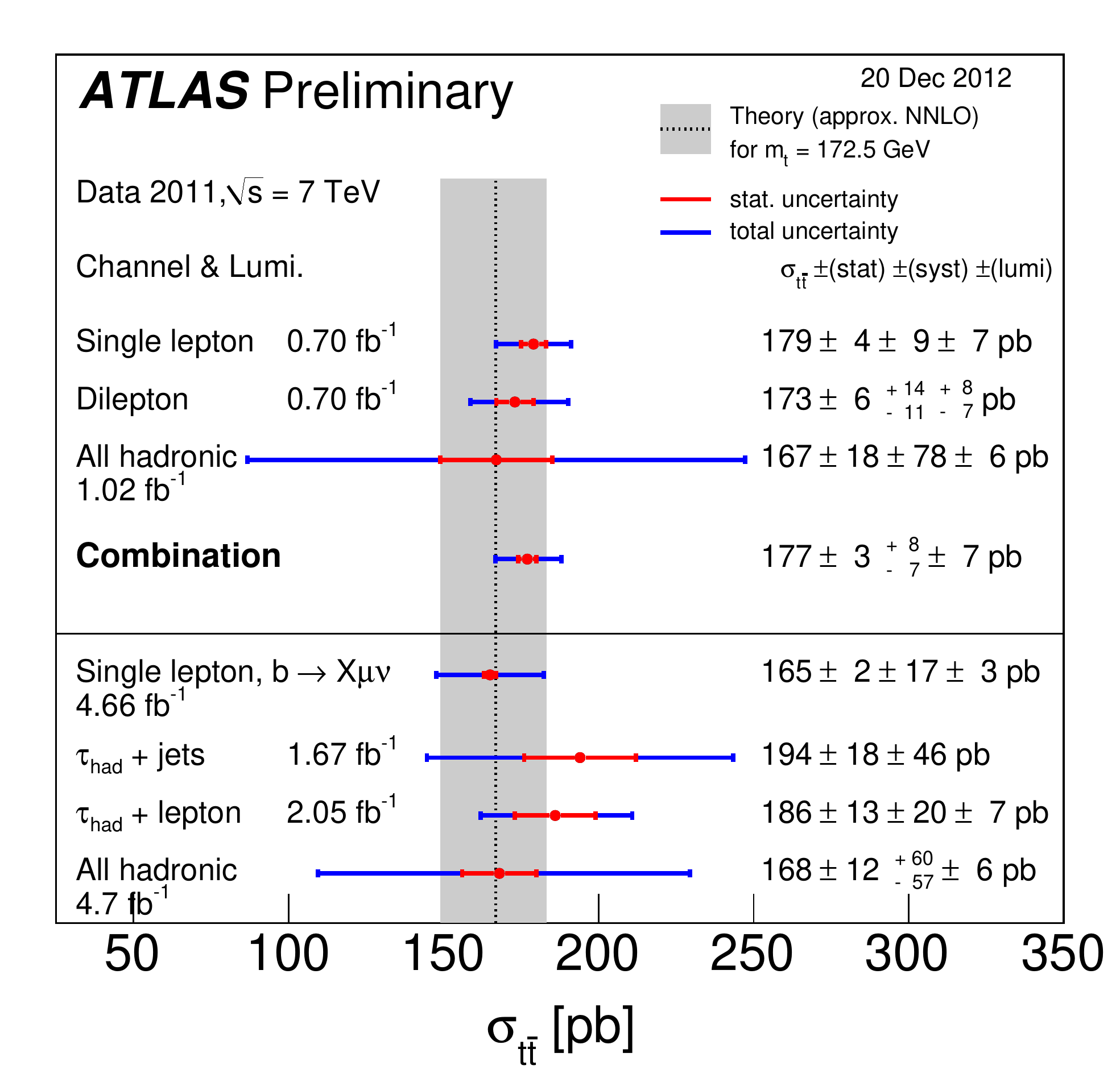}
 }
 \subfigure[CMS top cross section combination]{
 \includegraphics[scale=0.32, angle=0]{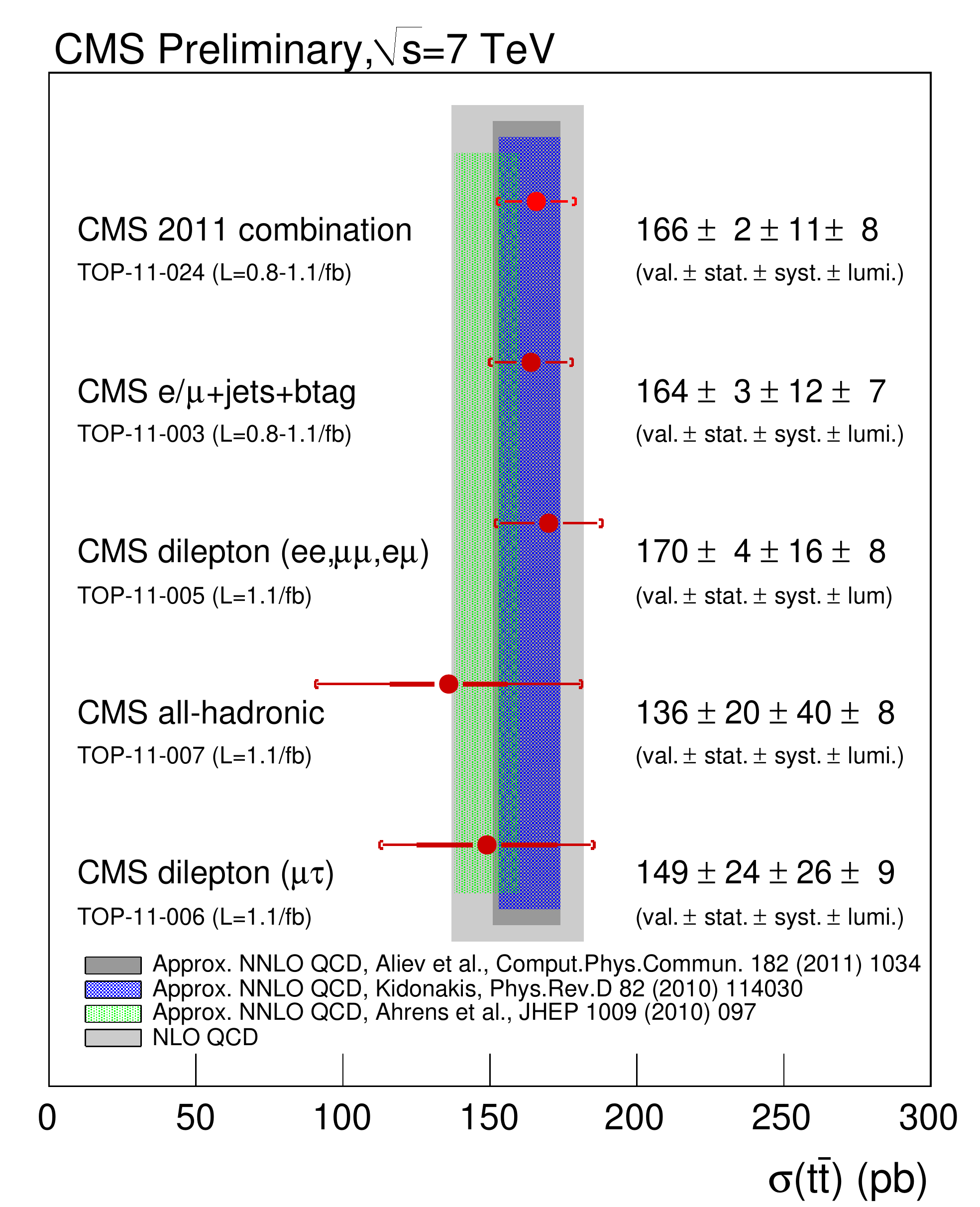}
 }
\end{center}
\caption{
Summary of measurements of the top mass and $t\bar{t}$ cross sections.  
}
\label{fig:atlas_sum1}
\end{figure}

Let us remind that the produced $t$($\bar{t}$)-quark decays almost exclusively in to a $b W$ pair. The $b$~quark hadronizes giving rise to a jet. The $W$~boson can decay {\em hadronically} into light quarks, which turn into jets, or {\em leptonically} into a pair composed by a charged lepton and a neutrino. The {\em semi-leptonic process} is defined by events in which one $W$ decays hadronically while the other one decays leptonically, i.e.
 \begin{equation}
\ttbar \rightarrow (b W) (b W) \rightarrow (b \quark \quark') (\bottom \ell \nu) 
\end{equation}
In the Standard Model  the fraction of semi-leptonic final states in  $\epem \rightarrow \ttbar$ is about 43\% compared with about 45\% fully hadronic final states.

A snapshot of the top-quark cross section and top mass measurements 
is given in Fig.~\ref{fig:atlas_sum1}.
These results indicate that:

\begin{itemize}
\item
Individual top-quark cross section measurements from ATLAS or CMS have started to reach a precision 
comparable with the corresponding 
Standard Model (SM) prediction at next-to-leading order (NLO) QCD 
(and especially those at the approximate NNLO QCD).
In many cases, individual measurements in certain channels are so far not sufficient for a stringent tests of the SM.

\item
Combined cross section measurements have reached the precision comparable with or smaller than the SM prediction.

\item
Overall experimental uncertainty is dominated by systematic uncertainties.
This continues to be true for combined results.


\end{itemize}


The first point can be obtained from the analysis of ATLAS and CMS data published during 2010-2013; the second point is a reflection of rather recent
developments that indicate the importance of combining multiple channels or results from multiple experiments.
For example, the measurements of top masses by CMS and ATLAS  have reached the relative precision of $\simeq 0.8\%$ \cite{Aaltonen:2012ra}, 
which is close to the similar Tevatron result, $0.54\%$ \cite{ATLAS-CONF-2012-095,CMS-PAS-TOP-12-001}. The uncertainties on
the measurements of $t\bar{t}$ cross section are quickly becoming comparable with NLO QCD predictions.
The combination of results is  significantly more difficult in the case of differential cross sections. 
The third point requires a deeper analysis which may help
to understand possible improvements in future measurements.

From the instrumentation point of view, top quark measurements 
require a comprehensive understanding of lepton identification, jet reconstruction, missing transverse energy and 
$b$-tagging.  These issues are discussed in great detail in ATLAS and CMS public documents, and some aspects continue to improve over time.  
But as we look ahead toward the future of the LHC, two other significant issues arise:  a dramatic increase
in pileup activity with rising instantaneous luminosity, and an appreciable production of ``highly boosted'' top quarks with $p_T \gg m_t$.

In the following sections of this paper, we discuss in detail the impact of these novelties on top quark measurements and new physics
searches, and explore possible improvements.  We divide this discussion into a first section dedicated to traditional measurements
of top quarks near their pair production threshold, and a second section dedicated to highly boosted top quarks.

Taking a somewhat longer-range view, in the third section we conduct a short survey of detector effects in the context of the ILC proposal.  There,
the promise of highly-detailed kinematic measurements must be evaluated given realistic limitations such as jet combinatoric ambiguities, 
selection of prompt leptons versus leptons from $b$-hadron decay, and measurement of $b$-jet charge.

\clearpage
\section{Top quarks near threshold}
\label{sec:sec_unboosted}
Top-quark reconstruction at low transverse energies is limited by a number of factors that determine total systematic uncertainty.
The largest instrumental uncertainties for top cross sections (total and differential), top-mass measurements 
and new physics searches involving top quarks come from the number of
sources:

\begin{itemize}
\item
Jet-energy scale, which is a dominant uncertainty for essentially all top-quark measurements. Typically, it accounts for 
$50\%$ of the overall uncertainty on cross sections.

\item
Jet-energy resolution uncertainty has a  systematic uncertainty on the top measurements
comparable with the jet-energy scale uncertainty.

\item
$b$-tagging efficiency uncertainty and mistag rate are the next largest contributors.

\item
Uncertainty on missing transverse-energy reconstruction is another important contribution.

\item
Lepton identification efficiency and reconstruction typically have the smallest uncertainties.

\end{itemize}
The above uncertainties  do not correlate with theoretical uncertainties and with 
our understanding of QCD backgrounds and other effects, which are beyond the scope of this review. 
The above discussion is sufficiently suggestive and indicates that any further progress in precision top measurements
that involve jet reconstruction  
can only come from a better understanding of low-$\ptjet$ jets (typically, $\ptjet<100$~GeV jets) and $b$-tagging.

\subsection{Fast Monte Carlo simulation}

For the detector-related top-quark studies for $pp$ collsions 
at a center-of-mass energy of $\sqrt{s}=14$~TeV, top quarks
were modeled  using the \pythia \cite{Sjostrand:2006za},  \herwig  \cite{Bahr2008} and {\tt MadGraph5} \cite{Alwall:2011uj} 
Monte Carlo (MC)  models.
The MC samples used in this article 
were processed through  a fast detector simulation
based on the  {\sc Delphes}~3.09 framework \cite{Ovyn:2009tx,deFavereau:2013fsa}
assuming the Snowmass detector geometry \cite{Anderson:2013kxz}.
The description of the MC samples are available from the Snowmass2013 web pages \cite{MCsnowmass,Avetisyan:2013dta}. The data 
are available in the ROOT \cite{root} and ProMC \cite{2013arXiv1306.6675C} formats. The latter format was
developed at Snowmass to mitigate the problems of limited storage on the web servers and limited network bandwidth.
More details about the {\sc Delphes} fast simulation for Snowmass2013 can be found in \cite{Anderson:2013kxz}. 

Jets were reconstructed with the {\sc FastJet}  package \cite{Cacciari:2011ma} using the anti-$k_T$ algorithm with distance parameter $R=0.5$. 
The simulation uses Particle-Flow (PF) jets reconstructed by clustering the four-momentum vectors of
particle-flow candidates. The particle-flow algorithm combines tracking
information with calorimeter clusters (for neutral hadrons and photons).
In case when no pileup correction is used, jets were constructed from all PF objects, irrespective their origin.
For pileup-corrected jets,  only jets that include  tracks originating from the primary vertex are used, and then they are combined
with neutral hadrons  obtained from the corresponding calibrated
electromagnetic  and  hadronic  energy.
Then, the jet-area method \cite{Cacciari:2007fd} was used to subtract pileup contribution 
\footnote{ For such simulations, we do not model out-of-time pileup.}. 
Since the pileup contribution
is negligible for charged hadrons from the primary vertex, only
the neutral contribution was corrected using the average energy density from pileup and the underlying event.

\subsection{Jets under high-pileup condition}

The future high-luminosity 
LHC plans to go beyond $\mmu\geq 100$ pileup events per bunch crossing, which will have a negative impact on 
many hadronic final-state observables.  
The most  vulnerable objects are low-$p_T$ jets and $b$-tagged jets, 
which are currently the main sources of systematic uncertainty for top-quark measurements.  
Below we will discuss this conclusion using the {\sc Delphes} simulation.

\begin{figure}[tbp]
\begin{center}
\subfigure[Jets in $t\bar{t}$ events without pileup removal]{
\includegraphics[scale=0.35, angle=0]{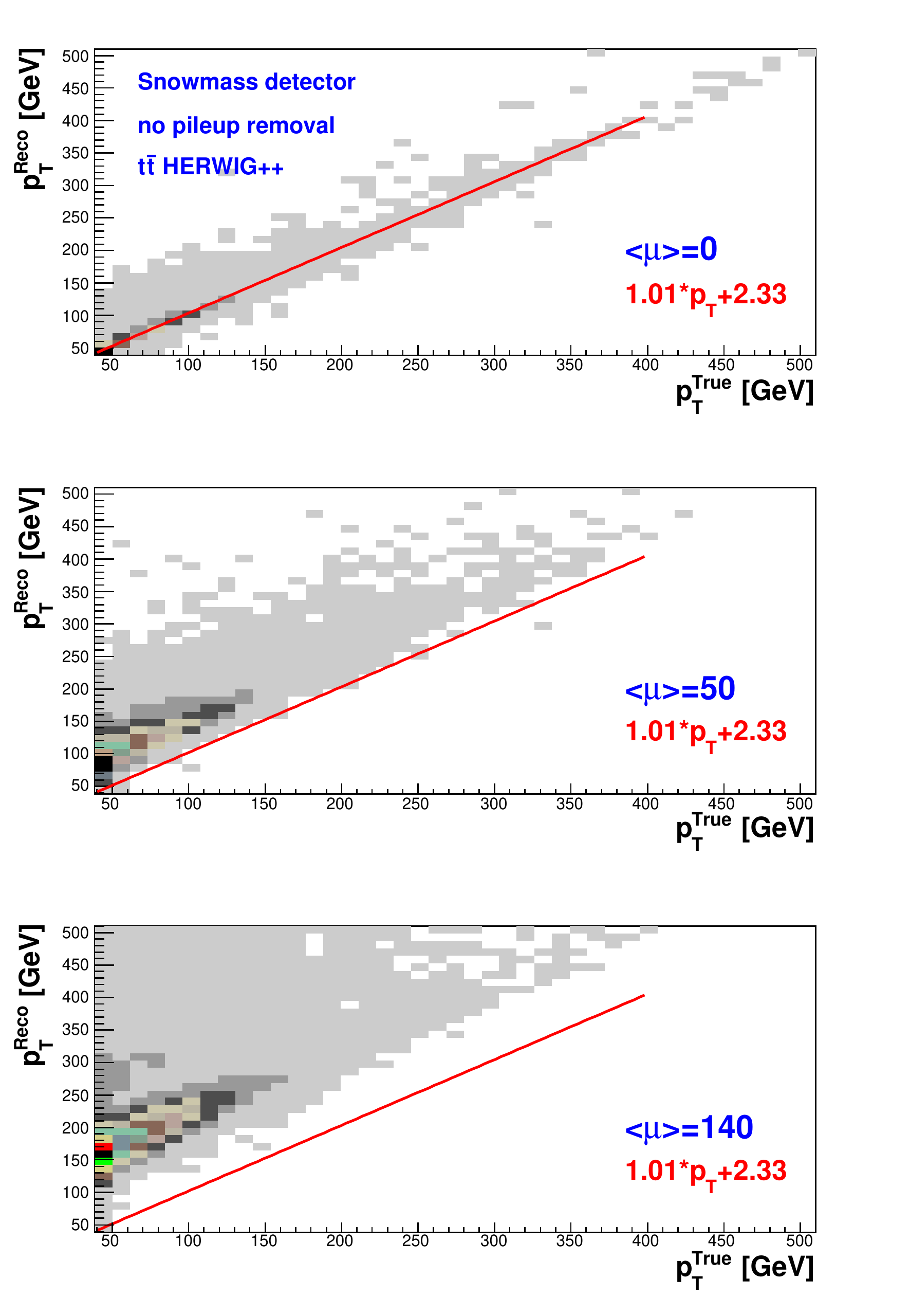}
}
\subfigure[Jets in $t\bar{t}$ events after pileup removal]{
\includegraphics[scale=0.35, angle=0]{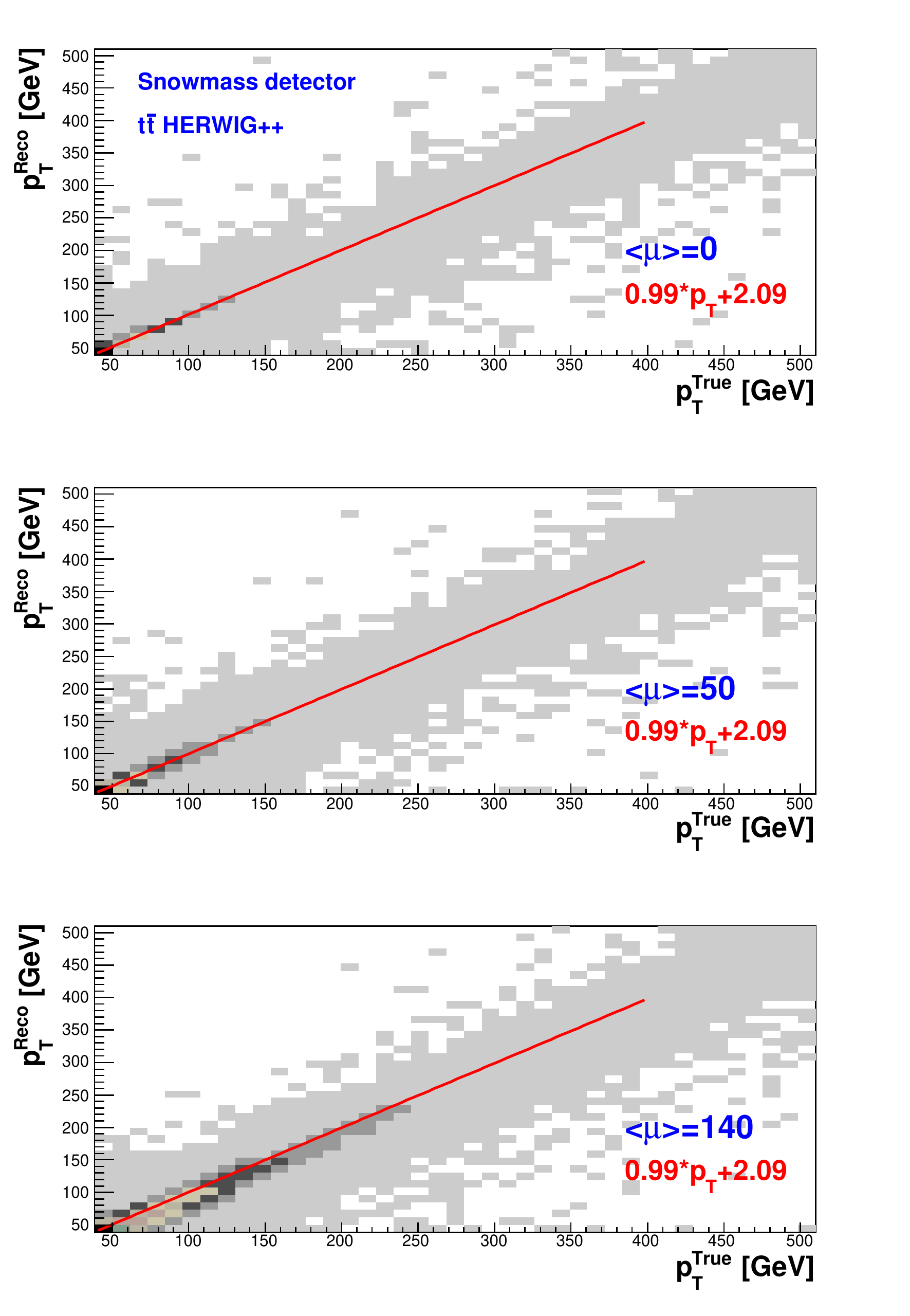}
}
\caption{
Reconstructed jet transverse momenta in $t\bar{t}$ events versus true jet transverse momenta
for (a) before and (b) after pileup removal and jet energy corrections.
Jets were reconstructed with the anti-$k_T$ jet algorithm using the size $R=0.5$.
The figures show different pileup scenarios:
$\mmu=0, 50, 140$ using the {\sc Delphes} fast simulation.  The line is a fit result using a first-order polynomial for $\mmu=0$; this line repeated for
$\mmu=50$ and $\mmu=140$ cases. 
}
\label{fig:jet_mu}
\end{center}
\end{figure}

The effects of pileup events include:
\begin{itemize}
\item
shifts in jet transverse 
energies by approximately 50 (120)~GeV for $\mmu=50$ (140), adding about one additional GeV for each pileup event;

\item
smearing of jet transverse momenta;

\item
a flux of low-$p_T$ fake jets. 
\end{itemize}

Figure~\ref{fig:jet_mu} shows the profile plot of jet transverse momenta for the reconstructed and truth-level jets for $\mmu=50$ and $\mmu=140$.
The studies were performed  without and with the pileup removal technique.
The reconstructed jets were matched to truth-level jets built using non-pileup particles.
The matching procedure was performed using a cone of size 0.1 in $\eta$ and $\phi$. 
The {\sc Delphes} simulation shows that 
a typical $30$~GeV jet from $t\bar{t}$ decays has a transverse momentum close to $60$ ($120$)~GeV for $\mmu=50$ (100) pileup events. 

Additional energy of $\sim 1$~GeV to jets from every pileup event assumes that no noise optimization is performed on the input signals. 
This is not true for realistic situations.  The pileup noise used for calorimeter cluster finding such that 
clusters with low energy significance relative to the actual pileup noise are suppressed. 
When the noise value is adjusted during clustering, then the pileup noise is significantly reduced. For example, for ATLAS, 
is about 20~GeV extra energy (on average) on top of R=0.4 jets for $\mmu=140$.

\begin{figure}[tbp]
\begin{center}
\subfigure[Jets in $t\bar{t}$ events without pileup removal]{
\includegraphics[scale=0.35, angle=0]{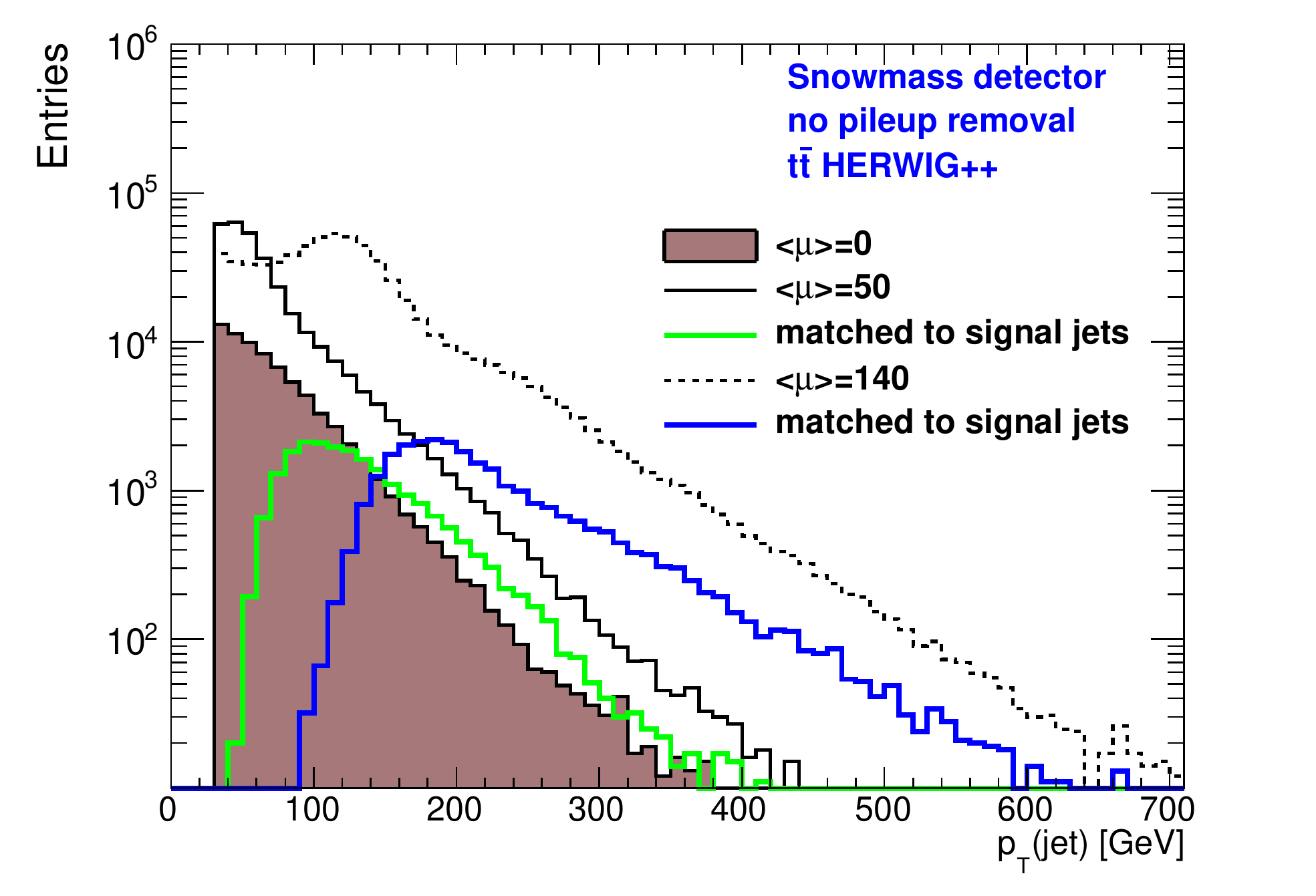}
}
\subfigure[Jets in $t\bar{t}$ events after pileup removal]{
\includegraphics[scale=0.35, angle=0]{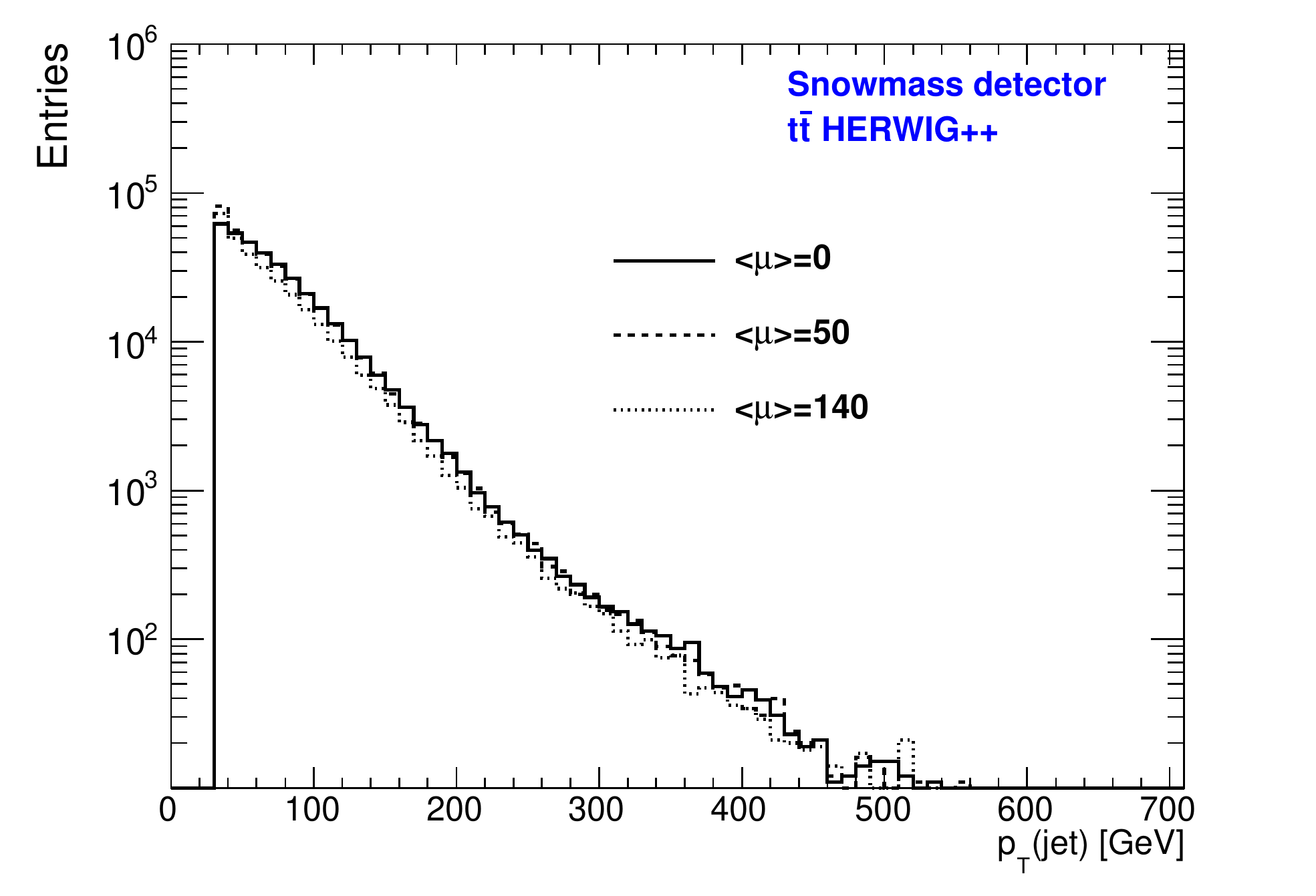}
}
\subfigure[Nr of jets in $t\bar{t}$ events without pileup removal]{
\includegraphics[scale=0.35, angle=0]{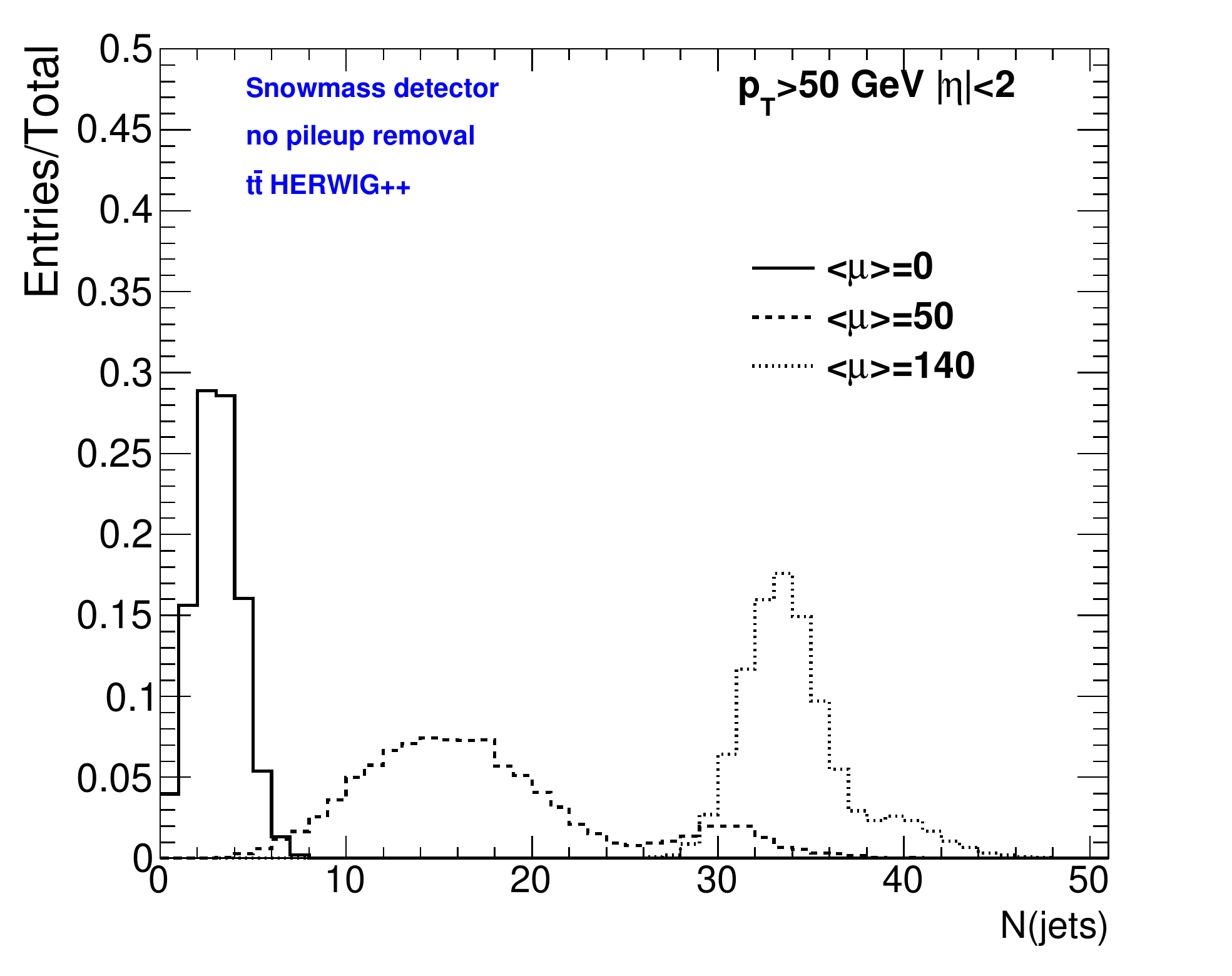}
}
\subfigure[Nr of jets in $t\bar{t}$ events after pileup removal]{
\includegraphics[scale=0.35, angle=0]{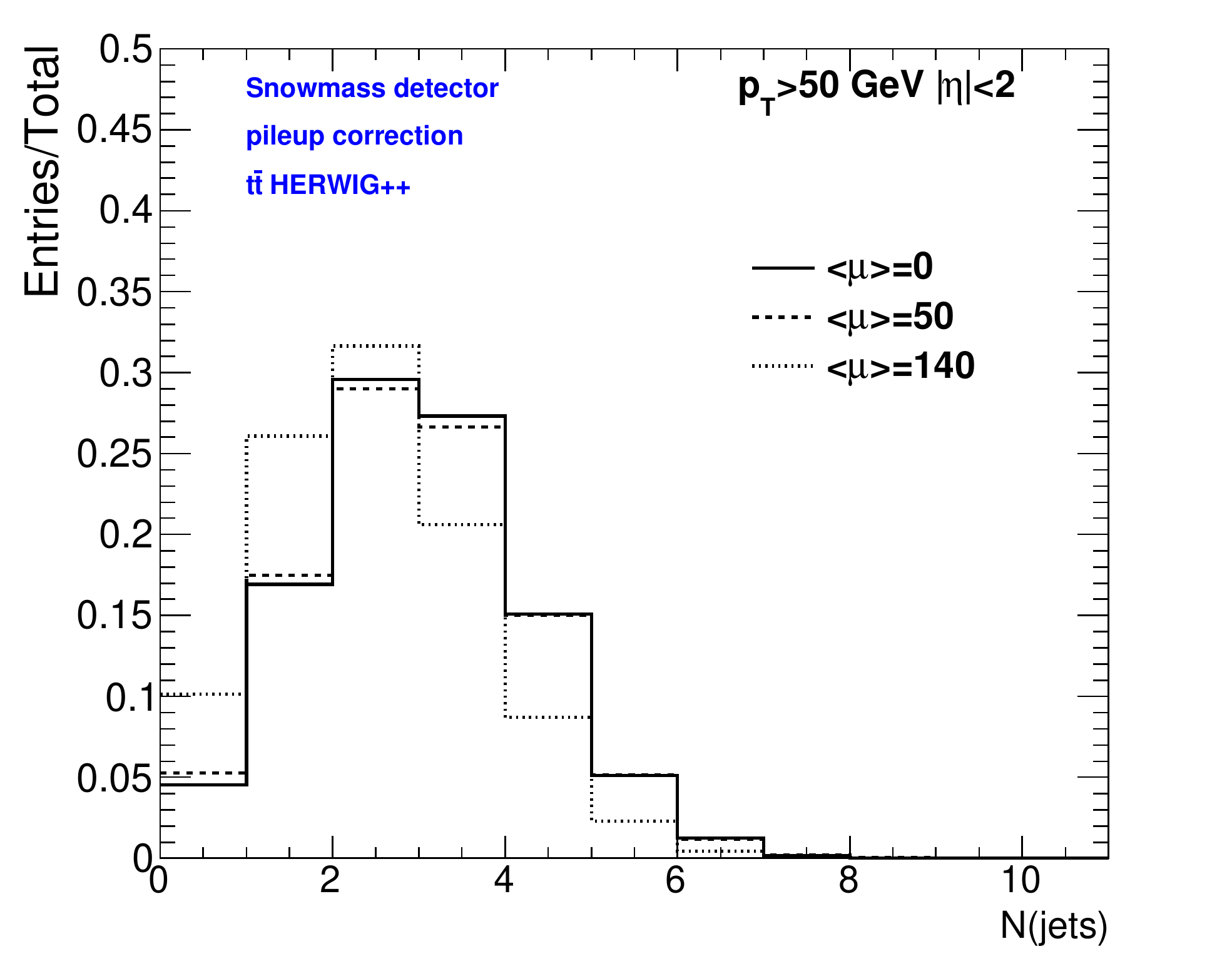}
}
\caption{
Reconstructed jet transverse momenta and numbers of jets in $t\bar{t}$ events before (a) and after (b) pileup removal.
The figures show different pileup scenarios:
$\mmu=0,50,140$ using the {\sc Delphes} fast simulation. For the transverse-momenta distribution, the cut $p_T>30$~GeV was
applied, while the jet multiplicity plots use  $p_T>50$~GeV selection. For all figures, $|\eta|<2$ was used.
}
\label{fig:atlas_mu_jetpt}
\end{center}
\end{figure}

Figures~\ref{fig:atlas_mu_jetpt}(a)-(b)  show the comparisons of jet transverse momentum distributions with and without pileup removal
based on the jet-area corrections, as it is included to the {\sc Delphes} simulation, which is closely follows the CMS approach to deal
with pileup. 
Figures~\ref{fig:atlas_mu_jetpt}(c)-(d) show the distributions for the number of jets at $p_T>50$~GeV before and after the pileup correction.
The numbers of the low-$p_T$ jets without the pileup removal can be as large as 50 for $\mmu=140$. 
The pileup reduction technique clearly leads  a substantial improvement for the average jet multiplicity, restoring
the original numbers of jets in $t\bar{t}$ events. Some discrepancy between $\mmu=140$ and $\mmu=0$ can be due 
insufficient tuning of the correction.  
Jets for the $\mmu=50$ and 140 have rather close characteristics to those shown for $\mmu=0$.   

\begin{figure}[tbp]
\begin{center}
\subfigure[PFlow-jets without pileup removal]{
  \includegraphics[scale=0.34, angle=0]{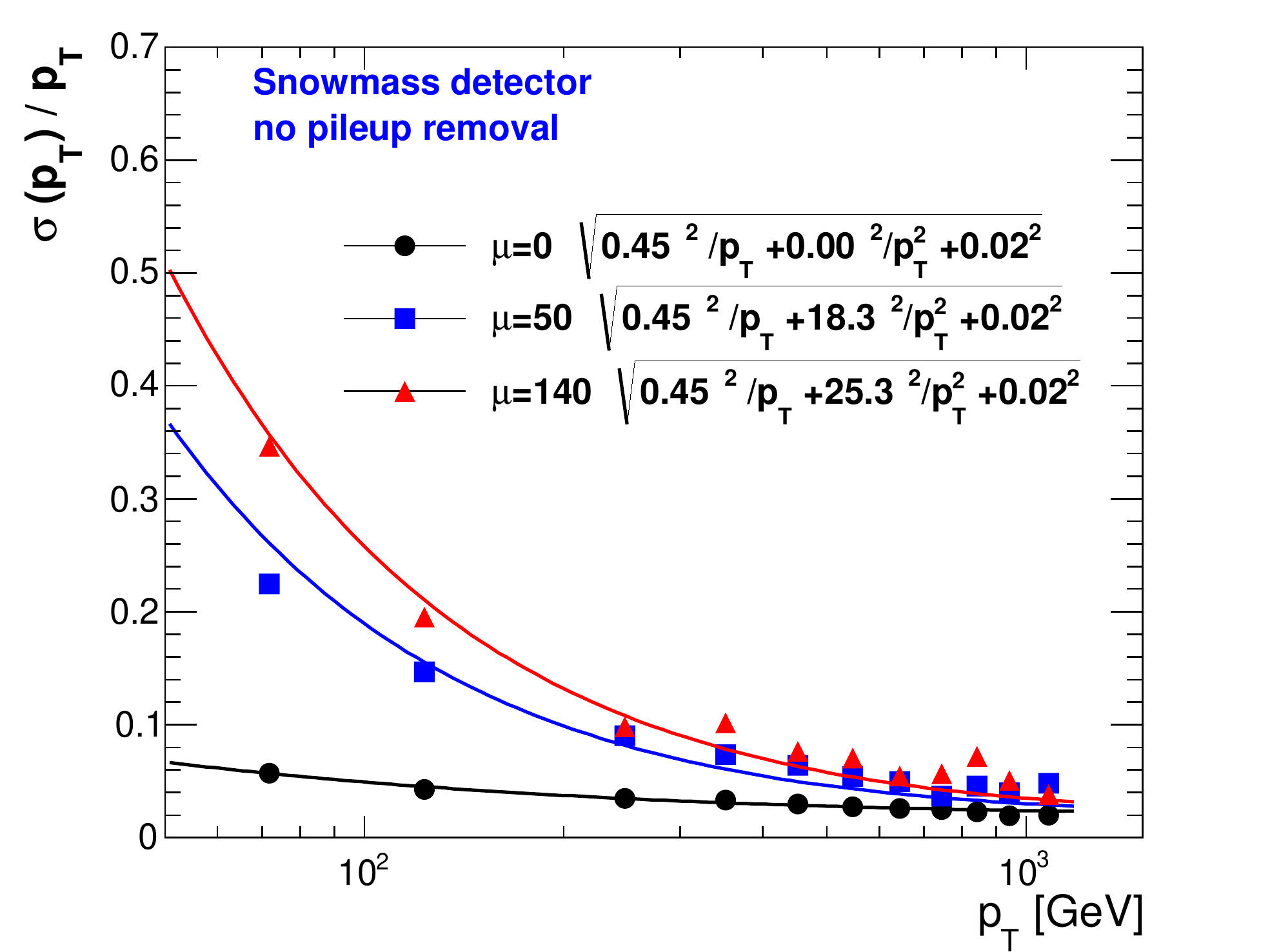}
  \includegraphics[scale=0.34, angle=0]{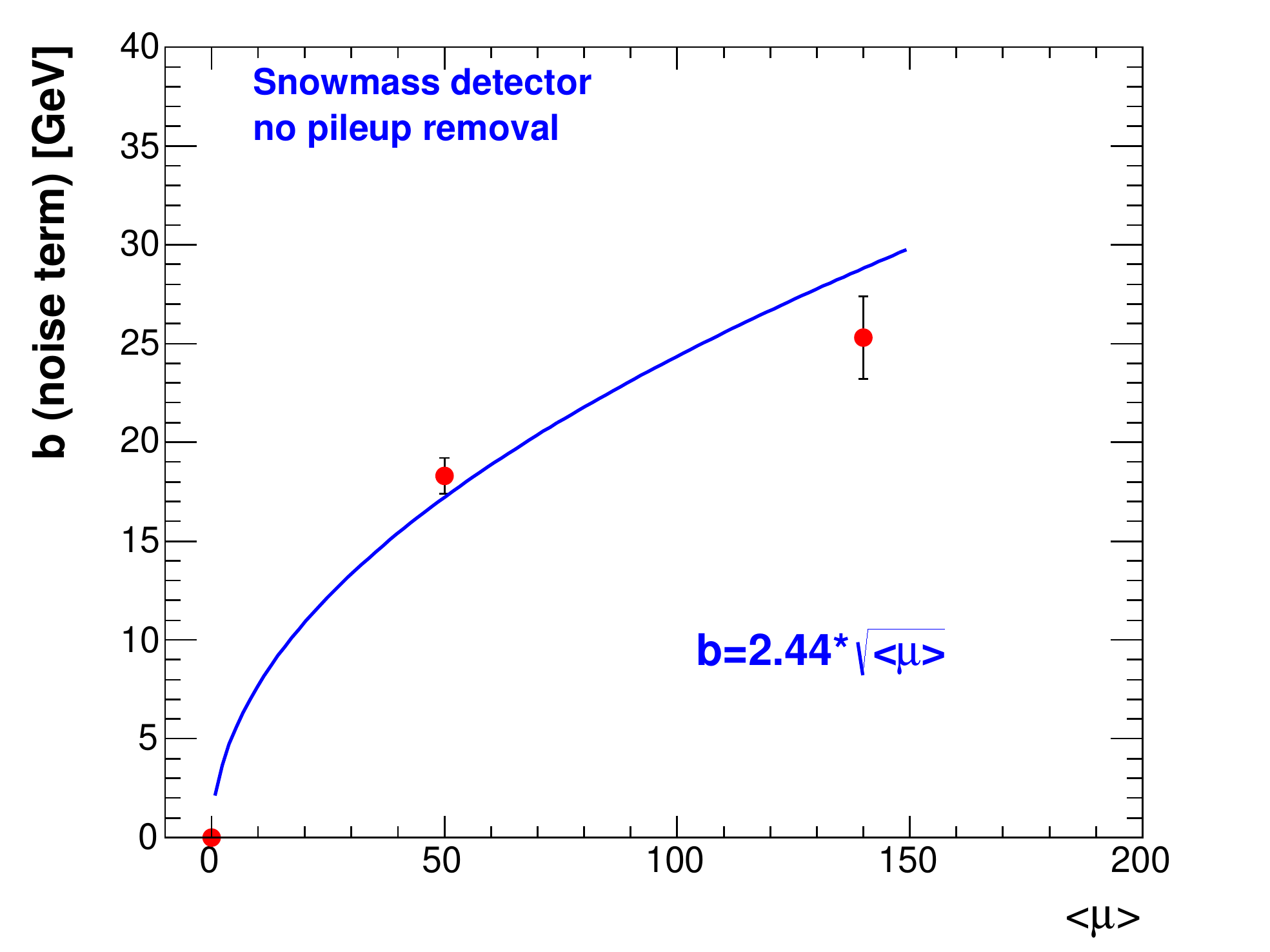}
}
\subfigure[PFlow jets with pileup removal]{
  \includegraphics[scale=0.34, angle=0]{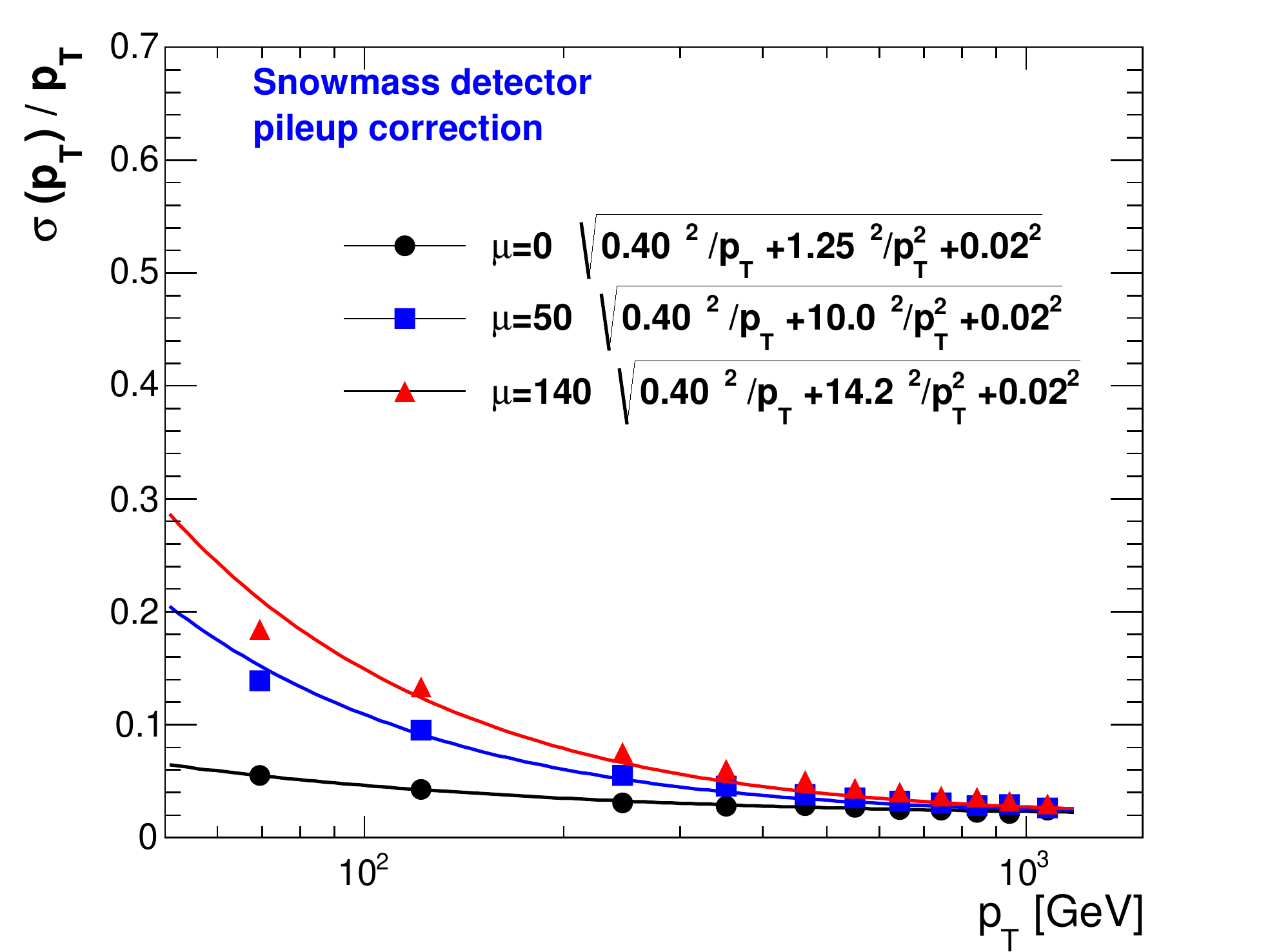}
  \includegraphics[scale=0.34, angle=0]{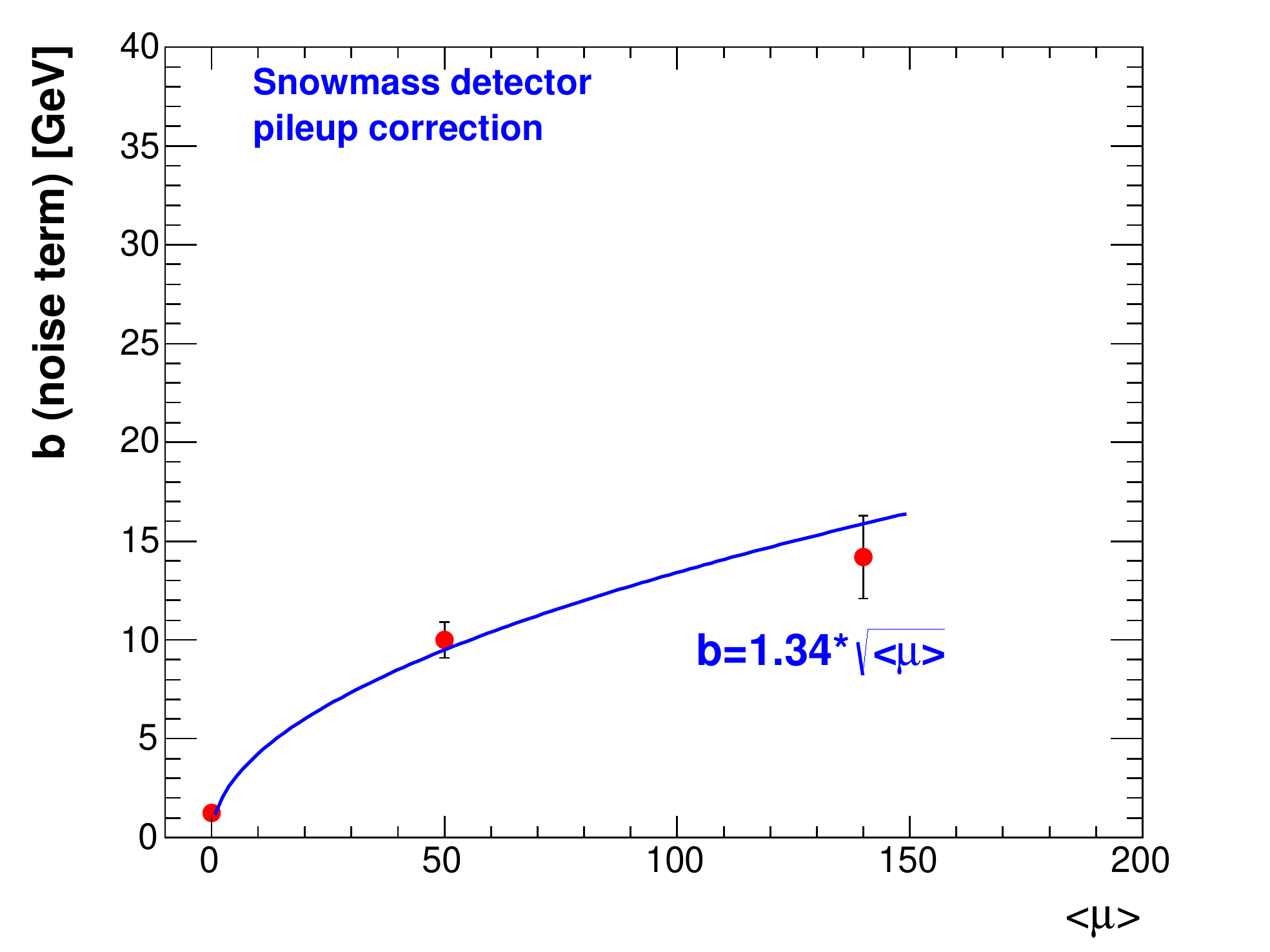}
}
\caption{
Jet energy resolution and the noise term dependence on $\mmu$  for different pileup scenarios
using the {\sc Delphes} fast simulation and the Snowmass detector. Jets are reconstructed with the anti-$k_T$ algorithm 
with R=0.5 using PFlow as inputs. (a) Without
pileup correction and (b) using the pileup correction based on the jet area correction.
}
\label{fig:jet_resolutions}
\end{center}
\end{figure}

The effect of pileup on jets  
can be studied by analyzing $p_T^{Reco}/p_T^{True}$ distributions in bins of $p_T^{True}$.
The fractional jet energy resolution can be described by the parametrization
$$
\sigma / p_T = \sqrt{a^2/p_T + b^2/p_T^2 + c^2}, 
$$
where $a$ (sampling) and $c$ (constant) terms are roughly independent of pileup \cite{Arnaez:1483523}.
The noise term, $b$, is expected to be pileup dependent.
Figure~\ref{fig:jet_resolutions} shows the fractional jet energy resolutions for the
{\sc Delphes} simulation for different pileup scenarios.
The same figure shows the dependence of the noise term obtained from the fits to the resolution as a function of $\mmu$, assuming $a$ and $c$ are fixed to the
$\mmu=0$ case. It can be seen that the pileup substantially changes the noise term which follows the $\sqrt{\mmu}$ trend.

From the above discussion, we can conclude that if identification of jets from the signal events
will be performed, a pileup subtraction technique should correct the signal jets by 200--400\%.
Such correction will increase jet uncertainties and lead to larger systematic uncertainties for jet cross
section and jet masses compared to 2010--2012 data. Methods and data-driven algorithms that are
dealing with pileup will be of much greater importance than ever before.

Let us give a quantitative model-independent estimate of the effect of pileup on top-quark measurements using PFlow jets. 
50 (140) pileup events lead to $\sim 40$ ($\sim 120$)~GeV extra energy for a 30~GeV jet.
Assuming that only the neutral component of a jet should be corrected for, 
such a correction is 
expected to be $\sim 20$ (60)~GeV for $\mmu=50$ (140). Assuming that the uncertainty on additional
energy due to pileup events is close to that for  
the jet measurement itself without pileup,
we conclude that the uncertainties on jets for the high pileup scenario is mainly driven by uncertainties on pileup corrections.
As an example, a $2\%$ jet-energy scale uncertainty for a jet measurement without pileup 
translates to a $3\%$ ($5\%)$ uncertainty for 50 (140) pileup events.  
This conclusion is still on the optimistic side, since many effects are not taken into account, such as out-of-time pileup
and uncertainty on the simulation of soft-QCD events.  The latter may invalidate the assumption that 60~GeV of energy coming from pileup
can be understood as well as the energy of a 60~GeV jet without pileup.  
Since jet reconstruction for top measurements accounts for the largest fraction of systematic uncertainty, we conclude
that systematic uncertainties for the SM top measurements will increase by a factor 1.5 (2.5) for the 50 (140) pileup scenario,
unless methods that are less reliant on jets are used.

Data-driven techniques based on $Z$+jets or $\gamma$+jets processes
may improve the assessment of the jet-energy scale compared to the above projection.
Such {\it in-situ} jet calibration will require  
a reliable reconstruction of  low-$p_T$  photons and leptons. For jets with $p_T>30$ GeV, photons and leptons should be reconstructed 
with $p_T>15-20$~GeV, which becomes challenging due to large prescale factors and effects of pileup on the isolation of such objects.  
One possible solution for a reliable usage for jets for high-precision top measurements  is to reconstruct top
quarks with relatively large $\pttop$ (for example, with $\pttop>200$), but this will reduce the available statistics. 

Another effect which is important for top reconstruction is jet energy resolution uncertainty, which can significantly 
contribute to systematic uncertainty on top reconstruction, and can have a size on the overall uncertainty  
as large as the jet-energy scale (for an example, see Ref.\cite{Aad:2012hg}). The effect of jet-energy resolution uncertainty is more difficult to assess using a 
model-independent consideration. The jet-energy resolution and associated uncertainties are expected to increase for high-luminosity runs.

\subsection{Top reconstruction under high-pileup conditions}
\label{sec:toprecPU}

Top reconstruction was performed using a procedure similar to the one discussed in \cite{ATLAS-CONF-2011-120}. An event selection was applied
by requiring at least four jets with $p_T >25$~GeV and $|\eta|<$ 2.5. At least one of the jets must be tagged by a $b$-tagging algorithm and there must be
at least two un-tagged jets in the event. We assume that the event contains at least one hadronically decaying top quark and that we can resolve two jets
originating from the $W$-boson and one jet originating from the $b$-quark.

The procedure starts by building up a list of $W$-boson candidates by pairing light jets in the event.  Combination with an invariant mass that falls
outside of a 60~GeV wide mass window around the $W$-boson mass are discarded. The candidate $W$-bosons jet pairs are combined with a $b$-tagged jet to
produce a list of candidate reconstructed top quarks. The combination with the highest $p_T$ is assumed to be the correct combination. It's important to
recognize that this procedure is biased towards selecting high-$p_T$ top candidates.

\begin{figure}[tbp]
\begin{center}
\subfigure[]{
  \includegraphics[scale=0.35, angle=0]{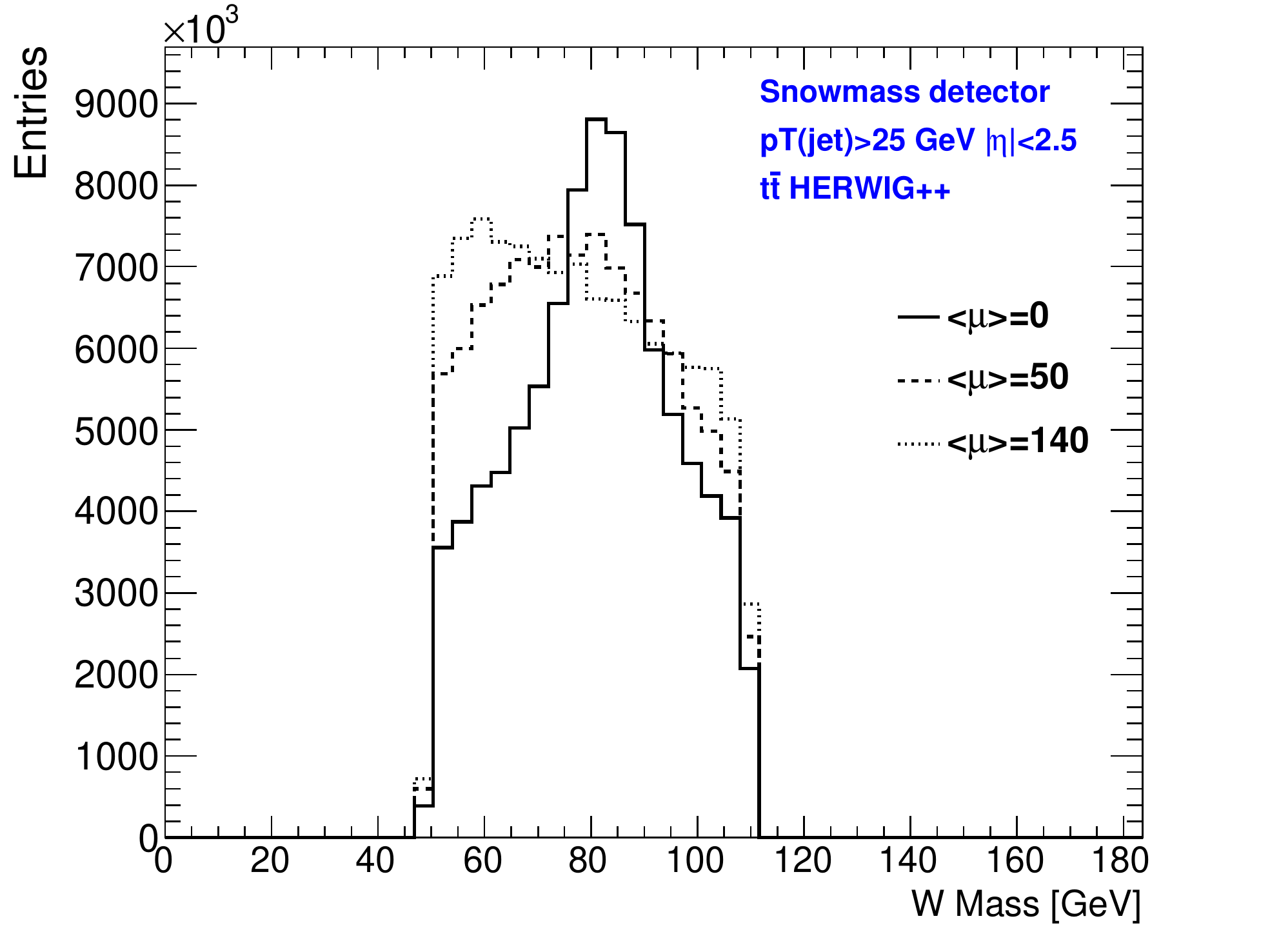}
}
\subfigure[]{
  \includegraphics[scale=0.35, angle=0]{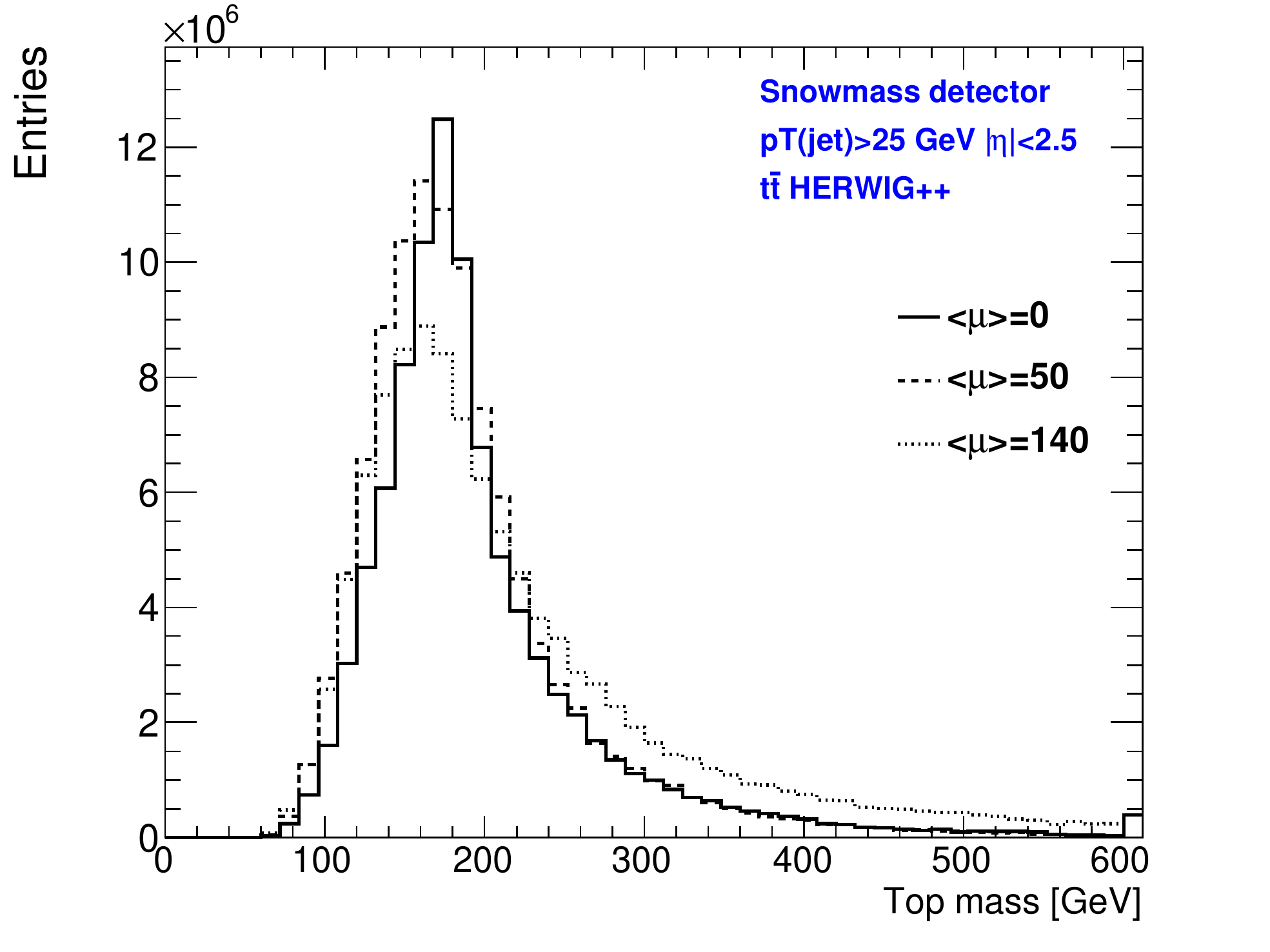}
}
\caption{Reconstructed $W$-boson and top quark masses for the three pileup scenarios.  Both distributions show a trend of lower reconstructed masses as
the number of pileup interactions increase.
}
\label{fig:topmm1}
\end{center}
\end{figure}

The distribution of reconstructed $W$-boson masses can be seen in Fig.~\ref{fig:topmm1}(a). As the number of pileup interactions increase, there is a
degradation in mass resolution of the light jet pairs and a systematic shift towards lower invariant mass values. Similarly, the mass of the reconstructed
top quark, shown Fig. \ref{fig:topmm1}b, shows a similar trend towards lower reconstructed masses as the number of pileup interactions increase.

\begin{figure}[tbp]
\begin{center}
\subfigure[]{
  \includegraphics[scale=0.35, angle=0]{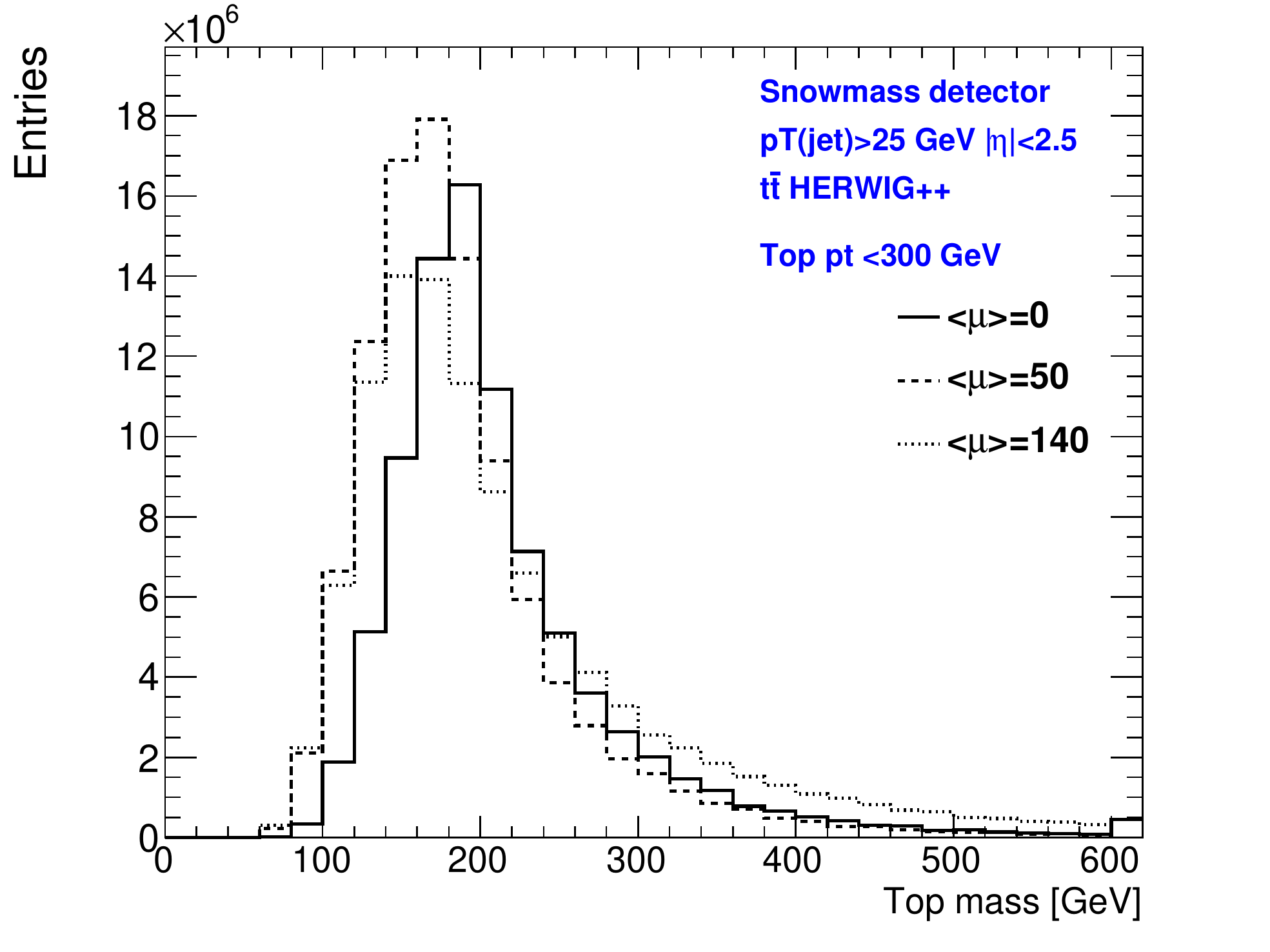}
}
\subfigure[]{
  \includegraphics[scale=0.35, angle=0]{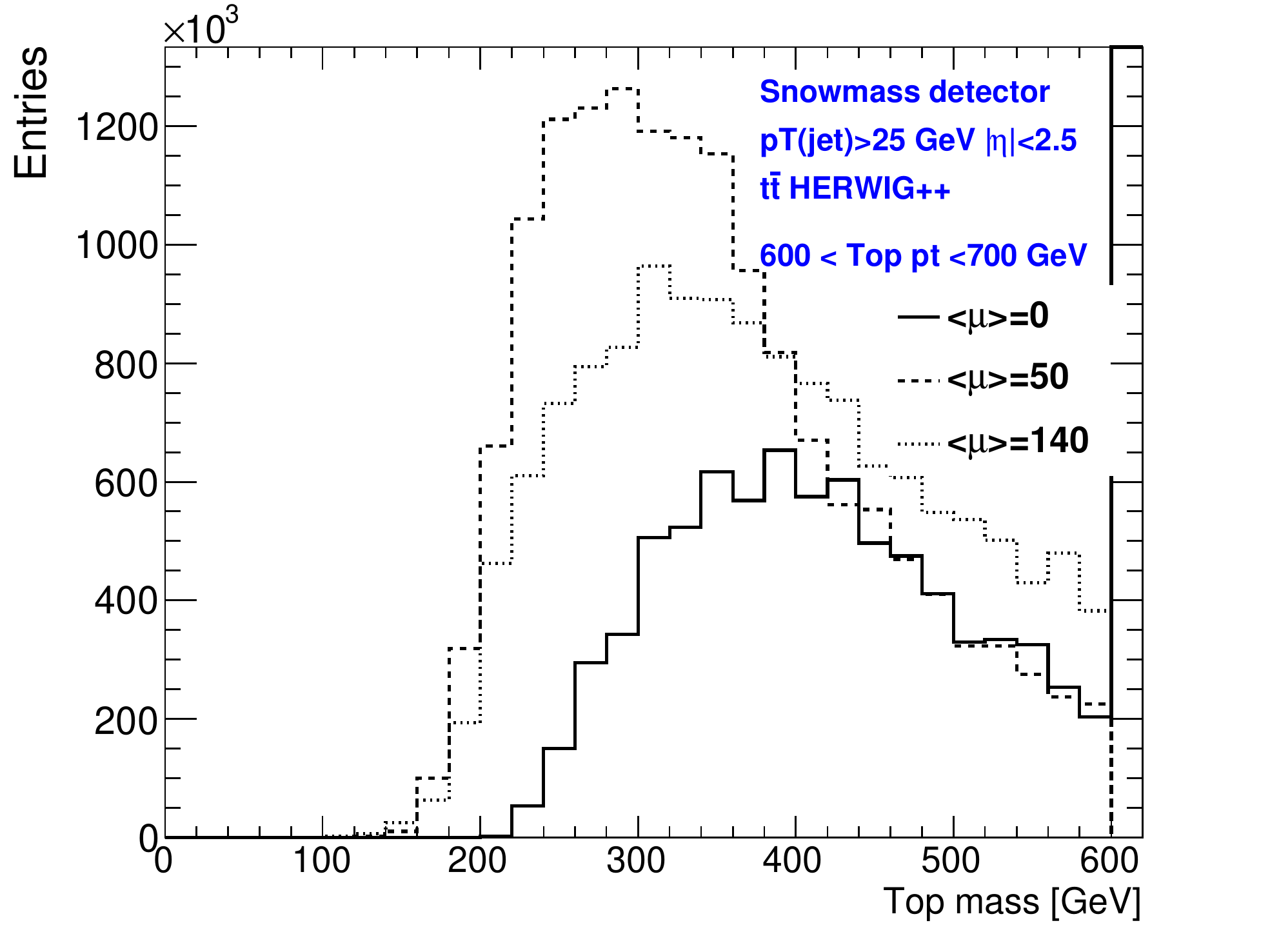}
}
\caption{Reconstructed top quark masses for a low and high $\pttop$ range.  For the low $\pttop$ range (a), the top quark reconstruction technique outlined in
Sec. \ref{sec:toprecPU} perform similarly for all three pileup scenarios.  For the higher $\pttop$ range, 
there is a significant shift in the central values
of the reconstructed top quark mass that is dependent on the number of pileup interactions.
}
\label{fig:topmm2}
\end{center}
\end{figure}

Figure~\ref{fig:topmm2} shows the reconstructed top quark masses for low and high $\pttop$ ranges.
For high-$\pttop$, the trijet mass is shifted with respect to the nominal top mass by 200 GeV (on average).  This indicates both a large contribution from
combinatoric background and the onset of the highly-boosted regime, where two or more quarks from a top decay are often merged into a single jet.  The mass also depends on the pileup scenario, trending to lower values for higher pileup and more aggressive subtractions.  Both the mass-shift and the pileup dependence indicate challenges that have to be faced for reconstruction of high-$\pttop$ using the standard approach.


\subsection{Summary}

Although different techniques to deal with large pileup 
still need to be assessed, a preliminary conclusion can already be drawn from these
studies. For the 2010--2012 experimental analyses, uncertainties in jet resolution, jet energy
and $b$-tagging were the largest contributors to the systematic uncertainty on top-related measurements.
For high-luminosity LHC runs with 50 or 140 pileup events per crossing, jet uncertainties can become only
more important.  We summarize our main conclusions below:

\begin{itemize}

\item
High-luminosity runs will bring us to the regime in which uncertainties 
on our understanding of pileup corrections for low- and medium-$p_T$ jets  will be the dominant factor, rather than instrumental 
uncertainties for signal jets as this was in the past.
Substantial energy corrections to the signal jets and additional contribution to jet energy resolution
will lead to additional uncertainties in detector modeling compared to 2010--2012 jet studies.
We expect that jet-energy scale uncertainty will increase by a factor two or more if no {\it in-situ} jet calibration
is used.
Data-driven techniques may improve the assessment of the jet energy scale compared to this projection. However, such techniques become
increasingly challenging since they require a reliable reconstruction of low-$p_T$  photons and leptons.
Jet-related uncertainties will also increase due to an increase of jet-energy resolution.


\item
For $\mmu > 100$, it is unlikely there will be a single top mass or $t\bar{t}$ cross section measurement  
based on low/medium-$p_T$ jets that will improve over the 2010--2012 measurements.  A combination of
multiple measurements by CMS and ATLAS should help to substantially ameliorate the growth of the uncertainties.  But, for
example, reaching the level of the theoretical cross section uncertainty will be highly challenging.
To reduce future systematic uncertainties at the LHC relative to their present-day levels, it may be necessary
to pursue less pileup-sensitive methods (e.g., relying more heavily on leptonic, semileptonic $b$ decays, and/or tracker-based observables).

\item
All searches that require a good understanding of low-$\pttop$  top quarks 
or low-$p_T$ jets will also be severely affected  by the new pileup environment.
These include various searches for new physics, as well as SM processes such as
$Ht\bar{t}$.


\item
A standard trijet mass for top reconstruction strongly depends on $\pttop$ due to combinatoric ambiguities and jet merging. 
For $\pttop>$700~GeV, the peak position is at 300--400 GeV assuming the same jet transverse momentum cuts as for low-$\pttop$ measurements.
This may limit our ability to identify top quarks at such large $\pttop$ using the traditional approach.

\end{itemize}


Because of the points made above, we conclude that the high-luminosity $pp$ collision runs with $\mmu > 100$ are unfavorable for 
high-precision SM measurements based on jets with $\ptjet$ below 100~GeV.
This conclusion will affect $t\bar{t}$ measurements near the threshold, single-top studies, studies
of processes with associated top production, as well as  searches for new physics 
that require detection of jets from low-$p_T$ tops.

\clearpage
\section{Highly-boosted regime}
\label{sec:sec_boosted}
%
%

A primary focus of the LHC is searches for particles produced
beyond TeV-scale energies.
However, the reconstruction of top quarks at very large transverse momenta is challenging.
The decay products become collimated, leading to failures of lepton isolation when the
$W$ decays leptonically, and substantial
jet overlaps when the $W$ decays hadronically.
$b$-tagging also becomes inefficient at large $p_T$.

\begin{figure}[tbp]
\begin{center}
\subfigure[Anti-$k_T$ R=0.5]{
\includegraphics[scale=0.52, angle=0]{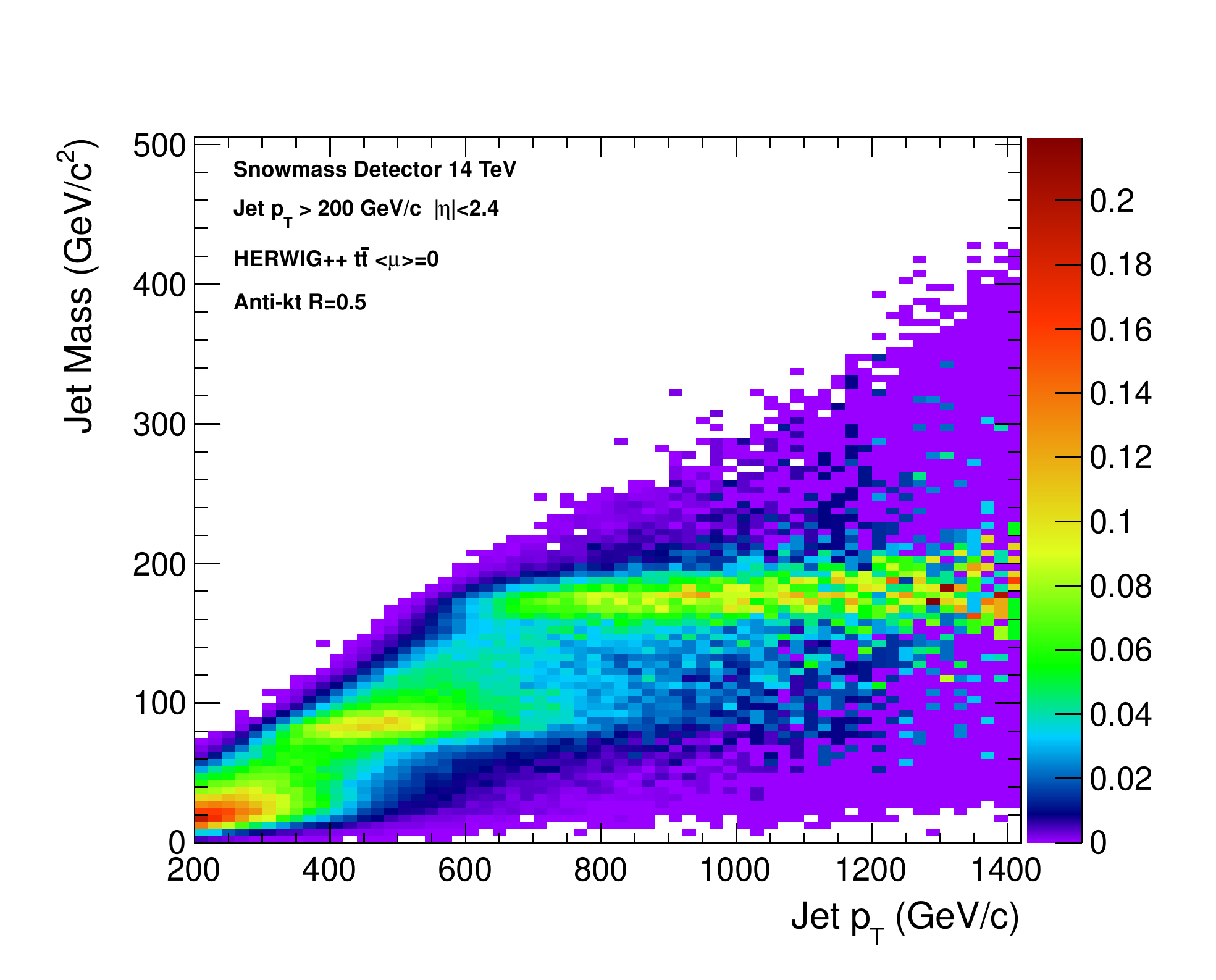}
}
\subfigure[Cambridge/Aachen R=0.8]{
 \includegraphics[scale=0.52, angle=0]{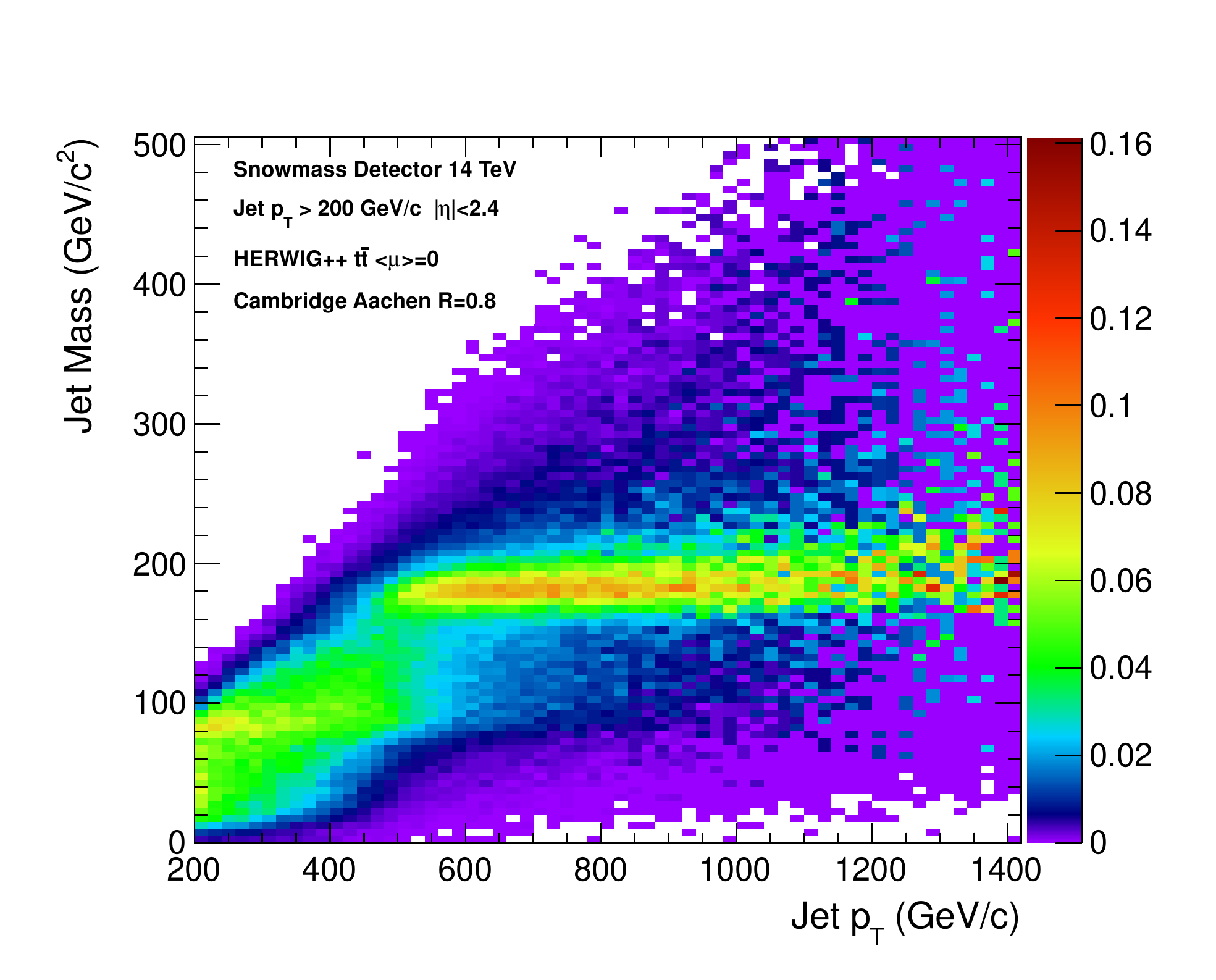}
 }
\end{center}
\caption{
Jet mass vs $\pt$ for two different algorithms and jet radii, in fully-hadronic $t\bar t$ events (without pileup) at $\pttop \gsim m_t$. The jet $\pt$ spectrum has been reweighed to be flat.
}
\label{fig:jetmass_2D}
\end{figure}


Top quarks that are so energetic that all of their decay products can fit into a standard-radius
LHC jet ($R=$0.4--0.6) are often called ``highly-boosted,'' or simply ``top-jets.''  Complete containment occurs for the 
majority of top decays when the transverse momentum exceeds roughly $4\times m_t$, or 700~GeV, as illustrated
in Figure~\ref{fig:jetmass_2D}.  Standard top
reconstructions then become impossible, and specialized techniques must be employed.
For leptonic decays, these include modified definitions of lepton isolation~\cite{Thaler:2008ju,Agashe:2006hk,Rehermann:2010vq,ATL-PHYS-PUB-2009-081}.
For hadronic decays, studies of detailed jet properties such as jet masses, jet shapes, and
jet substructure then become useful\cite{Agashe:2006hk,Lillie:2007yh,Butterworth:2007ke,Almeida:2008tp,Almeida:2008yp,Thaler:2008ju,
Kaplan:2008ie,Brooijmans:2008,Butterworth:2009qa,Ellis:2009su,ATL-PHYS-PUB-2009-081,CMS-PAS-JME-09-001,Almeida:2010pa,Thaler:2010tr,Hackstein:2010wk,Chekanov:2010vc,Chekanov:2010gv}.  In both cases, care must be taken to combat the
overwhelming QCD backgrounds, including lepton production in heavy flavor jets.

The crossover from threshold production into the highly-boosted regime is quite broad, and there is also an intermediate
``semi-boosted'' range between approximately (2--4)$\times m_t$, or 350--700~GeV.  In this regime, 
some or all of the top decay products may be merged, depending sensitively on the decay kinematics
and the radii used for jet reconstruction and isolation.  For some studies, it may be more effective
to simply contain all of the top decay products into a single ``fat jet'',  with a radius of 1.0 or larger.
Techniques similar to those useful in the highly-boosted regime can then be applied, though careful
attention must be payed to the effects of diffuse soft radiation such as pileup.
(For example, the {\sc HEPTopTagger}~\cite{Plehn:2009rk,Plehn:2010st} incorporates
filtering~\cite{butterworth-2008-100}, and is therefore well-adapted to this purpose.)

\begin{figure}
\begin{center}
\includegraphics[scale=0.38, angle=0]{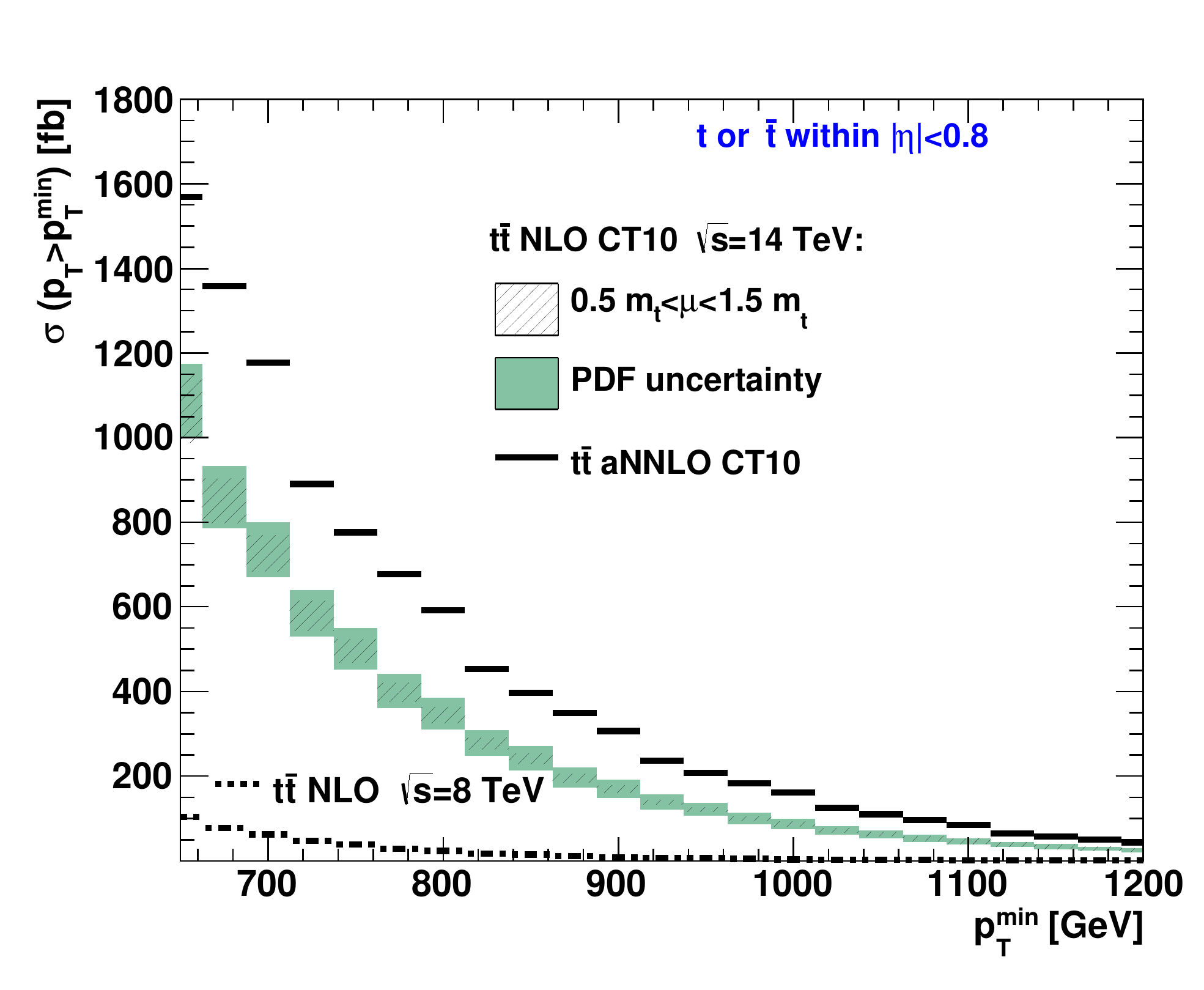}
\end{center}
\caption{
The NLO and approximate NNLO (aNNLO) cross sections
for the number of top quarks in the $t\bar{t}$ process as a function of transverse momenta
for  $|\eta|<0.8$.
The hatched area shows the renormalization scale uncertainty for NLO, while
the filled green area shows the PDF uncertainty.
The dashed line shows the NLO cross section for $\sqrt{s}=8$~TeV. See Ref.~{\protect \cite{Auerbach:2013by}} for detailed discussion. 
}
\label{fig:nlo}
\end{figure}

So far, the recent runs of the LHC have begun to probe into the semi-boosted region, using a combination of traditional
and fat-jet techniques (see \cite{Chatrchyan:2012ku,Aad:2012raa} for examples). 
True top-jets have not yet been accumulated with meaningful statistics,
but sizable samples are expected after the energy upgrade.  Figure~\ref{fig:nlo} shows the NLO and approximate NNLO cross sections
for QCD production of top pairs
in $pp$ collisions at $\sqrt{s}=14$~TeV \cite{Auerbach:2013by}.
For $\pttop>0.7$~TeV, 
hundreds of thousands of top quarks from QCD alone will be produced, 
assuming an integrated luminosity of $300$ fb$^{-1}$.
A possible future run at $\sqrt{s}=33$~TeV with comparable luminosity would further increase these rates by orders of magnitude,
and substantially extend the energy reach.  (For example, hundreds of $t\bar t$ pairs with $\pttop>3$~TeV could be observed.)
For the foreseeable future, the LHC will be in a unique position to undertake studies of 
tops produced at such high energies.

A large number of useful tools are now available for studying top-jets in the highly-boosted and semi-boosted regimes.  In the present
study, we will not undertake a review of these.  (Interested readers can learn more from the BOOST conference proceedings~\cite{Abdesselam:2010pt,Altheimer:2012mn}, where detailed descriptions and systematic comparisons of different techniques were undertaken.)  Instead, we will focus on only a small handful of the well-known methods for dealing with fully-hadronic decays, and attempt to gain some understanding of their robustness under the extreme conditions of high pileup and high energy.  We work exclusively in the highly-boosted regime, which is the most conceptually distinct from the traditional approaches described in the previous section.  However, we do note that specialized treatments of semi-boosted tops would also be interesting to study under high pileup conditions, as would highly-boosted semileptonic decays.  A dedicated understanding of possible optimization of high-$p_T$ $b$-tagging would also be very useful, but would likely require much more detailed studies using full detector simulations.


\subsection{Jet mass and area-based pileup subtraction}

\begin{figure}[tbp]
\begin{center}
 \subfigure[Jet masses]{
 \includegraphics[scale=0.32, angle=0]{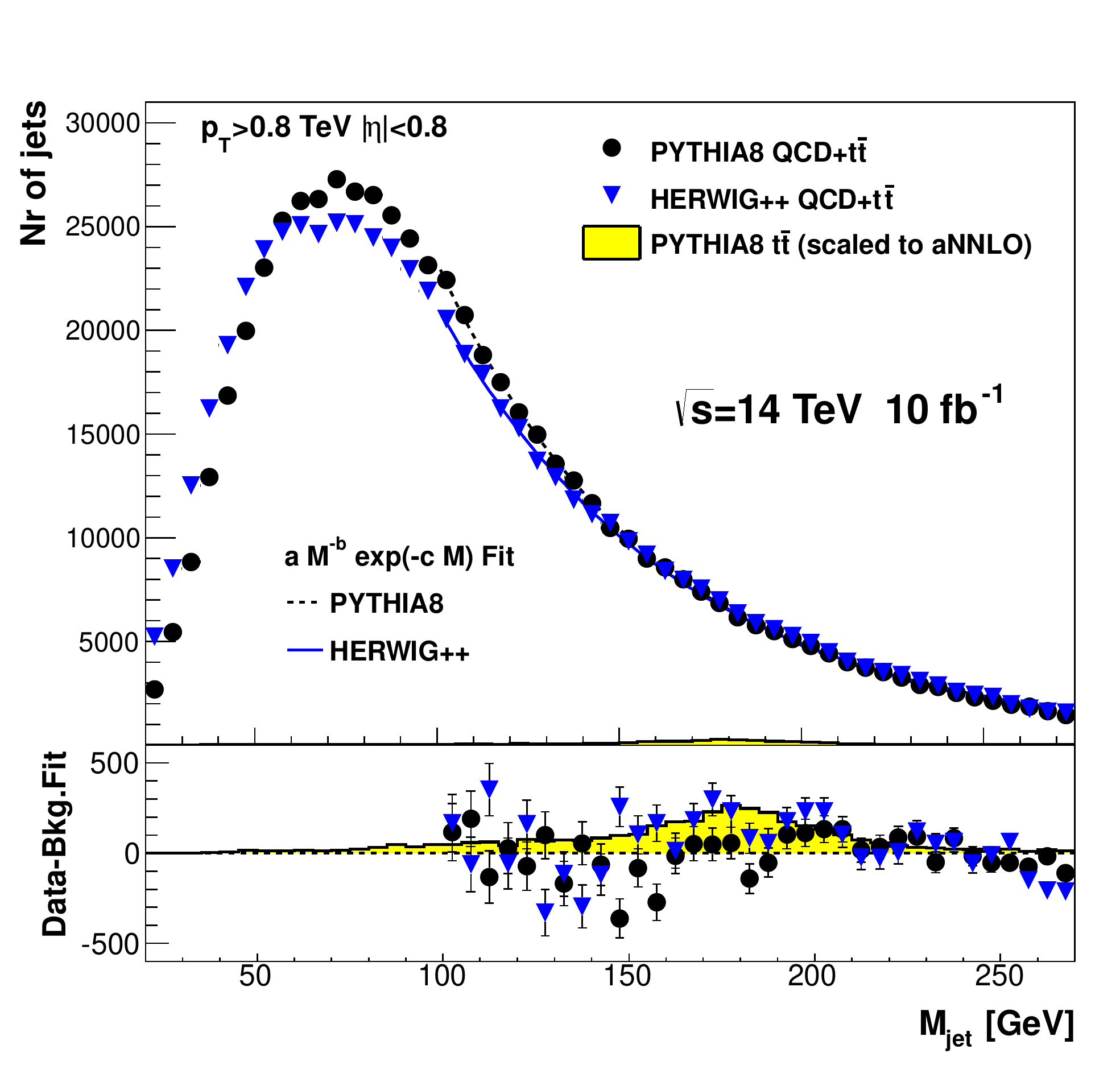}
 }
 \subfigure[Jet massses after $b$-tagging]{
 \includegraphics[scale=0.32, angle=0]{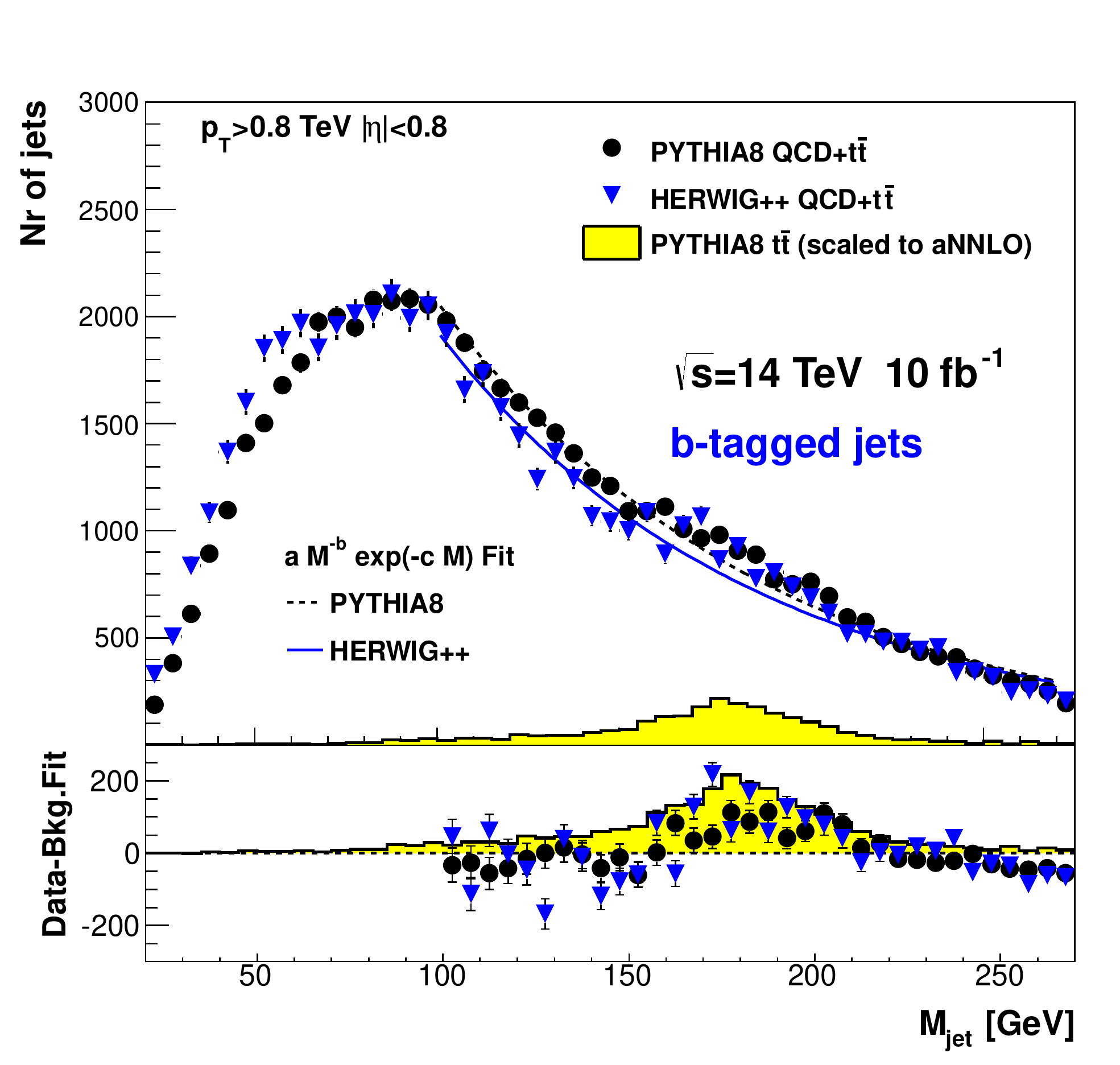}
 }
\end{center}
\caption{
(a) Expectations for the  jet mass distributions using \pythia and \herwig after the
fast detector simulation. For the simulation, top quarks were
added to light-flavor jets.  The $t\bar{t}$ component was scaled to the approximate NNLO normalization
\protect\cite{Kidonakis:2010dk,*Kidonakis:2012rm}.  The QCD dijet background
was scaled to the NLO inclusive jet cross section estimated with the NLOjet++ program {\protect \cite{Catani:1996vz,Nagy:2003tz}}.
A $\chi^2$ fit was performed using the background function $a\cdot \mjet^{-b} \cdot \exp{(-c\cdot\mjet)}$
in the mass range $100<\mjet<270$~GeV.
The \pythia expectation with the normalization from
the aNNLO for $t\bar{t}$ was scaled by a factor two.
(b) The same distribution using the $t\bar{t}$ signal yield predicted by the aNNLO
after applying the $b$-tagging for background and top jets.
The fit quality using the background function is $\chi^2/$ndf=2.7 for (a) and $\chi^2/$ndf=3.5.
}
\label{fig:jetmass}
\end{figure}

We start by considering the simplest observable sensitive to internal jet kinematics, the jet mass.  
As before, we use the {\sc Delphes} \cite{Auerbach:2013by} fast detector simulation, and we reconstruct highly-boosted tops using the standard jet anti-$k_T$ (R=0.5)
algorithm. 
Figure~\ref{fig:jetmass} shows the particle-flow masses for randomly-chosen jets with $\ptjet>0.8$~TeV in combined $t\bar{t}$ and QCD dijet processes at the 14~TeV LHC.  For this simple comparison, pileup was not simulated.
It can be seen that even the measurement of individual jet masses would by itself be on the verge of observing an SM boosted hadronic top quark signal, provided that the jet mass resolution and energy scale can be controlled to within 10\%. 
The jet mass distribution can be used to set a limit new particles which can lead to boosted top quarks.  With 10~fb$^{-1}$ of data, sources of new physics can be excluded if 
they lead to a top-quark cross section 
above 1184~fb defined in the fiducial region $\ptjet>0.8$~TeV.

Employing additional handles will make the top signal even cleaner.  For example, Figure~\ref{fig:jetmass}(b) demonstrates a clear signal if a factor two reduction of background using $b$-tagging can be achieved. This assumes a 40\% $b$-quark reconstruction efficiency, 10\% and 1\%
mistag rates due to $c$−quark and light-flavor jets, respectively.
The above exercise shows that in order to detect such inclusively produced boosted top quarks, 
jet masses should be understood within $10\%$ while $b$-tagging 
should have a factor of two rejection power against large non-$b$-jet backgrounds.
 
\begin{figure}[tbp]
\begin{center}
\subfigure[Jet masses for $\ptjet>0.6$ TeV]{
 \includegraphics[scale=0.32, angle=0]{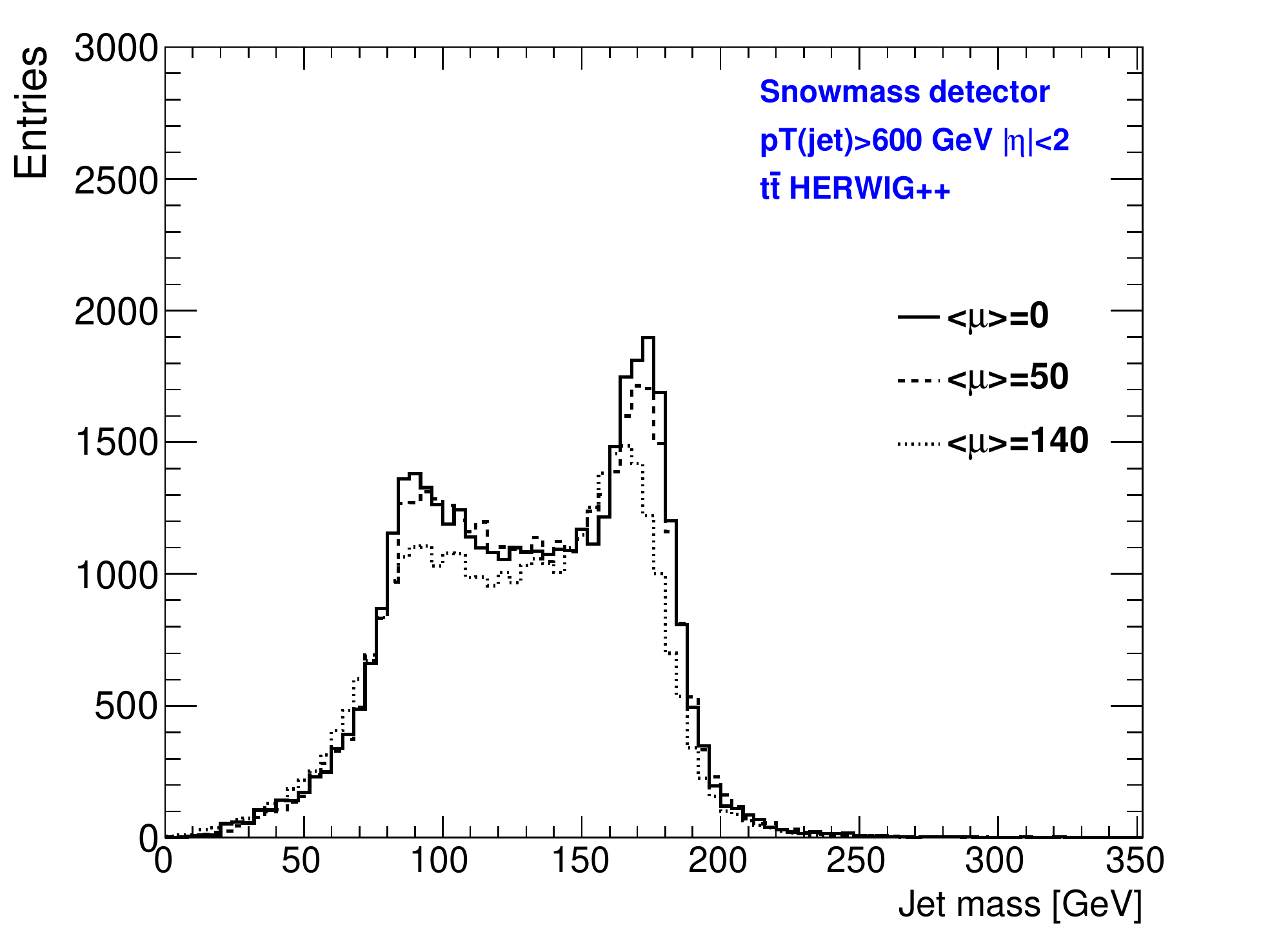}
 }
 \subfigure[Jet masses for $\ptjet>0.8$ TeV]{
 \includegraphics[scale=0.32, angle=0]{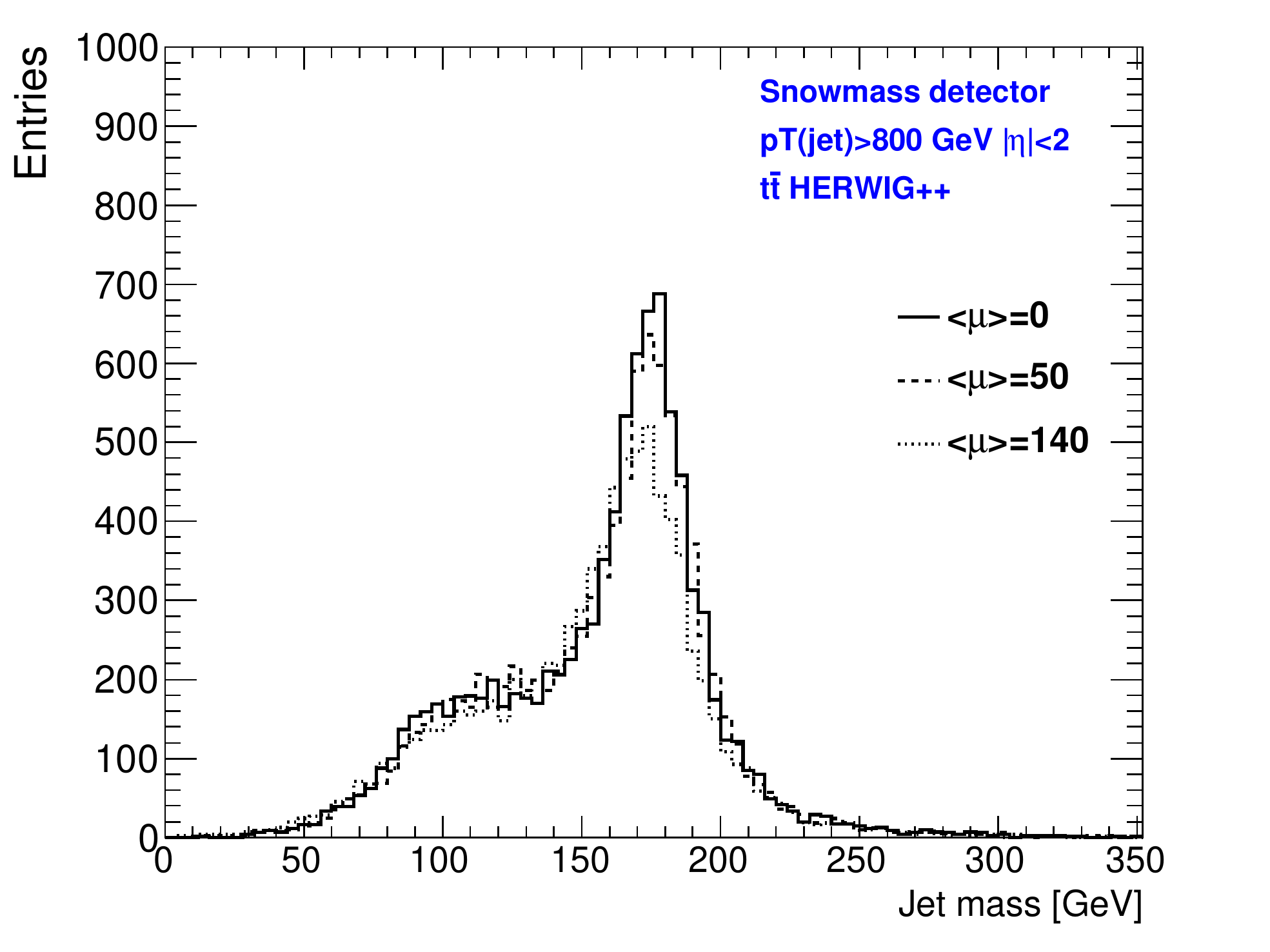}
 }
\end{center}
\caption{
Jet masses in $t\bar{t}$ events requiring jet transverse momenta (a) above 0.6 TeV and (b) above 0.8 TeV.
}
\label{fig:jetmass_pileups}
\end{figure}

The effect of different $p_T$ cuts and pileup scenarios on the masses of top-jets is shown in Figure~\ref{fig:jetmass_pileups}.
The jets were again reconstructed with the anti-$k_T$ (R=0.5)  jet algorithm, requiring $\ptjet>0.6$~TeV or $\ptjet>0.8$~TeV 
Pileup was subtracted using the jet-area method~\cite{Cacciari:2007fd}, which corrects the full four-vector of each jet.
A peak near 170~GeV is clearly observed for both choices of $\ptjet$ cut.  For the lower $\ptjet$ cut, a substantial fraction
of the events sit below the peak due to incomplete containment.  This includes a population of ``$W$-jets,'' visible as
a second peak at 80~GeV.  For the higher $\ptjet$ cut, most of the decays have become contained, and the top-mass peak is much stronger,
indicating that we have entered the highly-boosted regime.
These features persist when pileup is added, though the subtraction method tends to slightly over-correct, softening the 
mass distributions.  We will also comment on alternative pileup subtraction strategies below. 

\begin{figure}[tbp]
\begin{center}
\includegraphics[scale=0.4, angle=0]{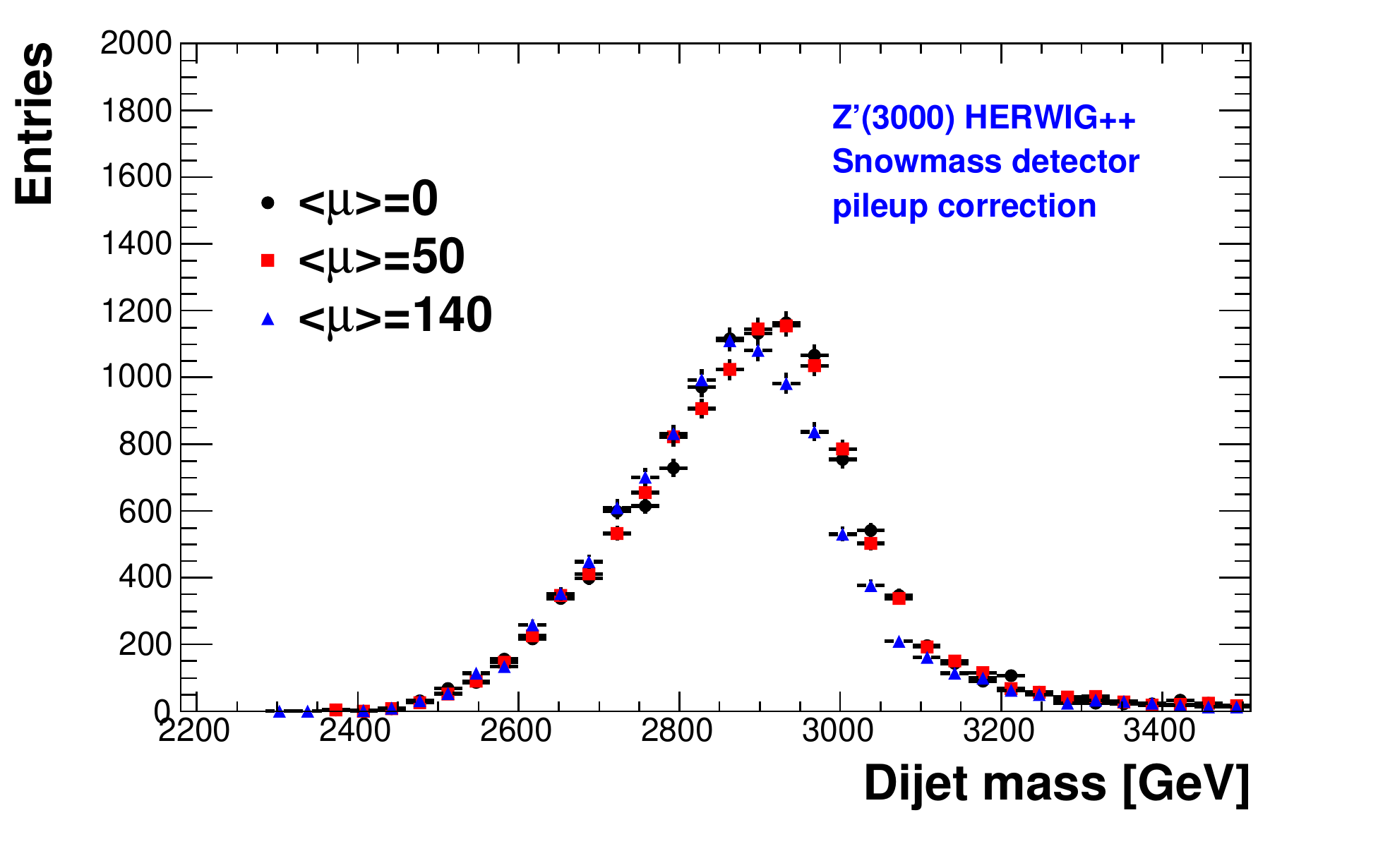}
\end{center}
\caption{
Dijet masses for $Z'\mathrm{(3 TeV)} \to t\bar{t}$ events using the Snowmass detector after correcting jets for pileup.
All three distributions have been obtained using 96,000 signal events.
}
\label{fig:zprime_mass1}
\end{figure}

Boosted tops are vital for exotic searches at multi-TeV masses.  Figure~\ref{fig:zprime_mass1} shows the {\it dijet} mass distribution in fully-hadronic $t\bar t$ produced in the decay of a narrow 3~TeV $Z'$ resonance, under the different pileup conditions with jet-area subtraction. 
The addition and subtraction of pileup leads to a small broadening of this distribution and a {\it downward} shift of up to 50~GeV for the peak position.  Assuming that such a modest effect can be corrected for, we conclude that the high-pileup scenario does not lead to substantial problems for the reconstruction of such states.  Pileup is a far more important problem for the reconstruction of the individual tops, which is crucial for discriminating this topology against generic QCD dijet events.

\begin{figure}[tbp]
\begin{center}
\subfigure[$\mmu=0$ scenario]{
 \includegraphics[scale=0.25, angle=0]{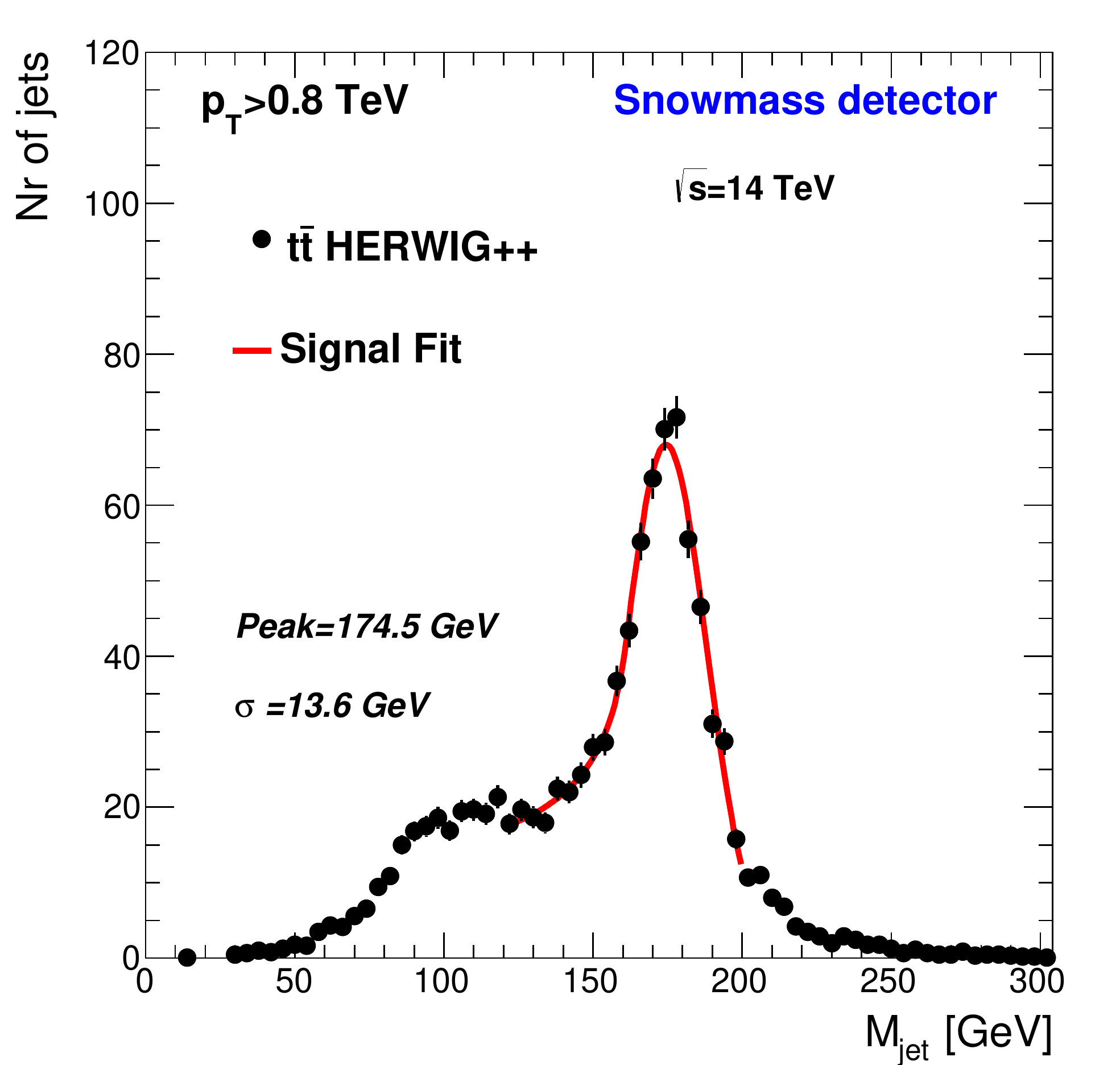}
 \includegraphics[scale=0.25, angle=0]{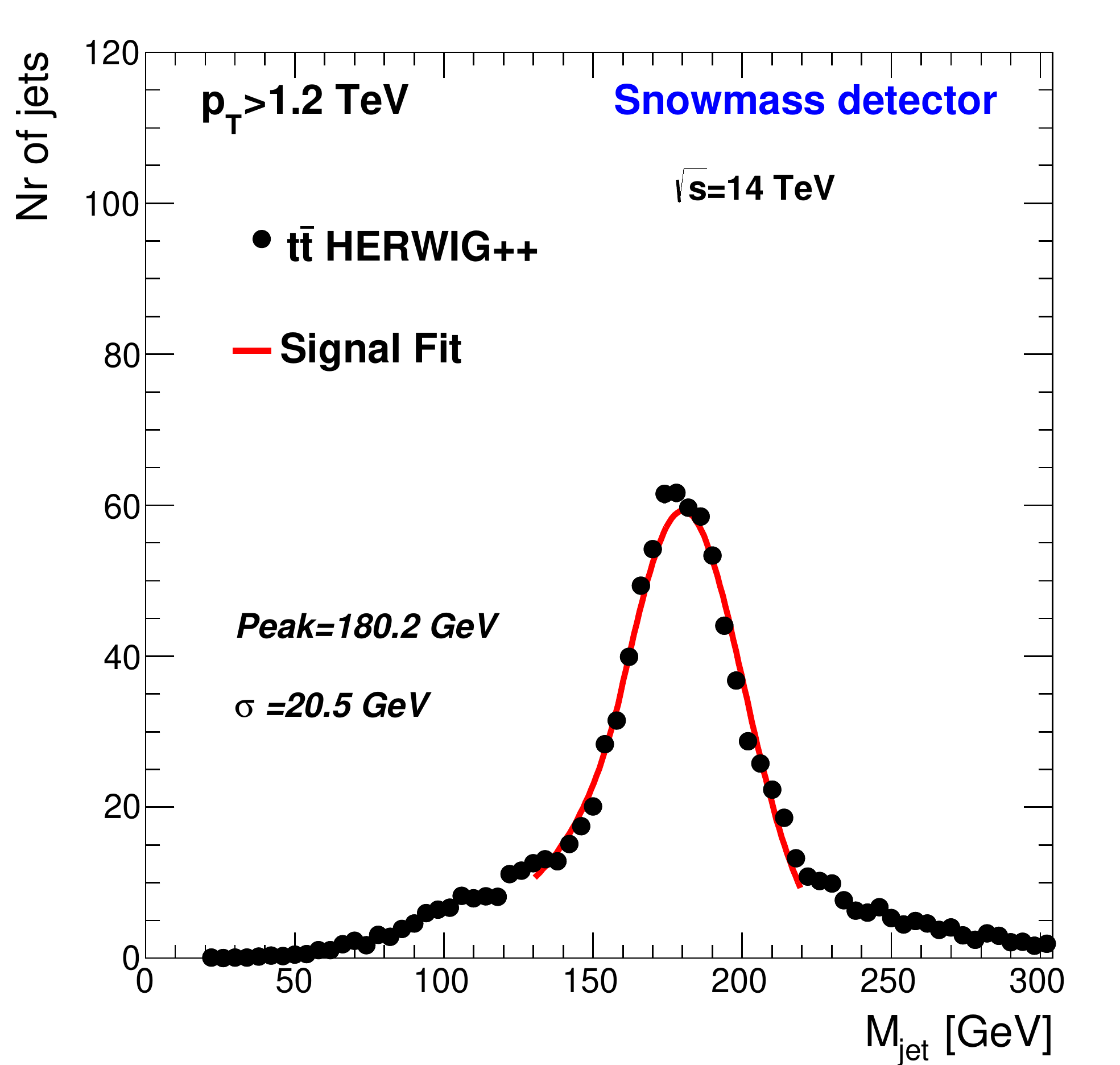}
 \includegraphics[scale=0.25, angle=0]{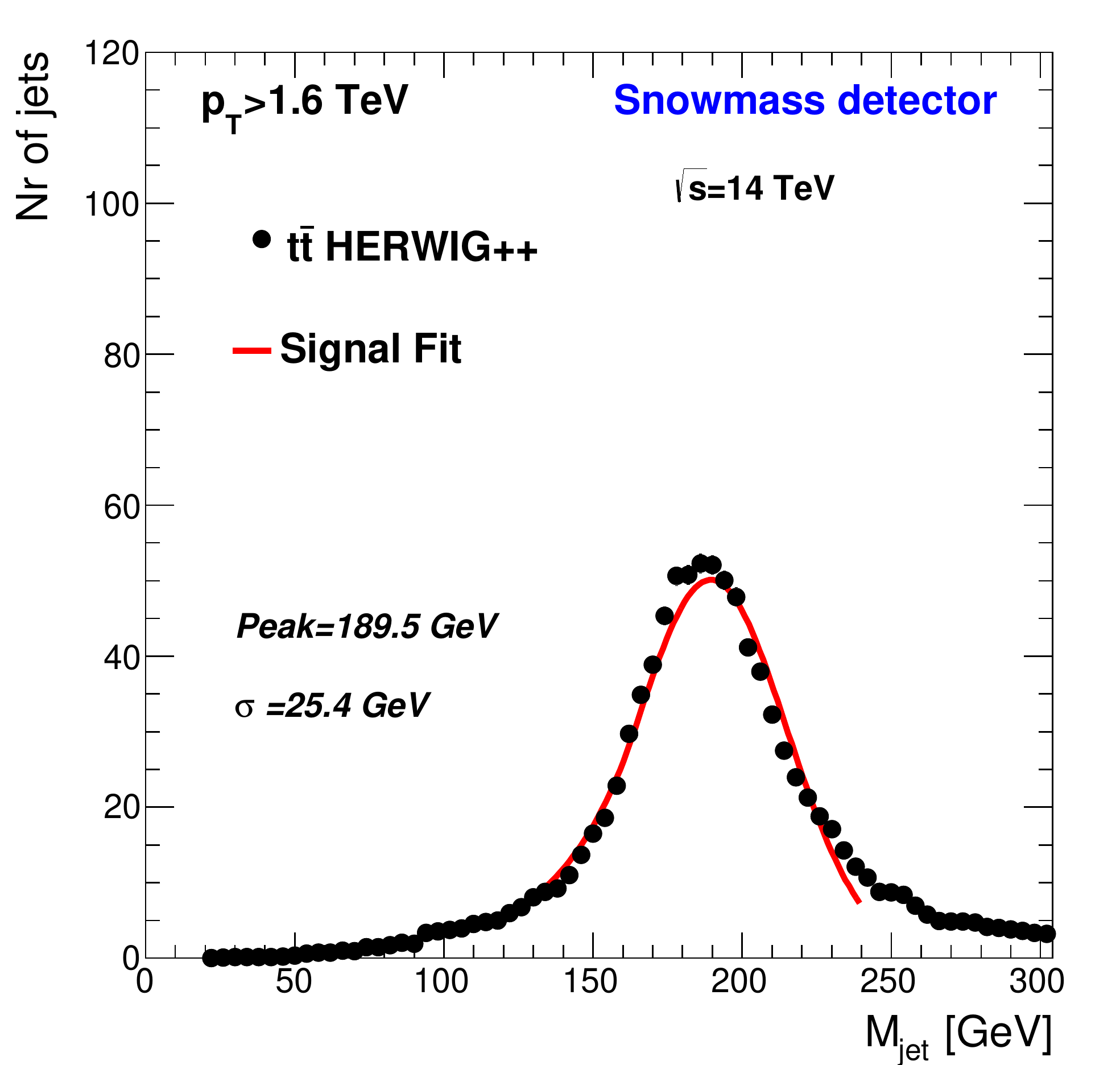}
}
\subfigure[$\mmu=50$ scenario]{
 \includegraphics[scale=0.25, angle=0]{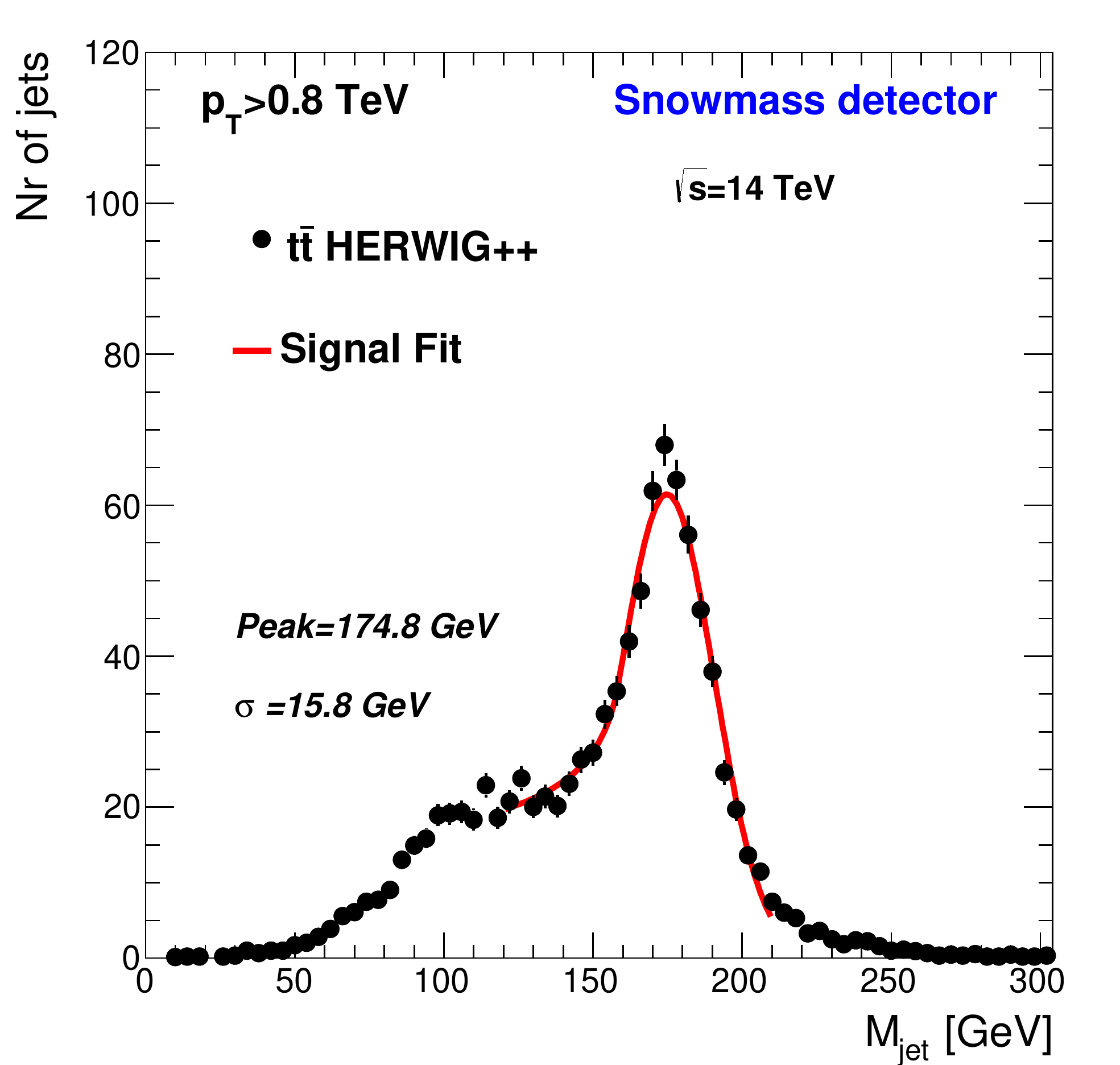}
 \includegraphics[scale=0.25, angle=0]{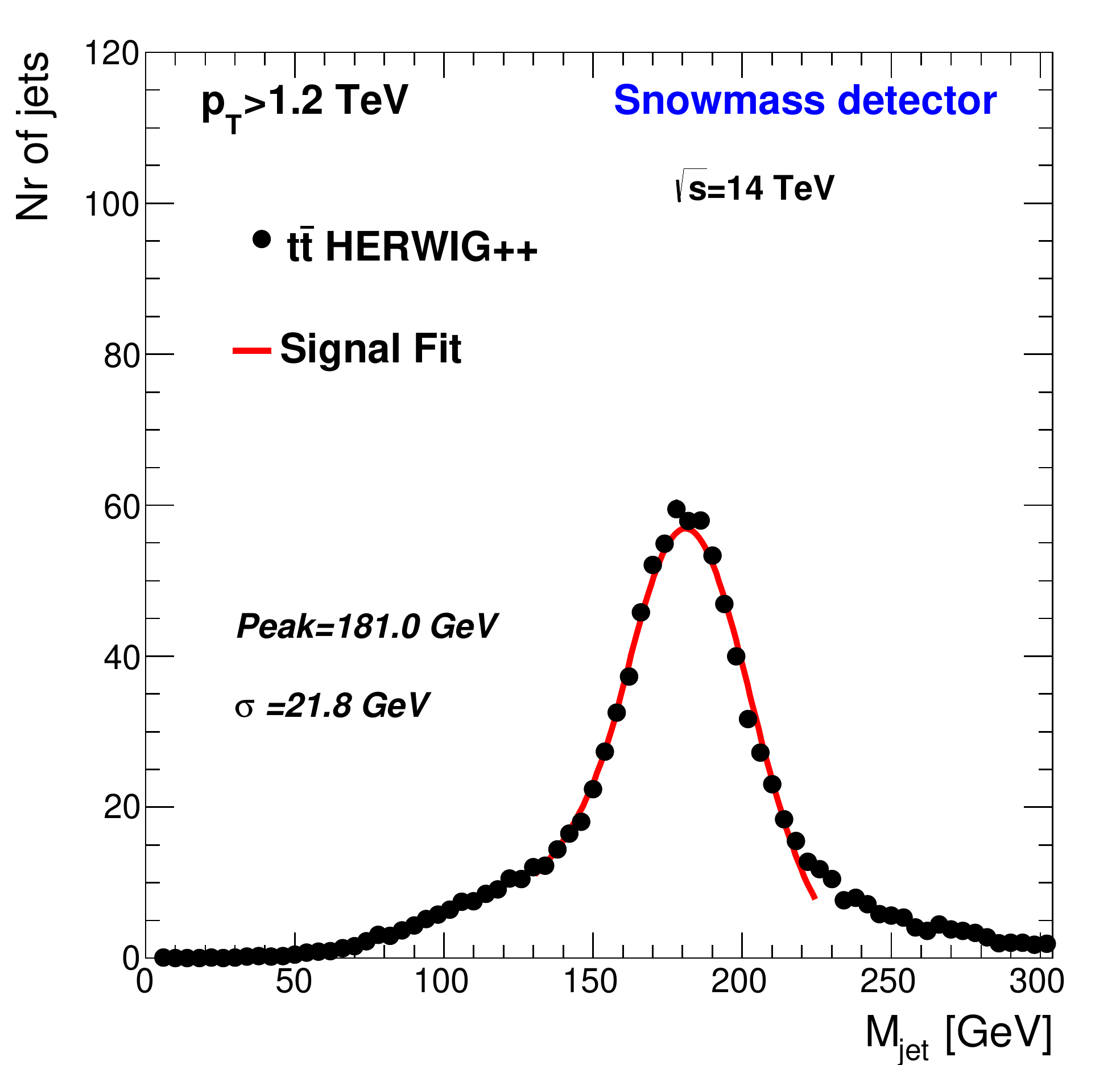}
 \includegraphics[scale=0.25, angle=0]{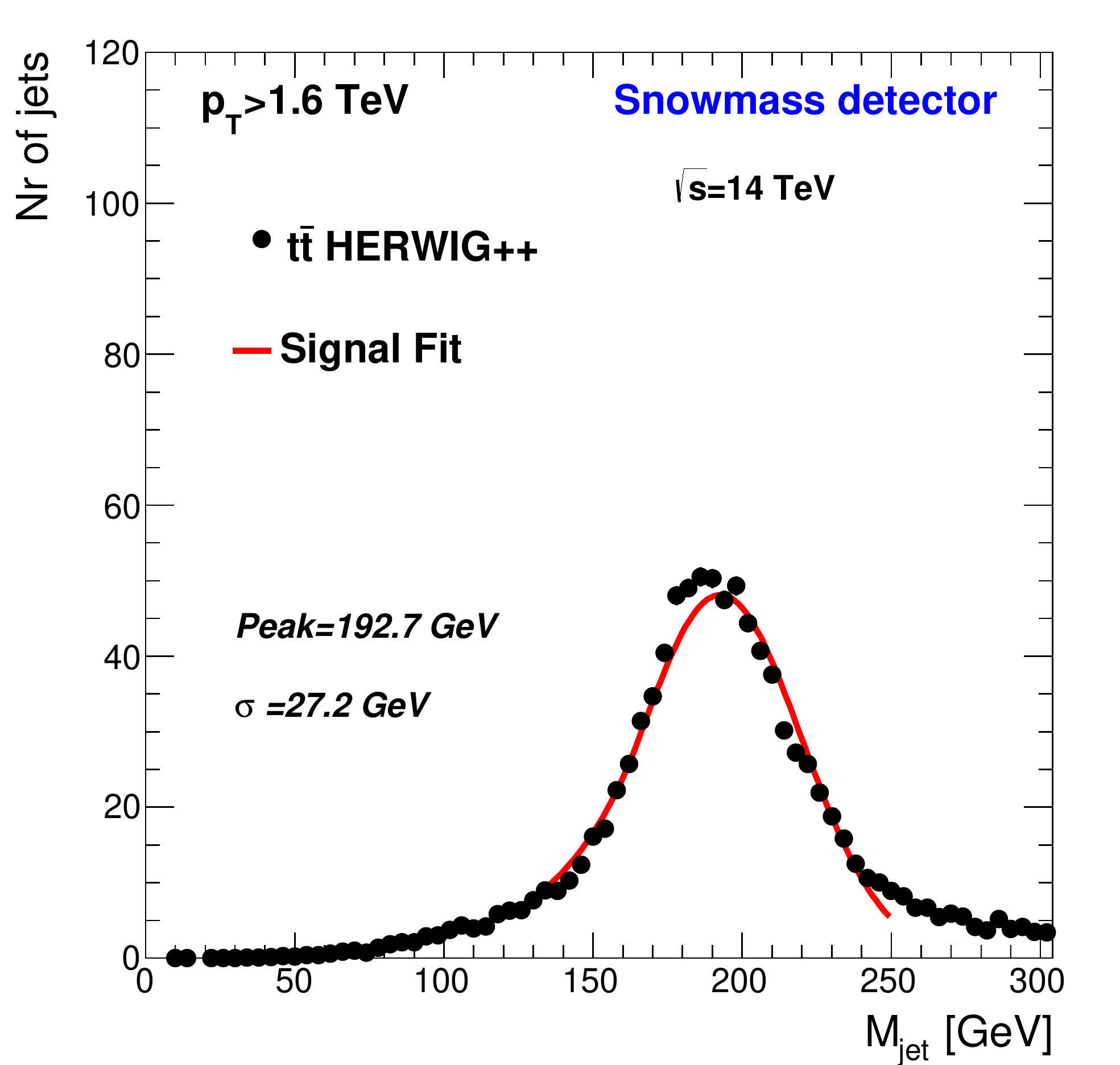}
}
\subfigure[$\mmu=140$ scenario]{
 \includegraphics[scale=0.25, angle=0]{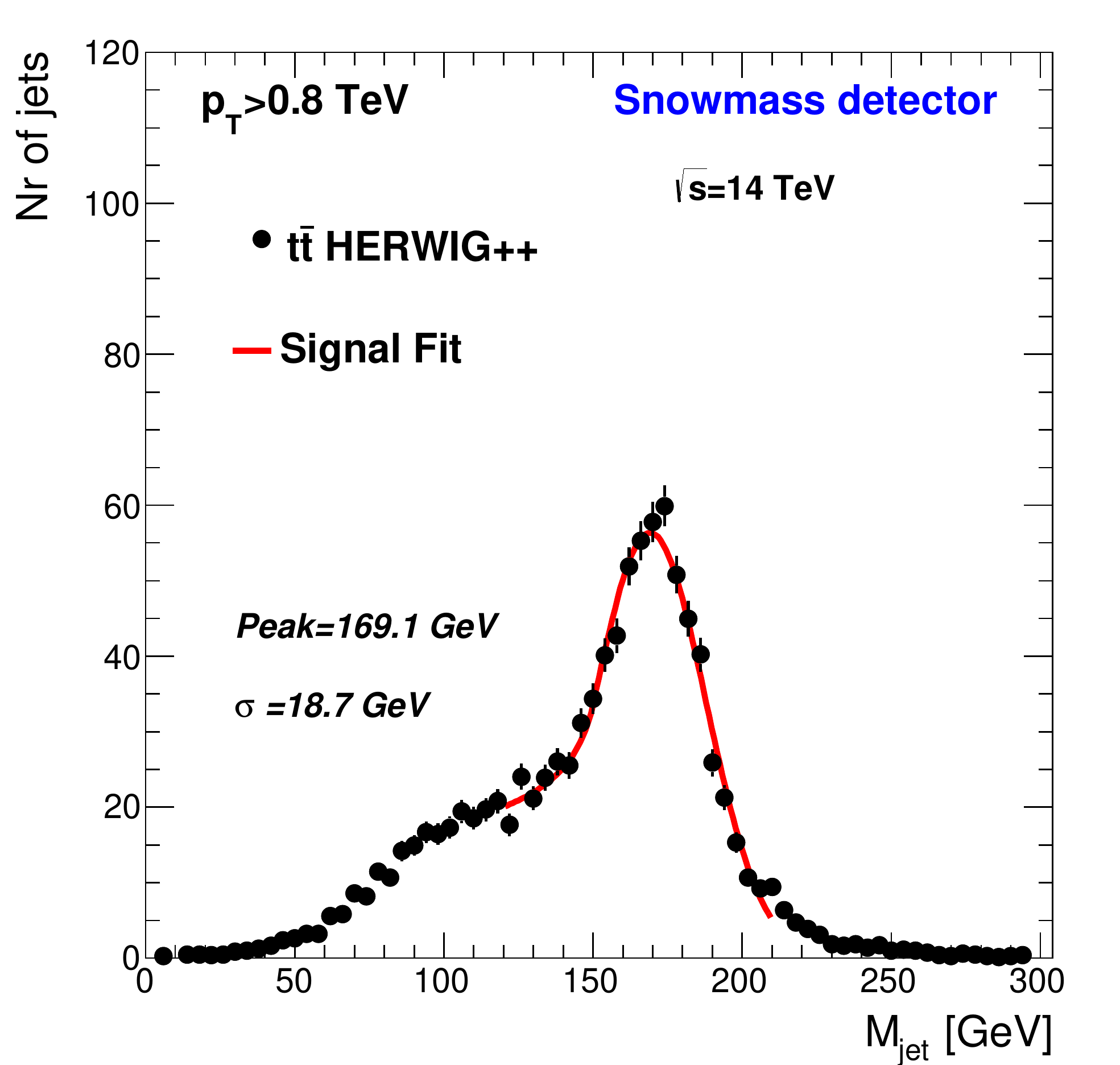}
 \includegraphics[scale=0.25, angle=0]{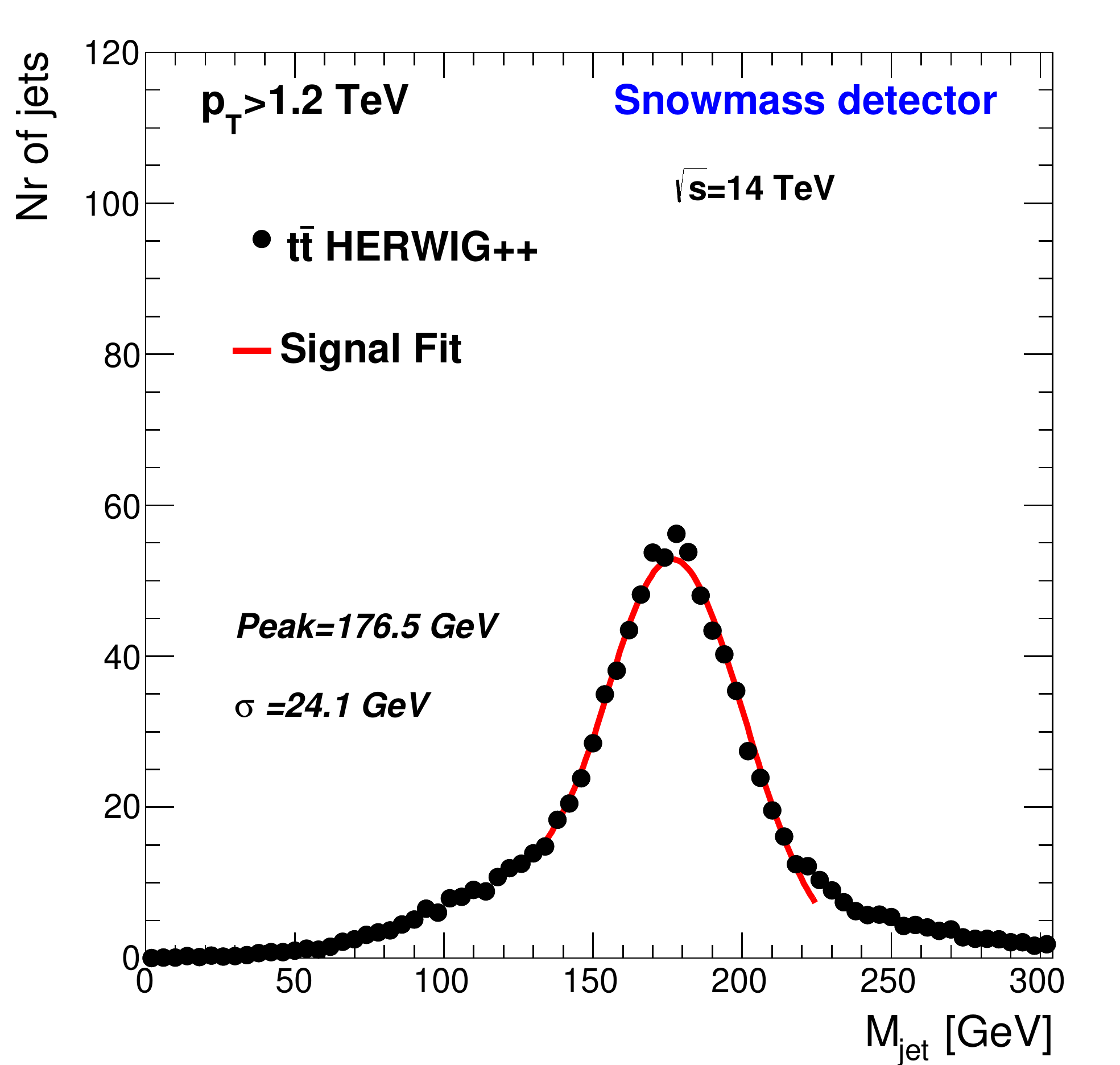}
 \includegraphics[scale=0.25, angle=0]{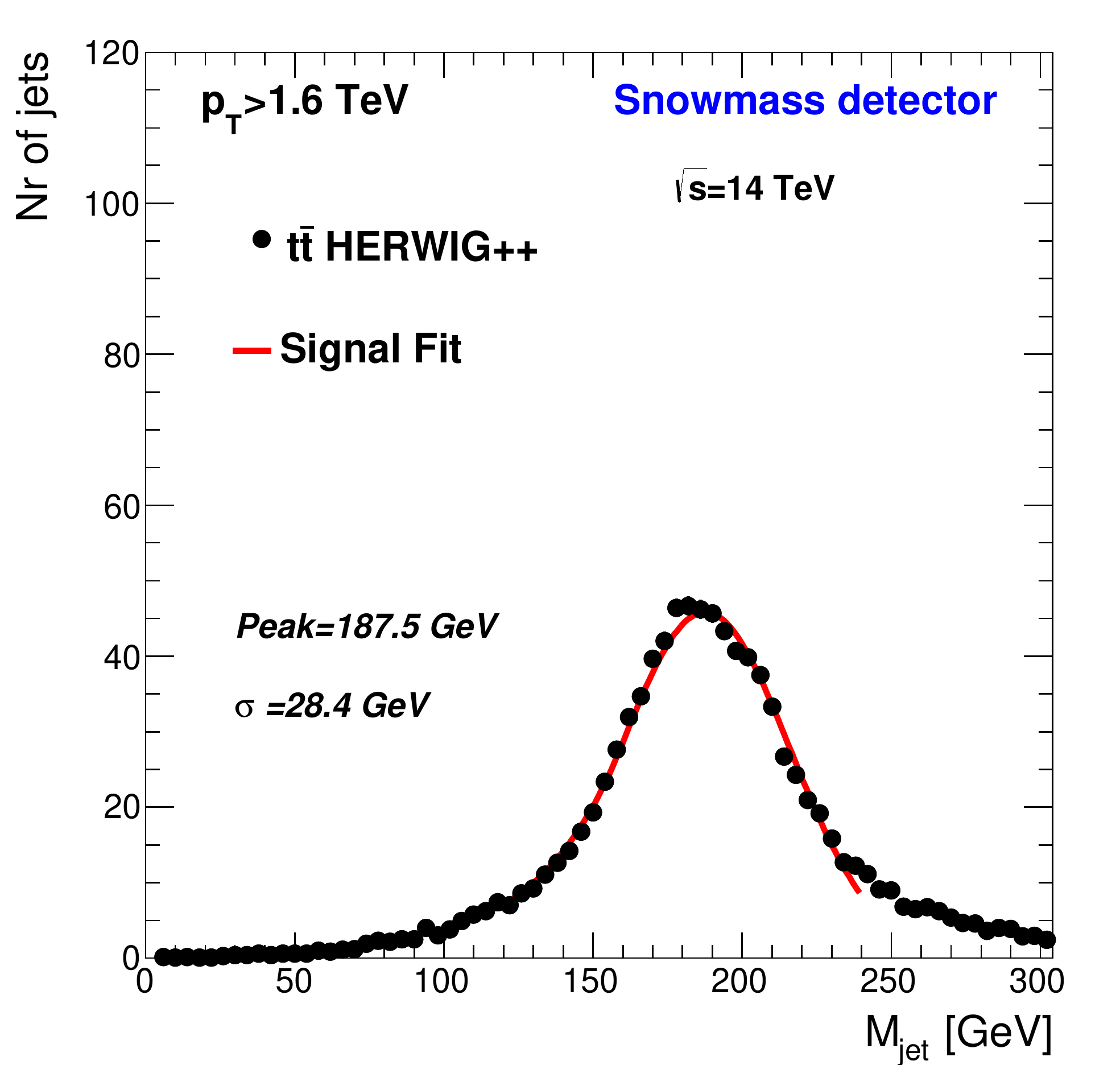}
}
\end{center}
\caption{
Jet masses for $t\bar{t}$ events for different $\ptjet$ and $\mmu$.
The core of the peak was  fitted using a
Crystal Ball function \protect\cite{Oreglia}.
All histograms have arbitrary normalisation (1000 events). 
}
\label{fig:jetmass_pileupsFit}
\end{figure}

Figure~\ref{fig:jetmass_pileupsFit} shows the top-jet mass distributions for $p_T$ cuts extending far into the boosted regime.  We fit each distribution with a Crystal Ball function~\cite{Oreglia}:
\begin{equation}
\left\{
\begin{array}{ll}
\frac{N}{\sqrt{2\pi}\sigma}\exp\Bigl(-\frac{(M-m_0)^2}{2\sigma^2}\Bigr) & \quad \text{for}~\frac{M-m_0}{\sigma}>-\alpha\\
\frac{N}{\sqrt{2\pi}\sigma}\Bigl(\!\frac{n}{\lvert\alpha\rvert}\!\Bigr)^n\exp\Bigl(-\frac{\lvert\alpha\rvert^2}{2}\Bigr)\Bigl(\frac{n}{\lvert\alpha\rvert}-\lvert\alpha\lvert-\frac{m-m_0}{\sigma}\Bigr)^{-n} & \quad \text{for}~\frac{M-m_0}{\sigma}\leq-\alpha,
\end{array}\right.
\label{cb}
\end{equation}
which has a Gaussian core (with mean $m_0$ and width $\sigma$) and a
power-law tail with an exponent $n$ to account for energy loss.  The parameter $\alpha$ defines the transition between the Gaussian and
the power-law functions.  The fit results indicate that jet mass resolution for large transverse momenta above 1.6~TeV is a factor two worse than for $\ptjet>0.8$~TeV.  The peak position also shifts upward by an amount comparable to increase in width.  Although increasing pileup also contributes to the broadening, we can see that by far the greater degradation comes from increasing the $p_T$.  The trend is a combination of ``instrumental'' effects simulated within the {\sc Delphes} detector model and physics noise such as gluons showering off of the top quark before it decays.  Below, we investigate both of these effects in more detail, and at even higher $p_T$.

\subsection{Jet grooming}

\begin{figure}[tbp]
\begin{center}
\subfigure[Trimmed Jet CA R=0.8 - recluster CA R=0.2 $\ptfrac$=0.03 ]{
 \includegraphics[scale=0.25, angle=0]{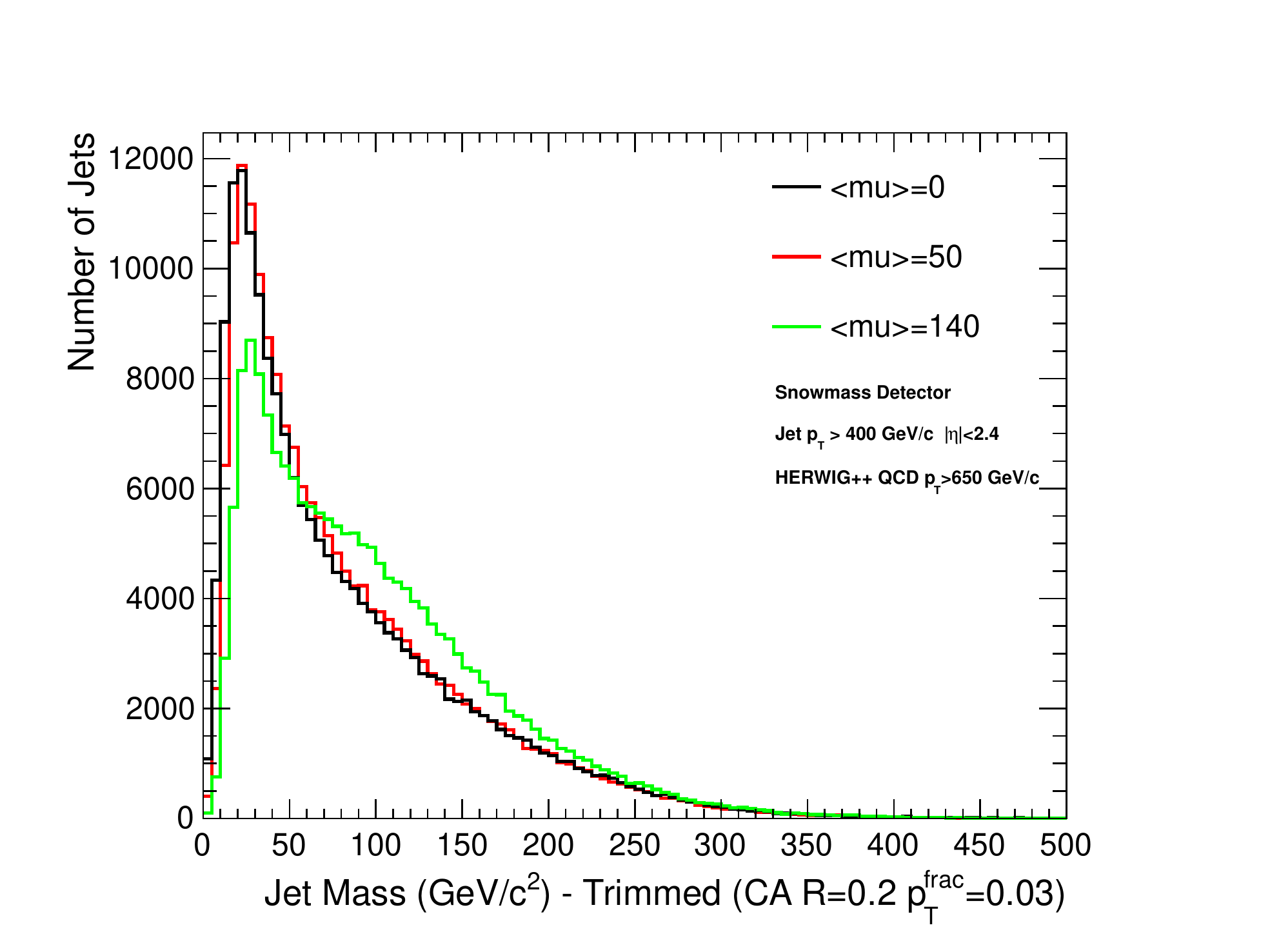}
 \includegraphics[scale=0.25, angle=0]{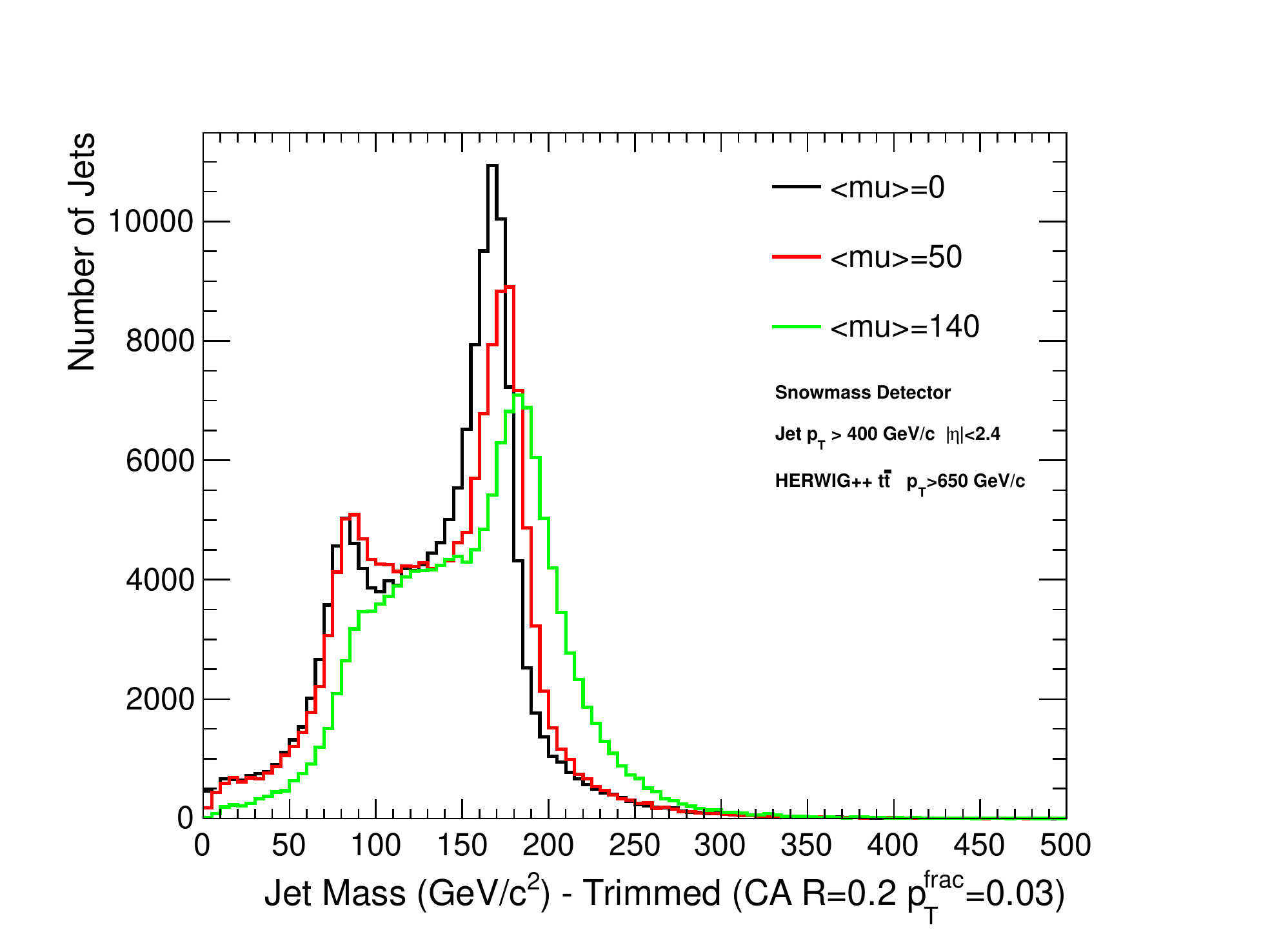}
}
\subfigure[Trimmed Jet CA R=0.8 - recluster CA R=0.2 $\ptfrac$=0.05 ]{
 \includegraphics[scale=0.25, angle=0]{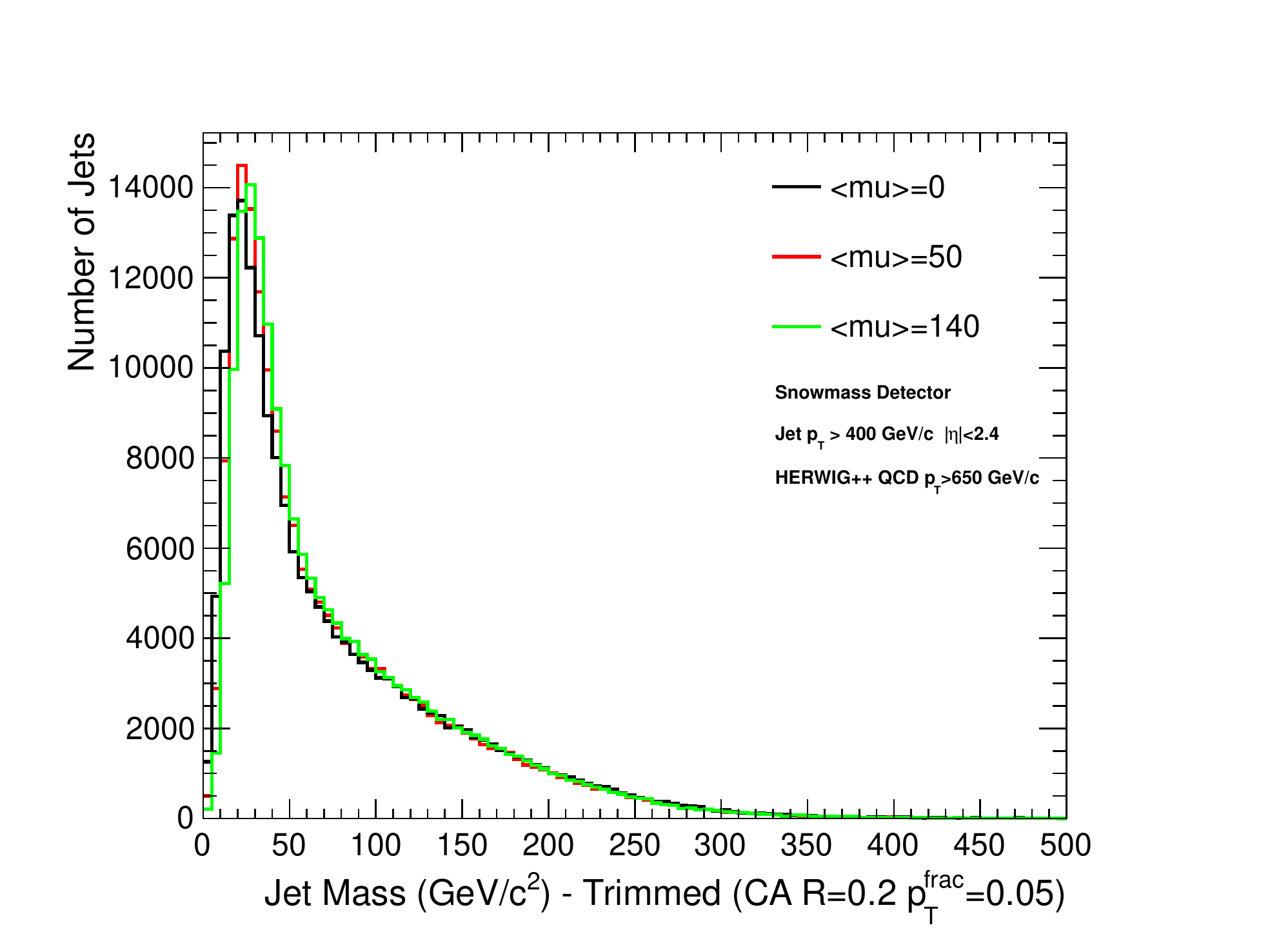}
 \includegraphics[scale=0.25, angle=0]{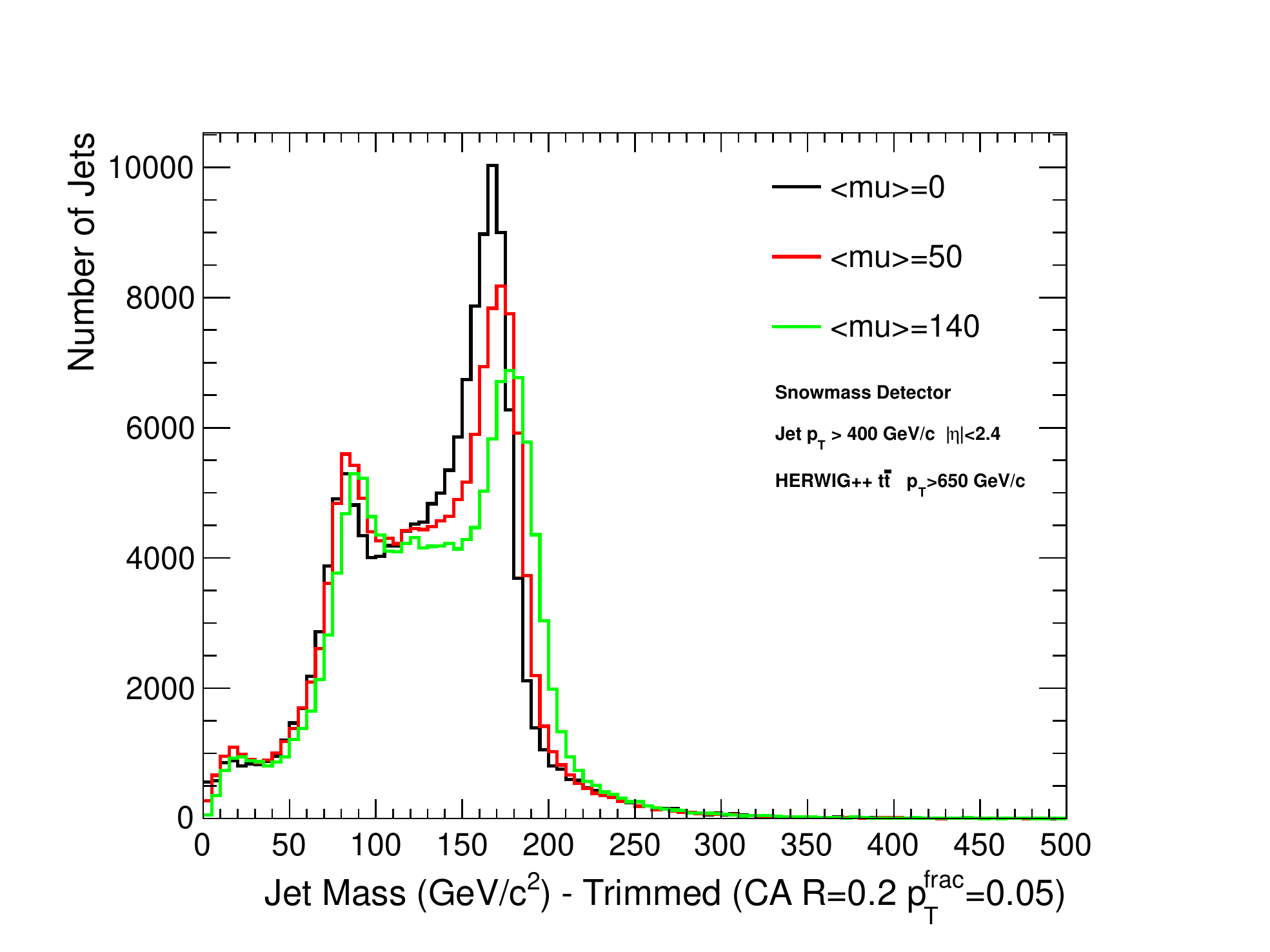}
}
\subfigure[Trimmed Jet CA R=0.8 - recluster CA R=0.2 $\ptfrac$=0.07]{
 \includegraphics[scale=0.25, angle=0]{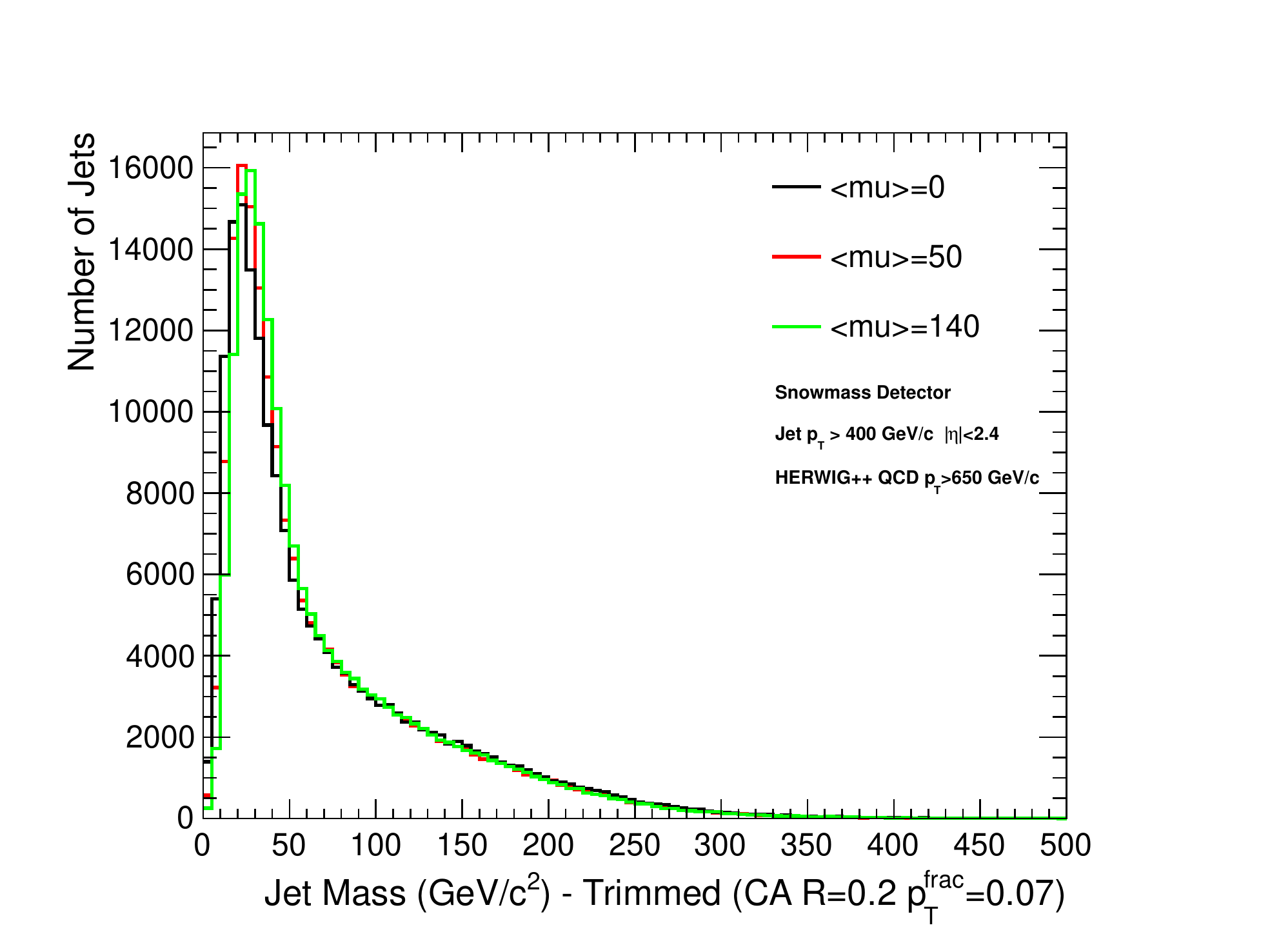}
 \includegraphics[scale=0.25, angle=0]{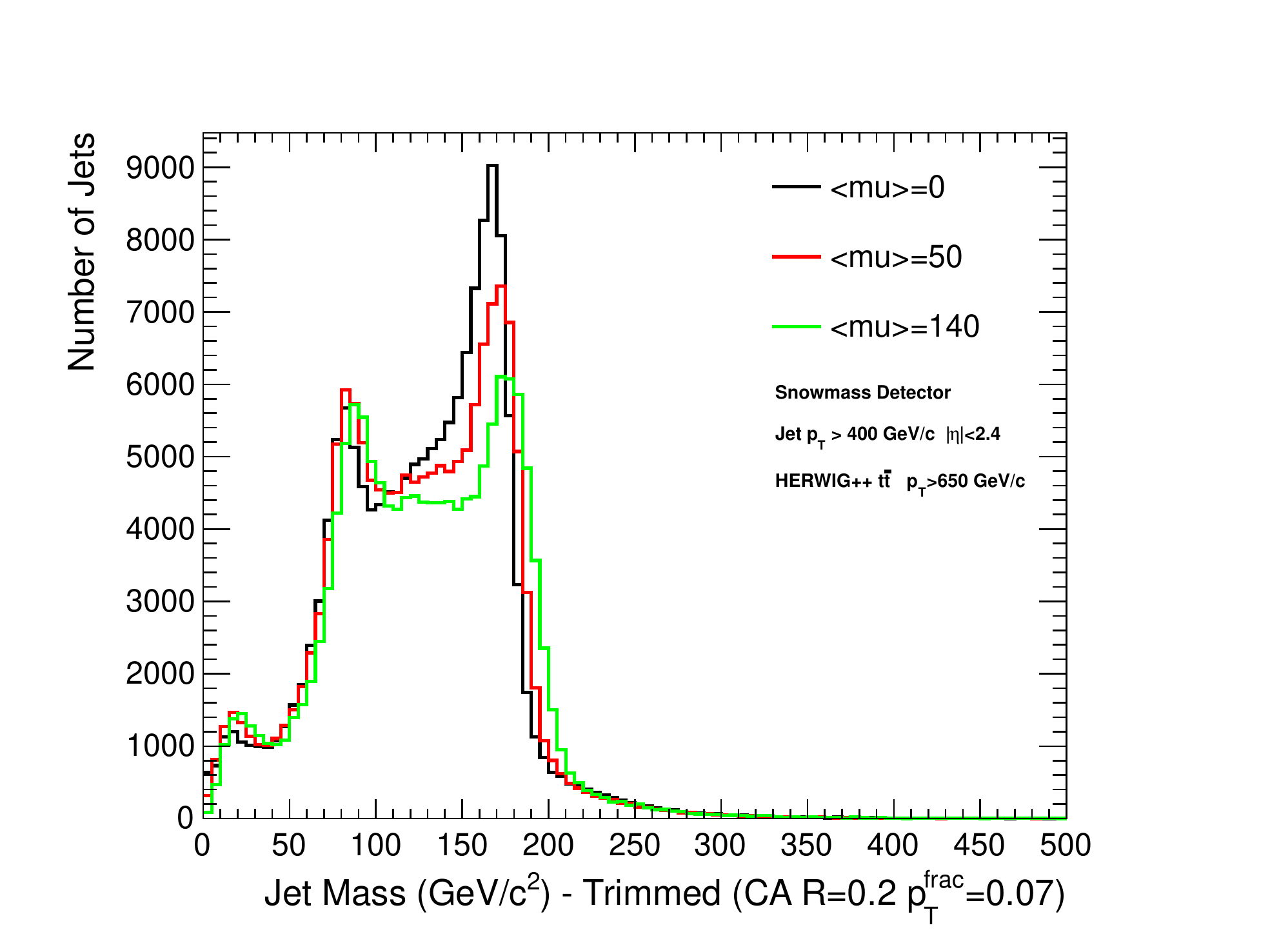}
}
\subfigure[Trimmed Jet CA R=0.8 - recluster CA R=0.3 $\ptfrac$=0.03]{
 \includegraphics[scale=0.25, angle=0]{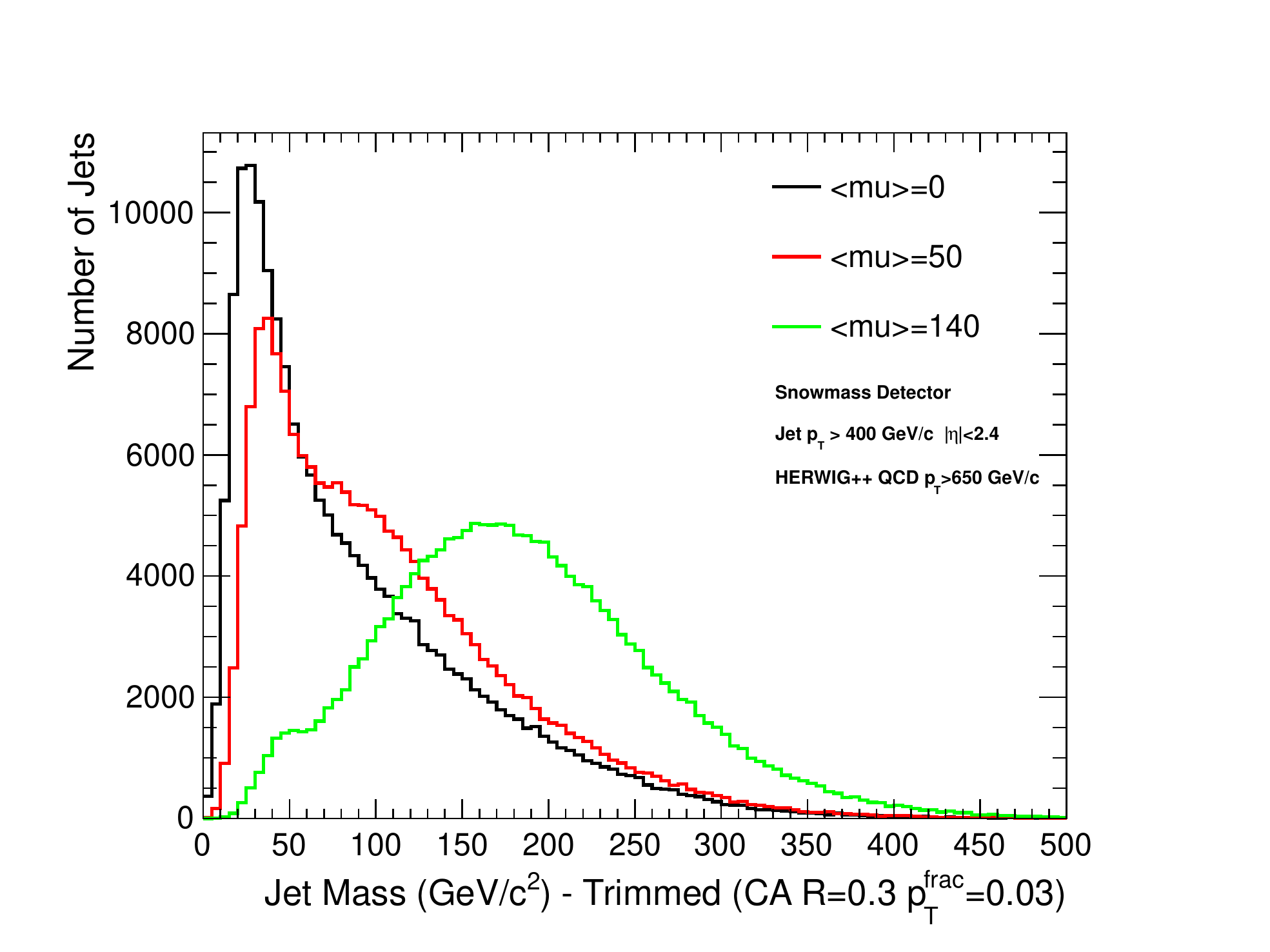}
 \includegraphics[scale=0.25, angle=0]{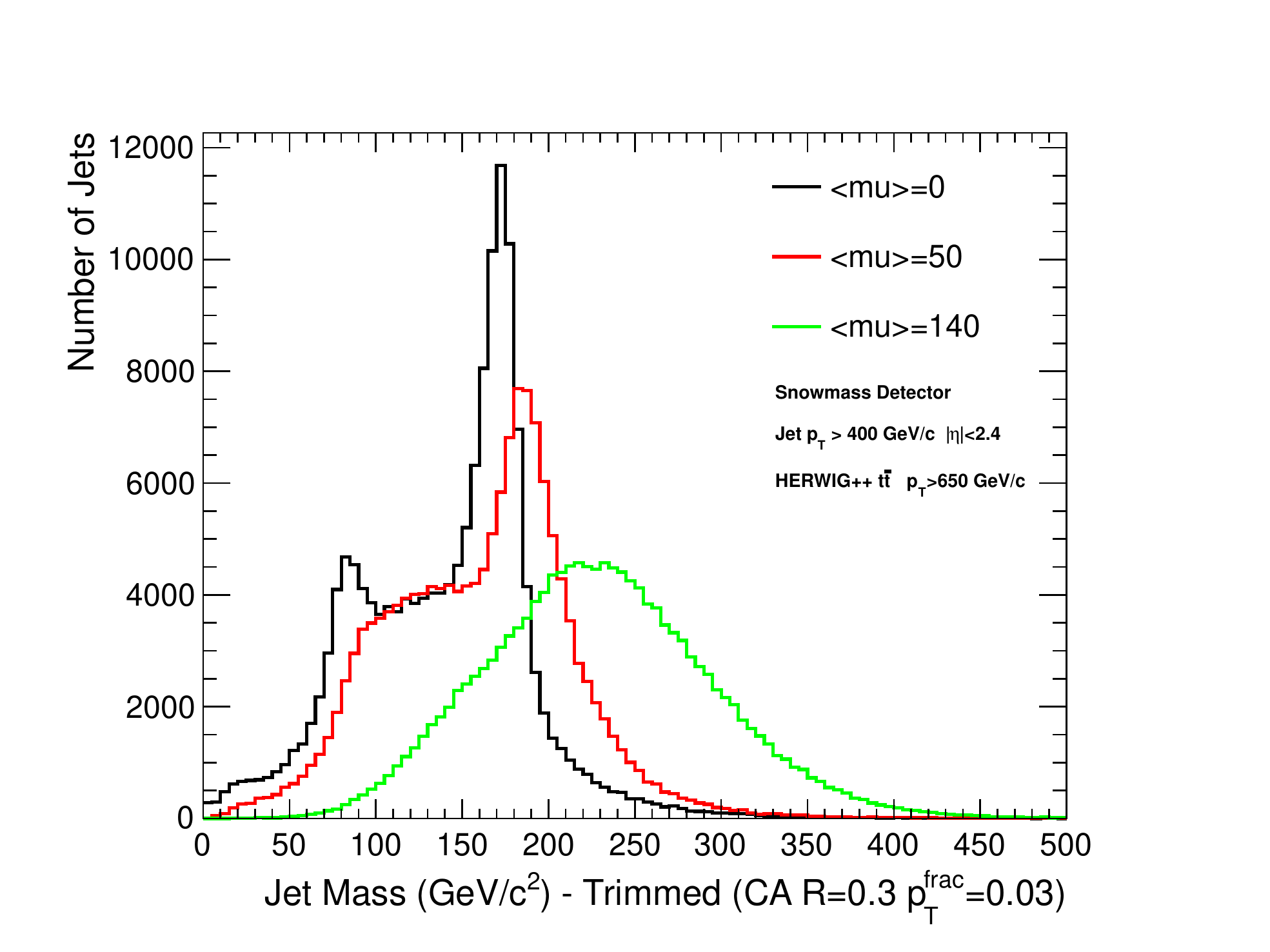}
}
\end{center}
\caption{
Trimmed jet mass for different trimmed jet parameters and for three different pileup scenarios. The C/A jet clustering algorithm is used with a distance parameter of $R = 0.8$. No area based pileup subtraction is used.  Trimmed subjets are reclustered with the C/A algorithm. Samples are generated with \herwig with $p_T>650$~GeV for QCD (left) and $\ttbar$ (right).   }
 \label{fig:jet_mass_compare_ca}
\end{figure}

\begin{figure}[tbp]
\begin{center}
\subfigure[Trimmed Jet CA R=0.8 - recluster KT R=0.2 $\ptfrac$=0.03 ]{
 \includegraphics[scale=0.25, angle=0]{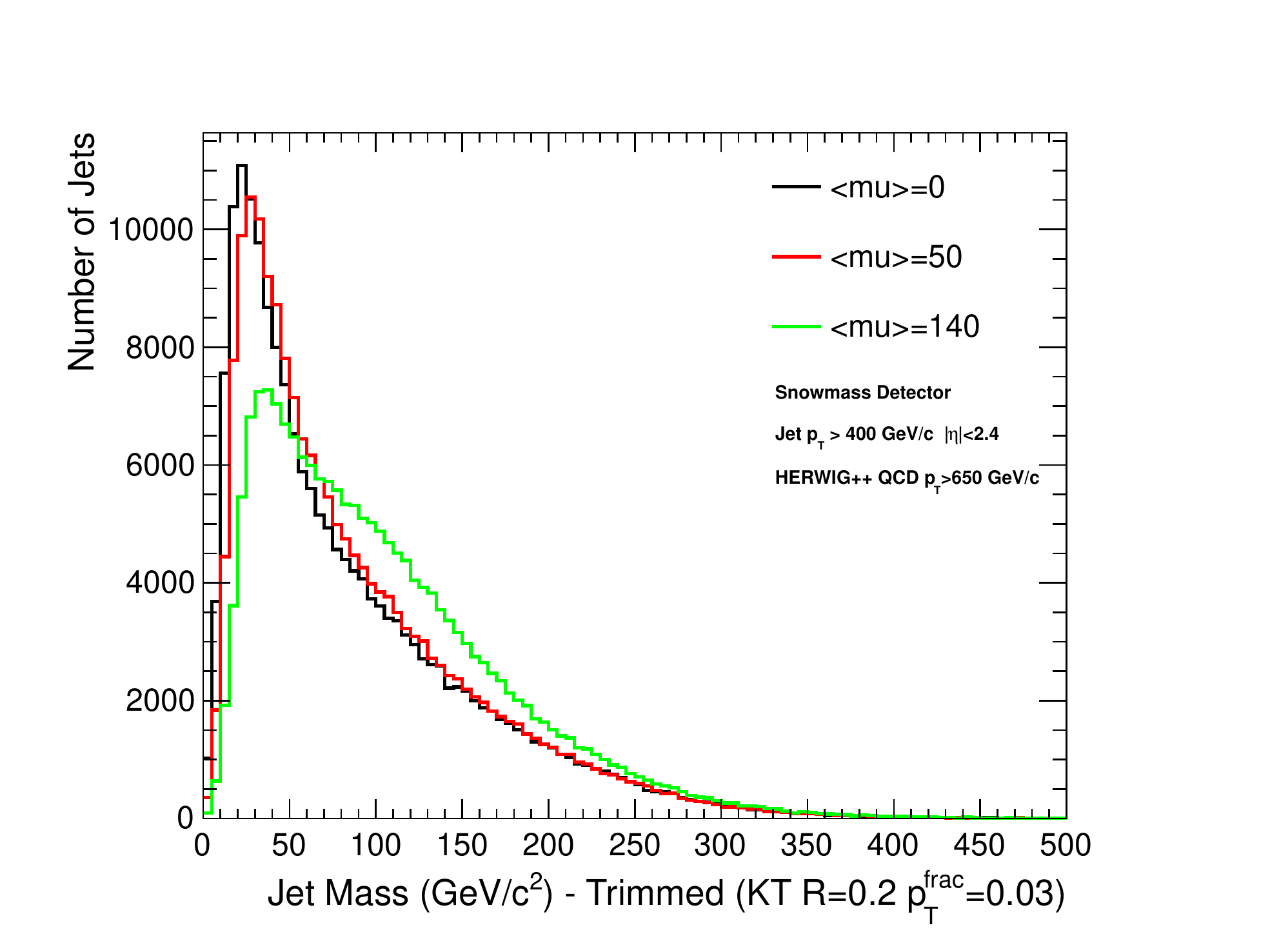}
 \includegraphics[scale=0.25, angle=0]{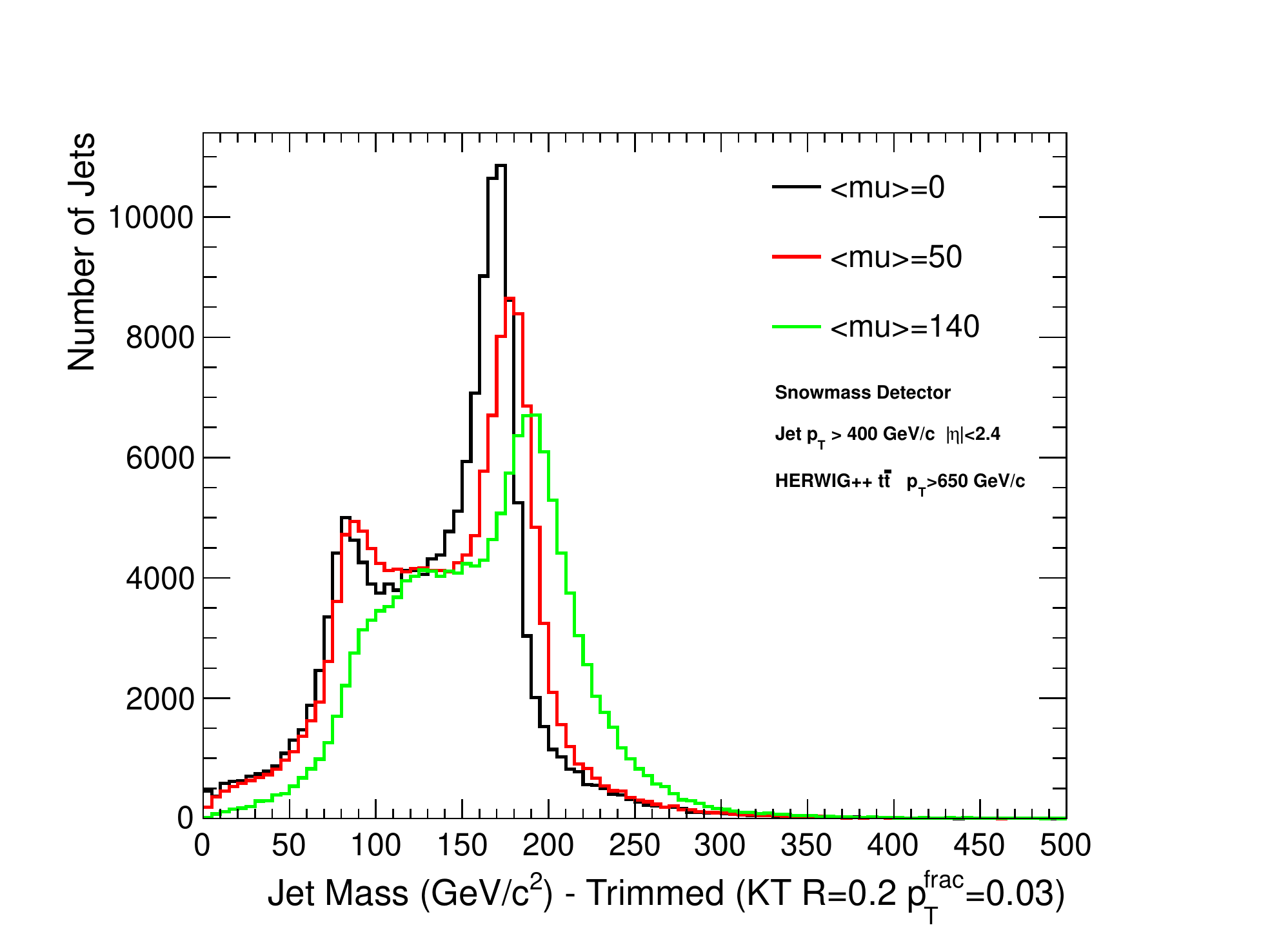}
}
\subfigure[Trimmed Jet CA R=0.8 - recluster KT R=0.2 $\ptfrac$=0.05 ]{
 \includegraphics[scale=0.25, angle=0]{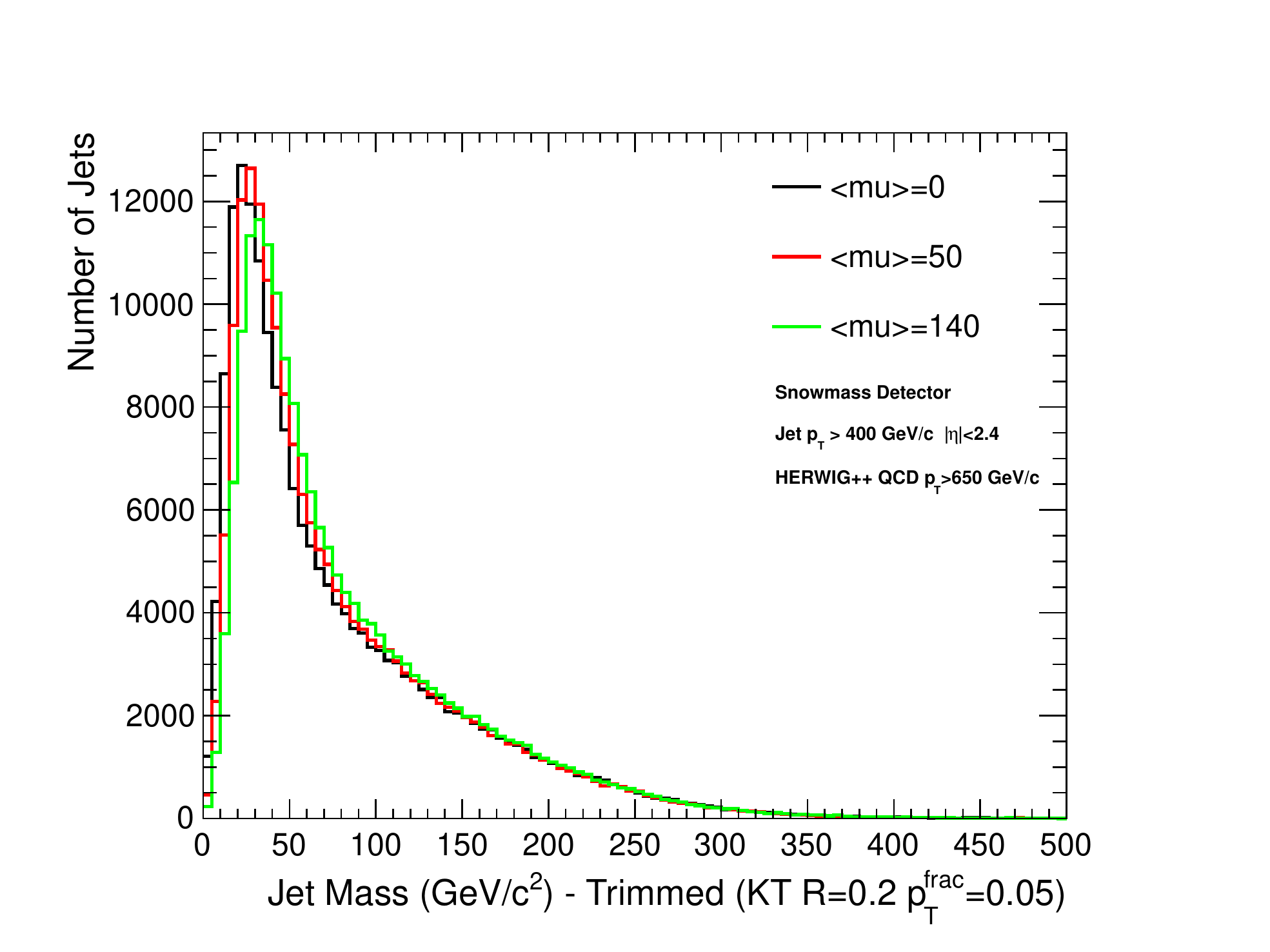}
 \includegraphics[scale=0.25, angle=0]{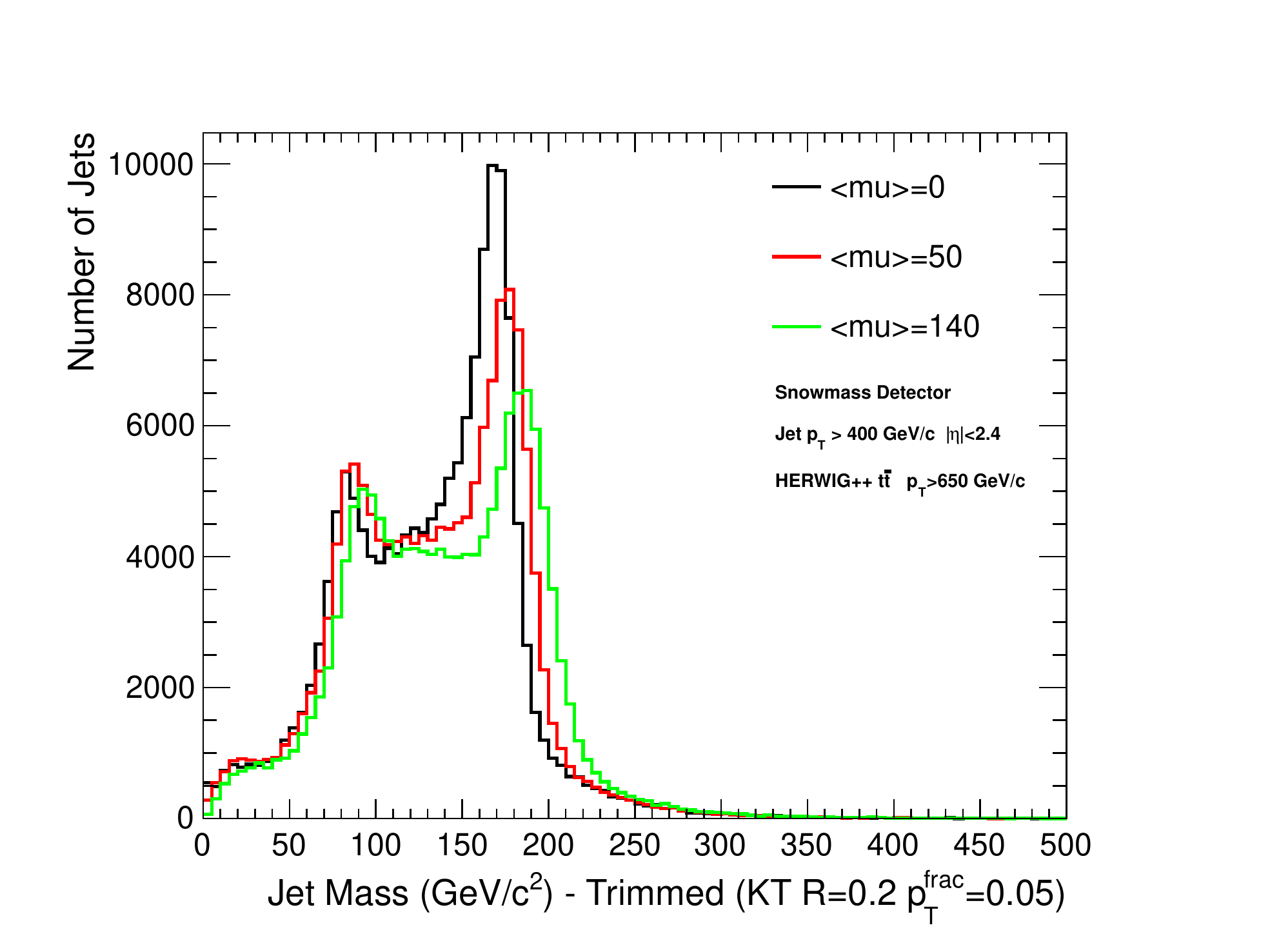}
}
\subfigure[Trimmed Jet CA R=0.8 - recluster KT R=0.2 $\ptfrac$=0.07]{
 \includegraphics[scale=0.25, angle=0]{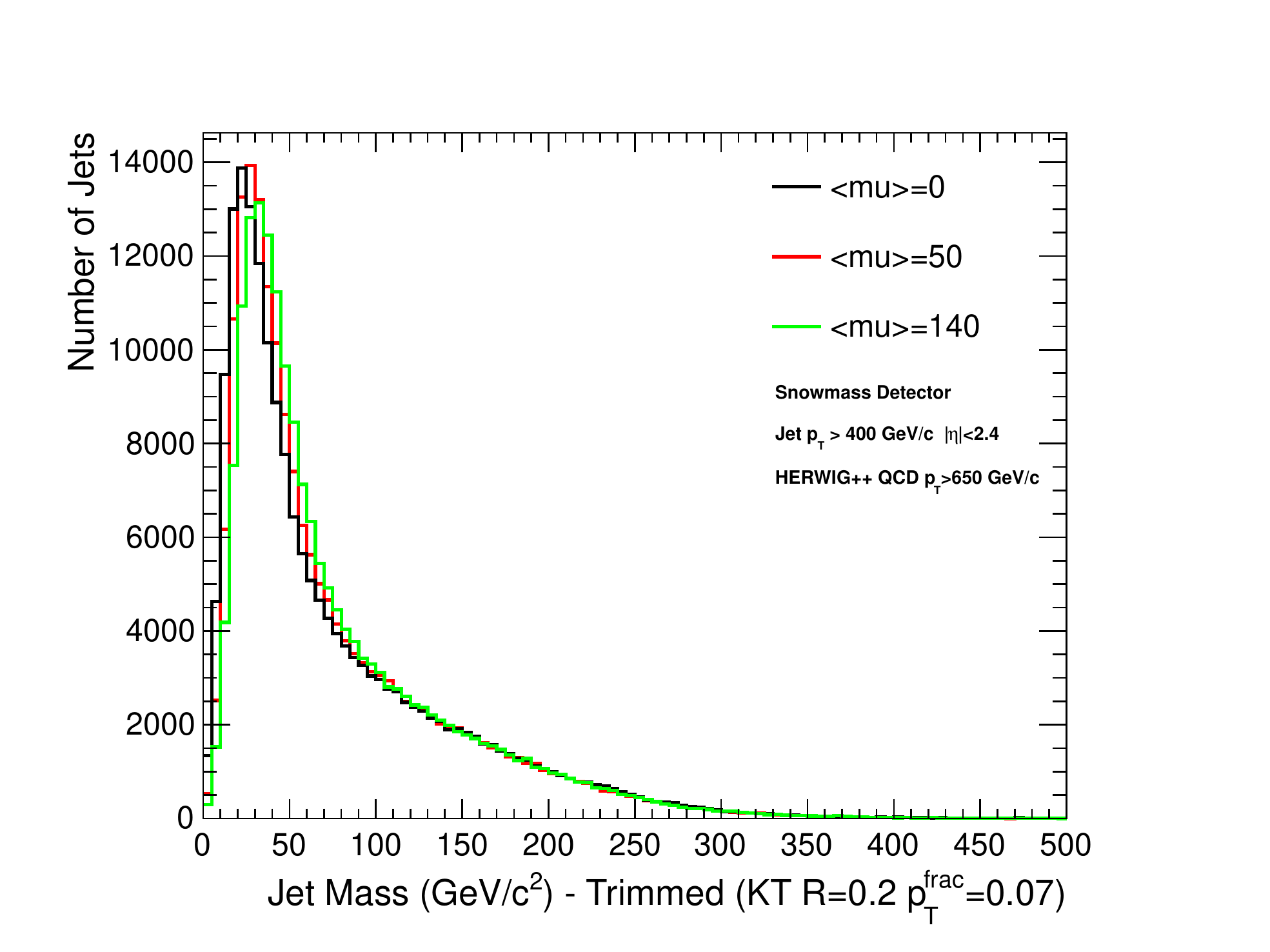}
 \includegraphics[scale=0.25, angle=0]{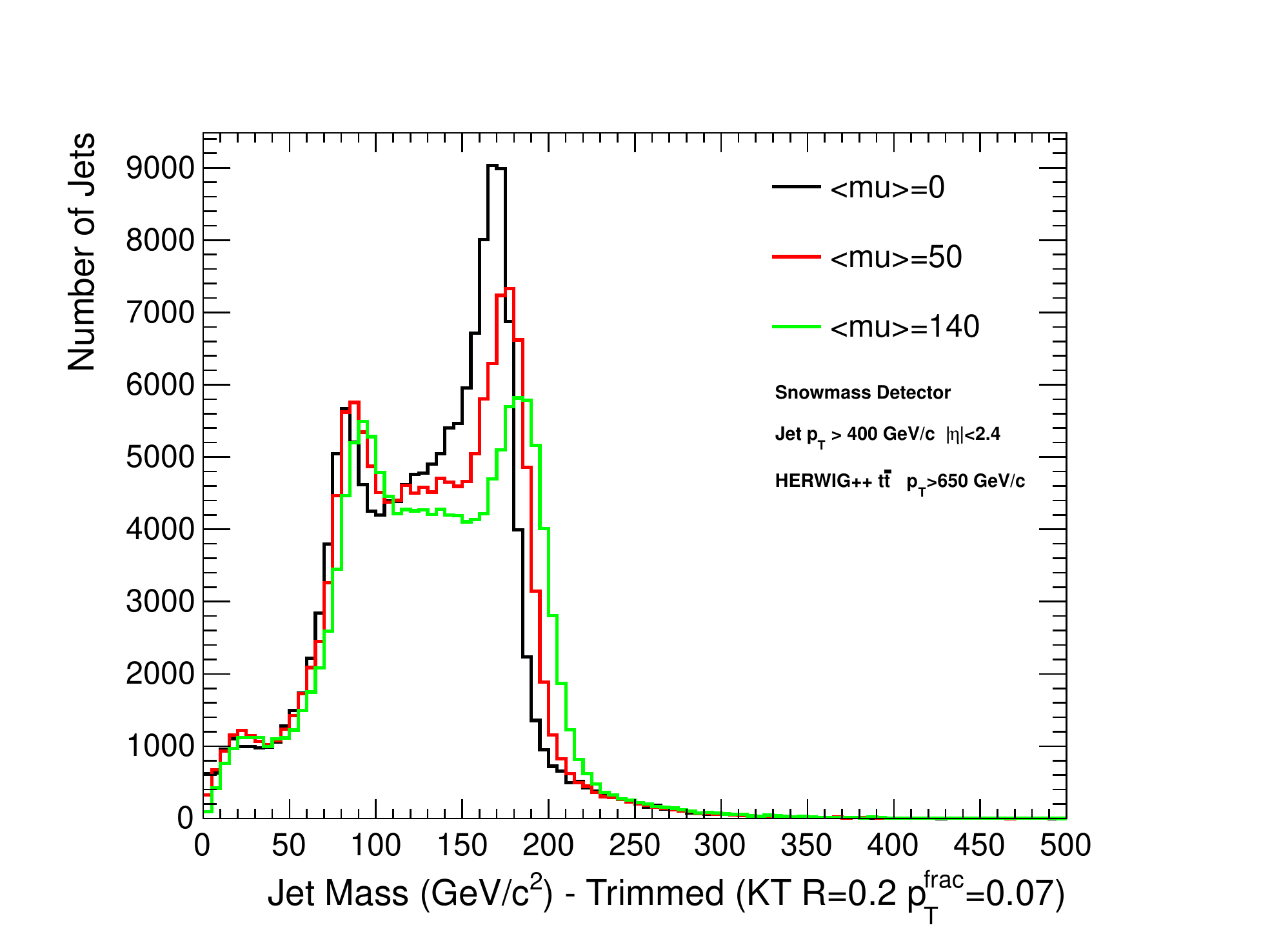}
}
\subfigure[Trimmed Jet CA R=0.8 - recluster KT R=0.3 $\ptfrac$=0.03]{
 \includegraphics[scale=0.25, angle=0]{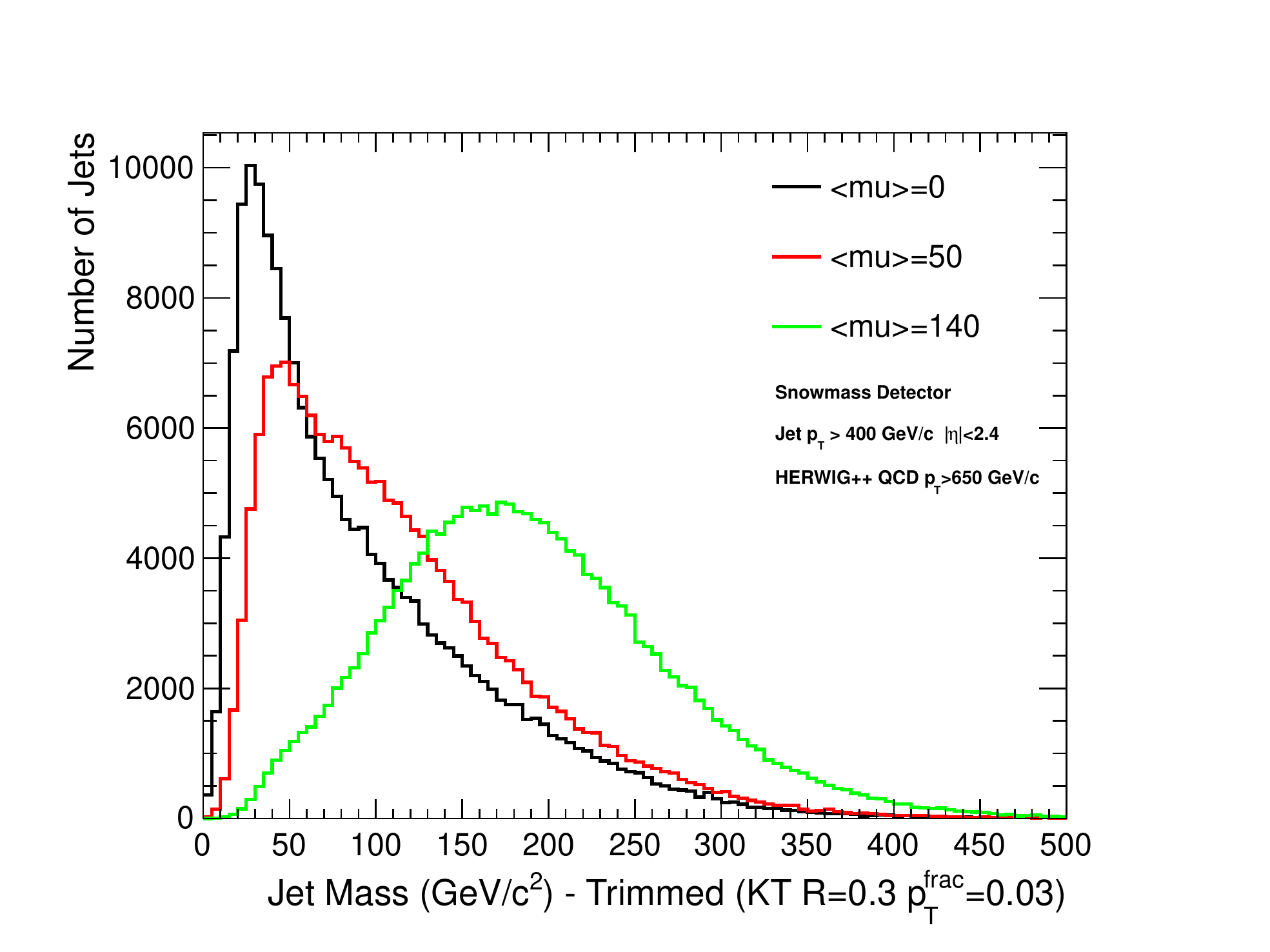}
 \includegraphics[scale=0.25, angle=0]{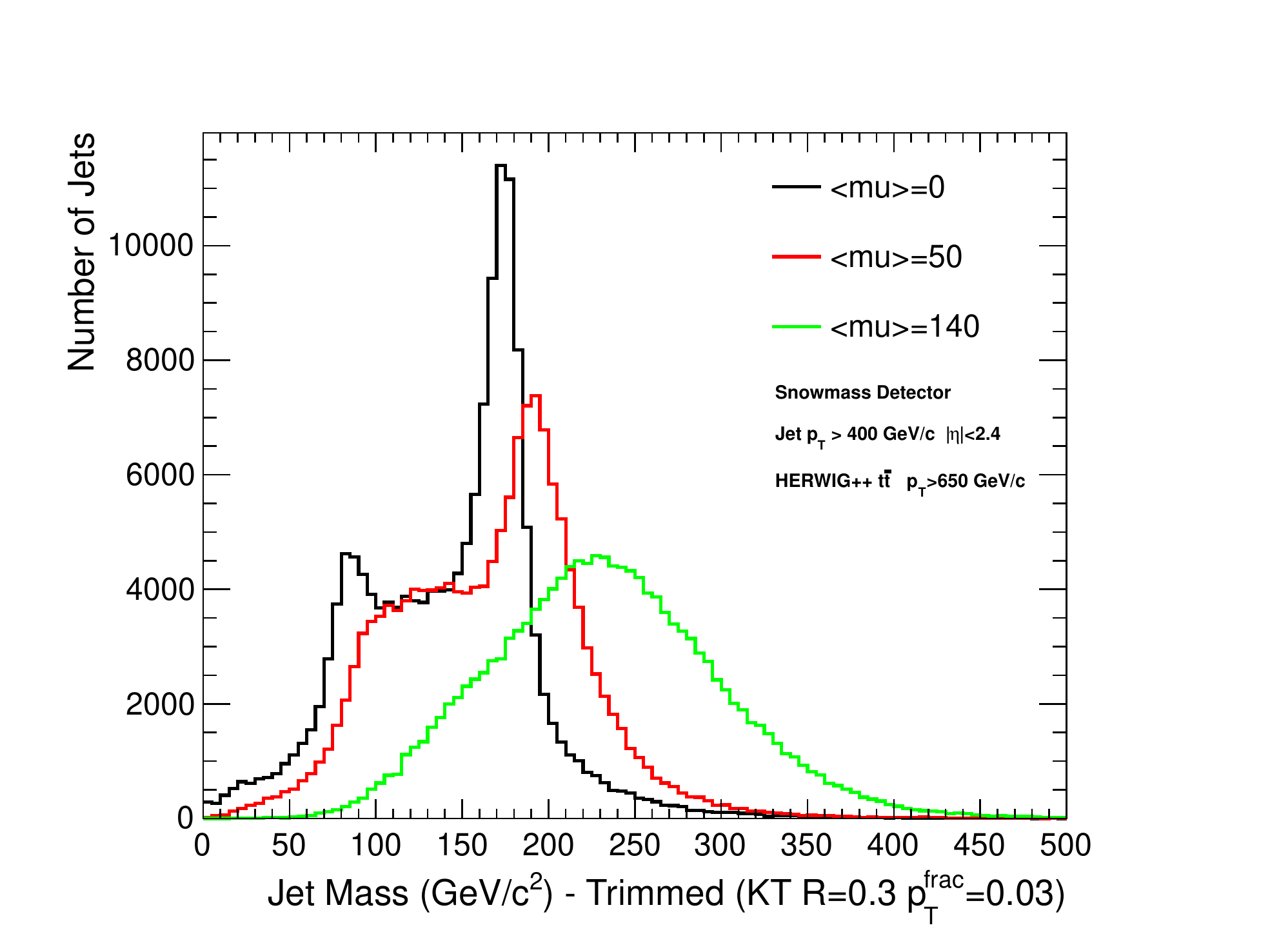}
}
\end{center}
\caption{
Trimmed jet mass for different trimmed jet parameters and for three different pileup scenarios. The C/A jet clustering algorithm is used with a distance parameter of $R = 0.8$. No area based pileup subtraction is used.  Trimmed subjets are reclustered with the $k_T$ algorithm. Samples are generated with \herwig with $p_T>650$~GeV for QCD (left) and $\ttbar$ (right).   }
 \label{fig:jet_mass_compare_kt}
\end{figure}

\begin{figure}[tbp]
\begin{center}
\subfigure[Charged hadron pileup subtraction (CHS) only]{
 \includegraphics[scale=0.38, angle=0]{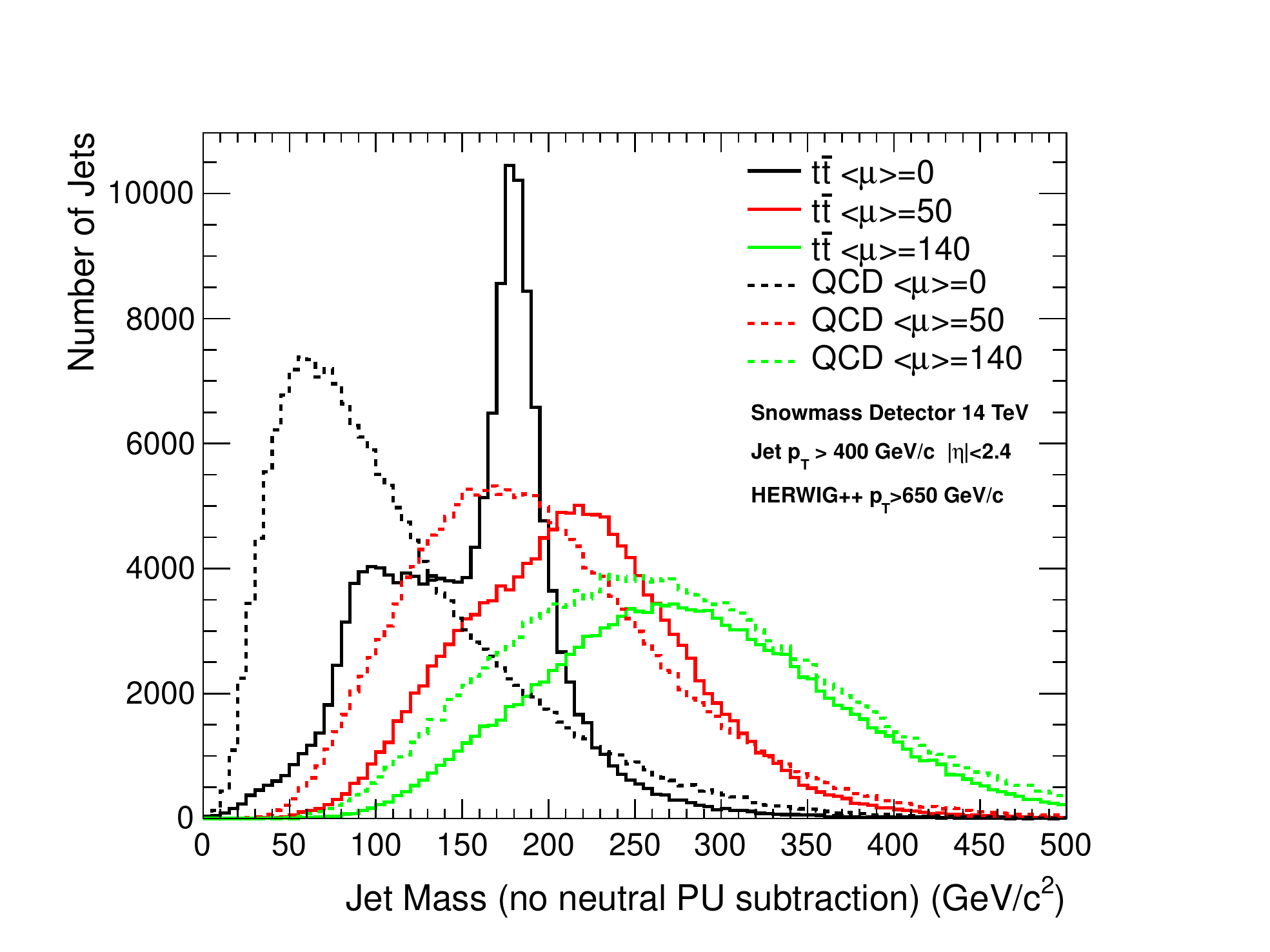}
}
\subfigure[CHS + area based pileup subtraction]{
 \includegraphics[scale=0.38, angle=0]{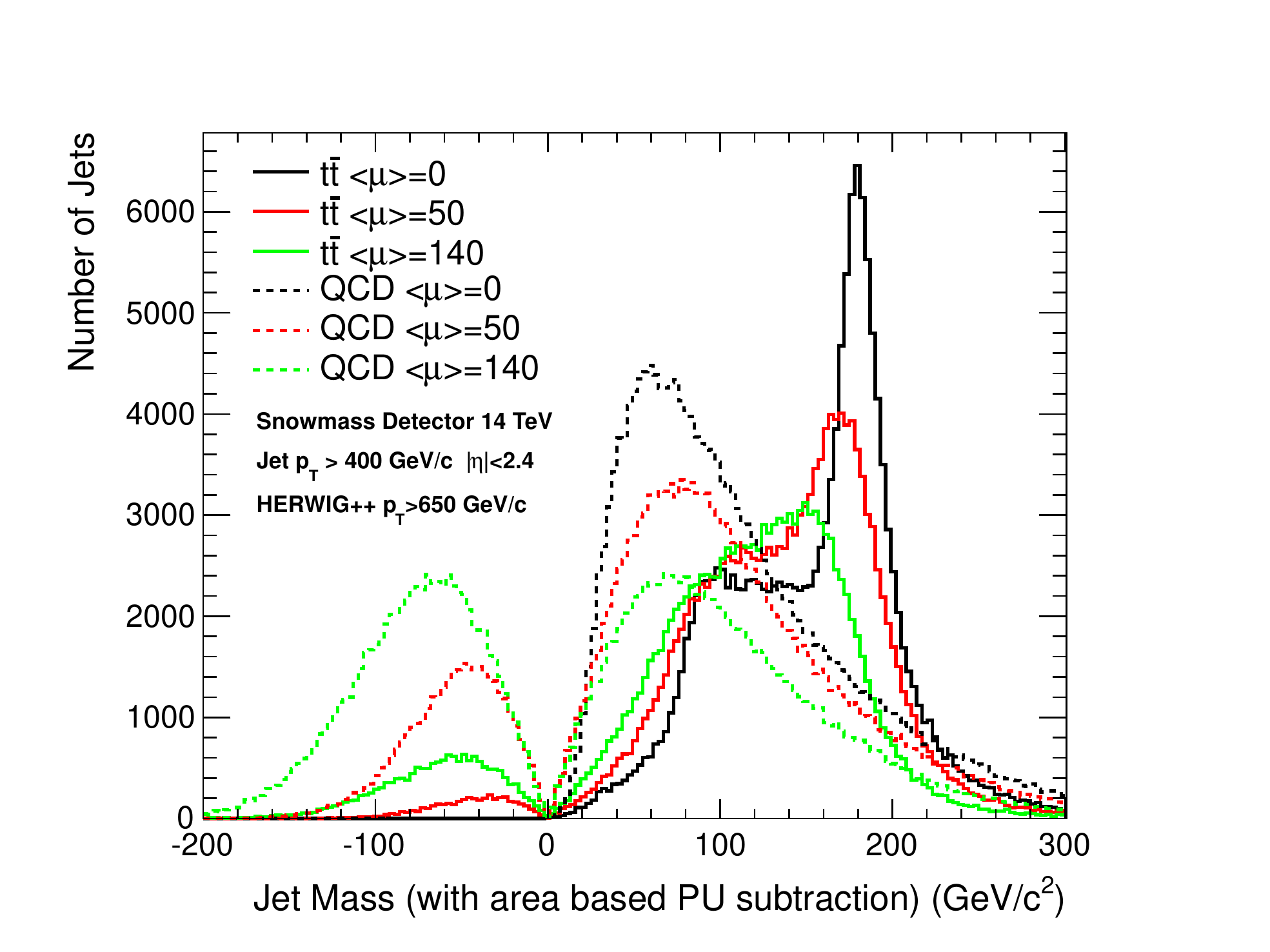}
}
\subfigure[CHS + Jet trimming (no area based correction)]{
 \includegraphics[scale=0.38, angle=0]{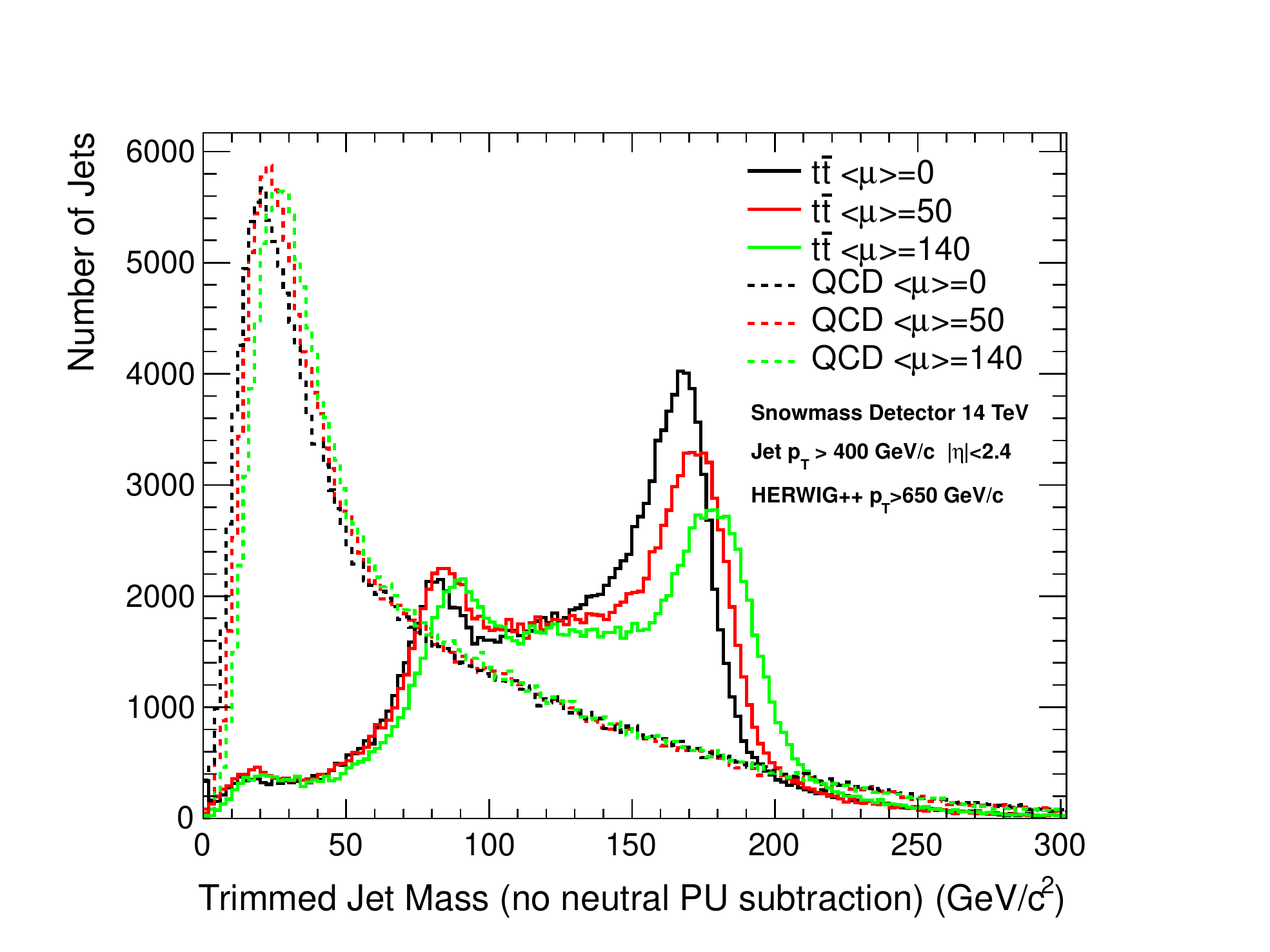}
}
\end{center}
\caption{
Comparison of reconstructed jet mass with different pileup mitigation techniques for three different pileup scenarios. The C/A jet clustering algorithm is used with a distance parameter of $R = 0.8$. The trimmed jet parameters are C/A $R=0.2$, $\ptfrac=0.05$.  Generated samples use \herwig $p_T>650$~GeV sample for QCD (dashed) and $\ttbar$ (solid).  }
 \label{fig:jet_mass_compare_pu_removal_qcd_ttbar}
\end{figure}

As an alternative to the standard area-based pileup subtraction, which corrects the total four-vector of a jet, a number of ``jet grooming'' algorithms have been developed.  These aim to selectively remove contaminating hadrons or calorimeter cells to improve jet mass resolution and to better reveal hard substructures.  Commonly-used algorithms include 
filtering \cite{butterworth-2008-100}, pruning \cite{Ellis:2009me} and trimming \cite{Krohn:2009th}.  (Detailed descriptions of the algorithms and their parameters can be found in the references.)  We focus here on trimming, which we have found to yield good results at very high pileup.
The trimmed jet mass for different trimming algorithm parameters is shown
in Fig.~\ref{fig:jet_mass_compare_ca} for reclustering the original jet with the Cambridge/Aachen (C/A) jet algorithm and in Fig.~\ref{fig:jet_mass_compare_kt} for the $k_T$ jet algorithm,
for QCD jets (left) and top jets (right).  In these plots, jets have first been clustered using C/A with $R=0.8$.  It can be seen that the trimmed jet mass can be relatively stable as a function of pileup.  For example, choosing C/A trimming with parameters $R=0.2$ and $p_T^{\rm frac} = 0.05$ (Figure~\ref{fig:jet_mass_compare_ca}b) leads to only a 10~GeV shift of the top peak for jets with $p_T \gsim 650$~GeV, even with 140 pileup events superimposed.  The QCD distribution is restored very nearly to the zero-pileup limit.

We can compare these results to both the uncorrected top-jet mass and the result of the area-based subtraction, as seen in Fig.~\ref{fig:jet_mass_compare_pu_removal_qcd_ttbar}.
Note that area-based pileup subtraction can result in spacelike jet four-momenta, and
in this situation the mass is computed via $m=-\sqrt{-(E^{2}-p^{2})}$. With 140 pileup events, the uncorrected top-jet mass is heavily broadened and its peak shifted up by about 100~GeV.  The QCD distribution is even more severely affected.  With the area-based subtraction (Fig.~\ref{fig:jet_mass_compare_pu_removal_qcd_ttbar}b), we see again a tendency to over-correct, and a significant population of ``negative mass'' events.  Figure~\ref{fig:jet_mass_compare_pu_removal_qcd_ttbar}c shows the $R=0.2$, $p_T^{\rm frac} = 0.05$ trimming case in closer detail, and represents a clear improvement over both the uncorrected jet mass and the area-based subtraction.  It might also be useful to pursue a hybrid method, that uses global event information to determine the event-by-event energy density of the pileup, coupled with a trimming procedure that is targeted to remove it (rather than using a trimming energy parameter that is a fixed fraction of the uncorrected jet $p_T$).  Regardless, we can conclude that a high-pileup environment is manageable for top-jets at the low end of the highly-boosted regime, though there remains some room for further improvements.

\subsection{Jet substructure at  $p_T > 1$~TeV}

Over the course of the next run of the LHC, and at future high-energy hadron colliders such as a 33~TeV LHC, it will be possible to observe top-jets at transverse momenta far above 1~TeV.
Such high energies are well into the highly-boosted regime, where jet substructure and jet shapes are often discussed as useful tools to study all-hadronic decays and to help reduce
the large QCD and $W$+jets backgrounds~\cite{Agashe:2006hk,Lillie:2007yh,Butterworth:2007ke,Almeida:2008tp,Almeida:2008yp,Thaler:2008ju,
Kaplan:2008ie,Brooijmans:2008,Butterworth:2009qa,Ellis:2009su,ATL-PHYS-PUB-2009-081,CMS-PAS-JME-09-001,Almeida:2010pa,Thaler:2010tr,Hackstein:2010wk,Chekanov:2010vc,Chekanov:2010gv}.
At $p_T$ above 1~TeV, pileup is not expected to be a major concern, and for the following studies we set $\mmu$ to 0.  (E.g., as in the previous subsection, pileup might first be removed using a jet grooming procedure, though it will ultimately be necessary to study the detailed interplay of pileup removal and jet substructure.)  Instead, the major issues become detector granularity and hard radiation from the top quark before it decays.

The {\sc Delphes} detector simulation that we have used for most of this report applies a particle-flow style reconstruction to the event (EFlow), though with photon energy still treated at the same spatial resolution as the hadronic neutrals.  Since the ATLAS and CMS ECALs are 4--5 times better segmented than the HCALs, this could in principle be improved even further.  However, a potentially important effect that is not modeled by {\sc Delphes} is the failure of precision tracking in the cores of very high-$p_T$ jets, both due to the higher hit density and the fact that straighter tracks yield poorer momentum measurements.  Such effects could only be modeled in full ATLAS and CMS detector simulations.  The calorimetry, by contrast, should only improve as the energy increases.  Therefore, in this subsection, we conservatively set aside the possibility of using the tracker, and focus only on calorimeter quantities\footnote{The pattern of hits in the tracker would still yield valuable directional and energy information, even if track reconstruction proves unreliable.  This might be especially useful given the generally non-projective geometry of the calorimeters, the spread in vertex longitudinal positions, and the nonzero radius of particle showers within the calorimeter material.  We do not attempt to model any of these effects, nor any of the potential compensating benefits of adding back in the tracker hits.}. 

The default {\sc Delphes} barrel calorimeter granularity for Snowmass 2013 is close to $0.087 \times 0.087$ in the $\eta$-$\phi$ plane.  This is adequate to achieve good mass resolution on combinations of widely-separated jets, but the masses of individual top-jets can be significantly degraded.  It is therefore crucial to at least combine in some of the higher spatial resolution information from the ECAL.  We model this in a very simplified manner, by running dedicated {\sc Delphes} simulations with a 2$\times$ refinement in the $\eta$ and $\phi$ granularities.  Each tower is therefore broken up into four.  We reassemble the hadronic towers back to the original calorimeter grid, and keep the ECAL towers at the refined resolution.  There are then several ways to exploit the added information in the ECAL towers.  We use one of the simplest:  the four ECAL towers associated to each HCAL tower are vector-summed, and the resulting direction is assigned to the complete ECAL+HCAL tower energy.\footnote{{\sc Delphes} also does not model energy deposits in the ECAL from hadrons, whether by direct interaction with the ECAL material or by secondary photons produced from interactions with the tracker material.  The net effect of these processes is a substantial increase in the total fraction of jet energy captured by the ECAL, from roughly 30\% to something closer to 50\% (80\%) in CMS (ATLAS).  This only improves the association of the ECAL energy depositions with the full energy flow pattern.}

\begin{figure}[tbp]
\begin{center}
 \includegraphics[scale=0.4, angle=0]{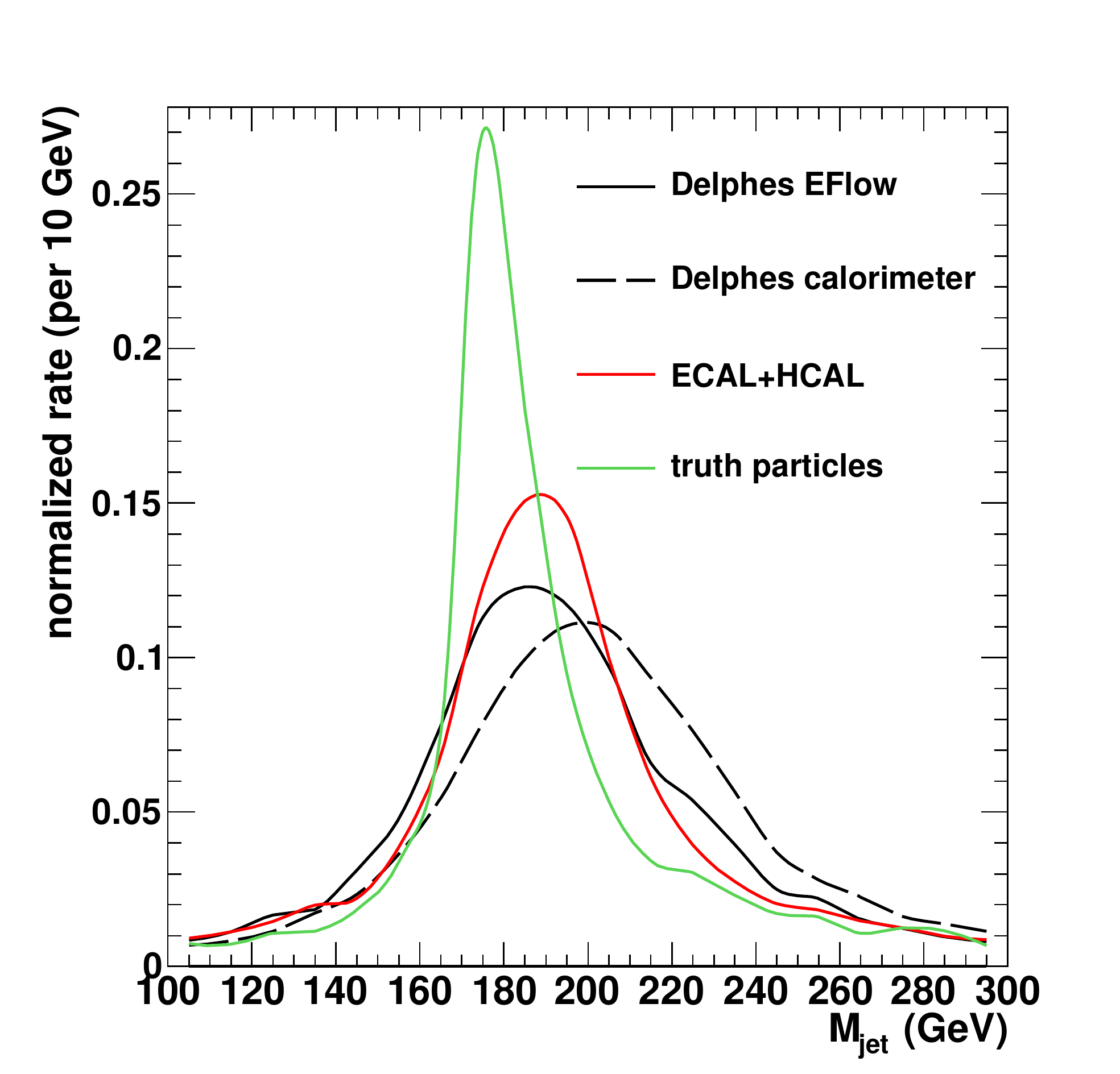}
\end{center}
\caption{
Top-jet mass distributions for $p_T$(jet)$ > 1.6$~TeV, using the anti-$k_T$ algorithm with $R=0.5$.  Different curves correspond to different treatments of the {\sc Delphes} detector, described in the text.  No pileup events have been overlayed, and no four-momentum subtractions have been applied.   
}
\label{fig:detector_comparison}
\end{figure}

Figure~\ref{fig:detector_comparison} shows the distribution of anti-$k_T$ (R=0.5)  top-jet masses for several treatments of the detector, taking only jets with $p_T > 1.6$~TeV.  It can be seen that the original {\sc Delphes} calorimeter has worse mass resolution than the EFlow model, but that adding in the ECAL information improves the calorimeter resolution even beyond EFlow.  This not only indicates the possible improvement from taking a finer-grained view of the photons in the event, but also that the photons trace the energy flow as faithfully as the tracks.  The figure also shows the underlying particle-level distribution, thereby indicating the relative degree of mass smearing from the detector model versus from physics.

As demonstrated earlier in Fig.~\ref{fig:jetmass_pileupsFit}, the mass resolution degrades with increasing $p_T$.  Roughly speaking, there are three components contributing  to this trend.  The first is the detector itself, as just illustrated.  The second is diffuse soft radiation in the jet.  This more generally would include GeV-scale pileup particles, but it also includes a higher-energy contribution from showering of the primary event, such as FSR off of the top quark before it decays.  The third component to the mass broadening is the hard region of these emissions, for example those which could result in the formation of additional hard jets in the event.  These emissions are often correlated in direction with the top quark, and can become merged into the top-jet after clustering.  The net effect of the extra hard radiation is a broad high-mass tail.  The effect can already be seen in Figs~\ref{fig:jetmass_pileupsFit} and~\ref{fig:detector_comparison}.  While there is no true physical dividing line between the ``soft'' and ``hard'' components of the radiation, we can call ``hard'' any emission that causes the jet mass to fall far outside of the core distribution.

\begin{figure}[tbp]
\begin{center}
 \includegraphics[scale=0.4, angle=0]{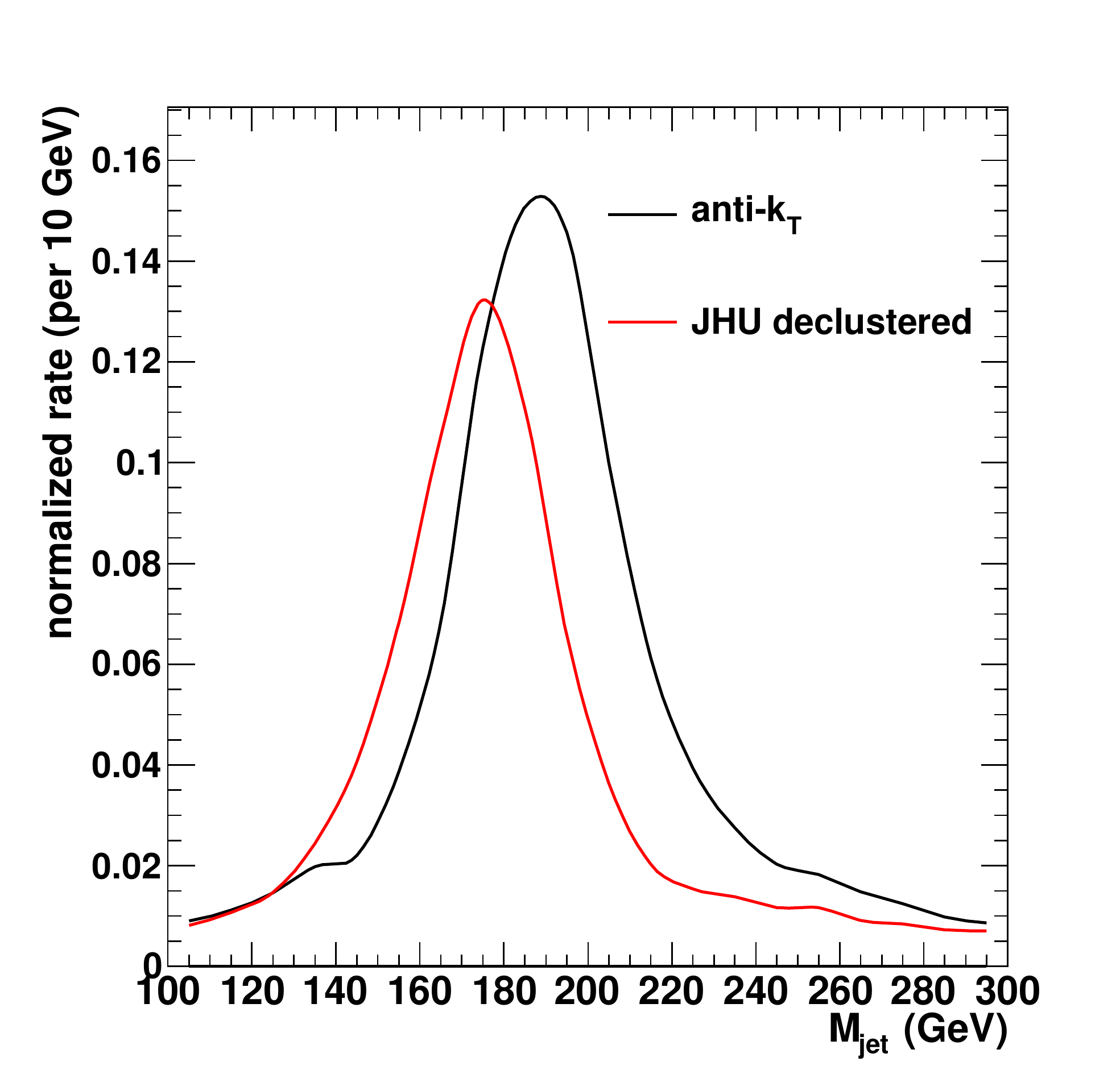}
\end{center}
\caption{
Top-jet mass distributions for $p_T$(jet)$ > 1.6$~TeV, using the ECAL+HCAL model.  The black curve is the anti-$k_T$ (R=0.5) jet mass.  The red curve is sum of subjet masses found with the JHU top-tagger, run on the original jet constituents with $\delta_p = 5\%$ and $\delta_r = 0.085$.  The rate for finding 3 or 4 subjets is 80\%.
}
\label{fig:declustering}
\end{figure}

Many substructure-based top-taggers automatically subtract contaminating radiation from within a top-jet, and we start by demonstrating how effective this is in the context of the JHU top-tagger in the absence of pileup.  For these studies, we recluster the constituents of the anti-$k_T$ (R=0.5) jets with the Cambridge/Aachen algorithm.  We then run the declustering stage of the JHU tagger with parameters $\delta_p = 5\%$ and $\delta_r = 0.085$, and place no cuts other than the requirement of successful identification of at least three subjets.  The $\delta_r$ parameter here serves as a collinear cutoff in $\eta$-$\phi$ space, and is smaller than has been used elsewhere, reflecting the higher $p_T$ scale and the use of finer-grained detector elements.  Figure~\ref{fig:declustering} shows a comparison between the masses of anti-$k_T$ (R=0.5) jets before and after declustering.  The declustering refines mass resolution in the core of the distribution and corrects the peak back to the true top mass.  The ``price'' for this refinement is an approximately 20\% loss in efficiency, due to top-jets that fail to properly decluster according to our criteria.  In addition to the top/W mass and decay angle distributions, successful declustering is one of the ways that the JHU top-tagger discriminates top-jets from ordinary QCD-jets.

\begin{figure}[tbp]
\begin{center}
 \subfigure[$p_T$(jet) $> 1.6$~TeV (LHC14)]{
 \includegraphics[scale=0.32, angle=0]{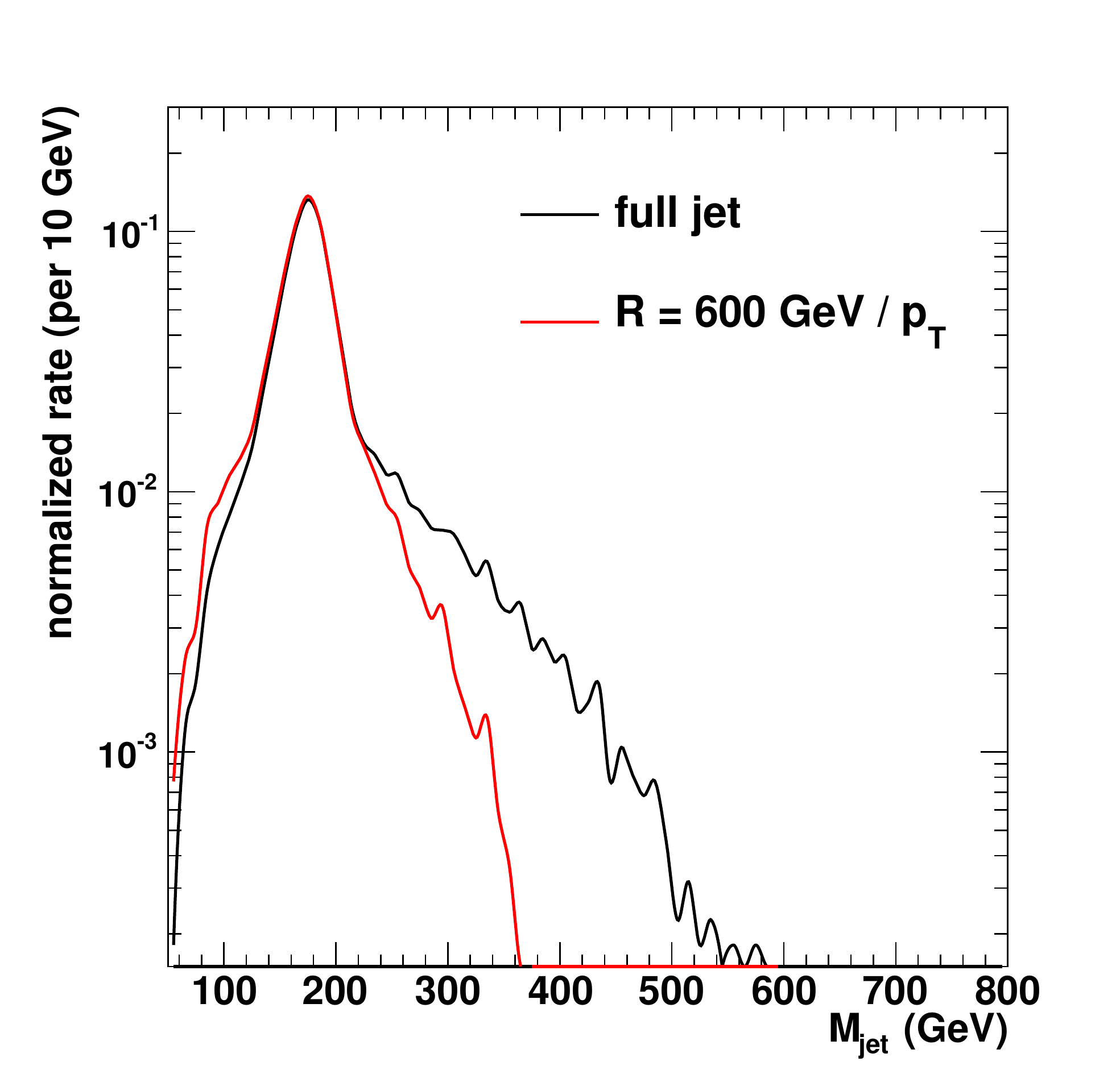}
 }
 \subfigure[$p_T$(jet) $> 3$~TeV (LHC33)]{
 \includegraphics[scale=0.32, angle=0]{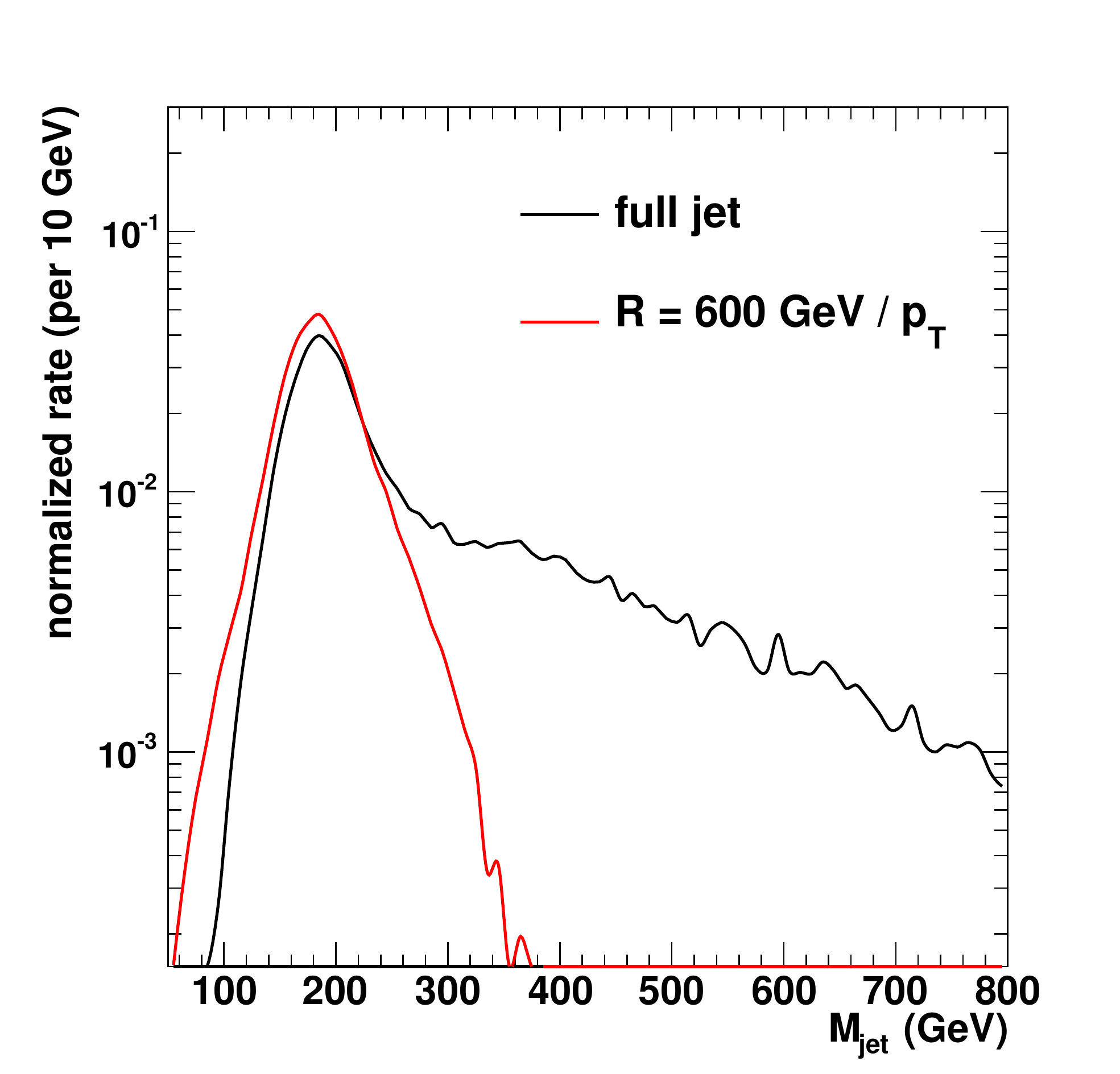}
 }
\end{center}
\caption{
Top-jet mass distributions after JHU subjet-finding for (a) $p_T$(jet)$ > 1.6$~TeV at the 14~TeV LHC, and (b) $p_T$(jet)$ > 3$~TeV at the 33~TeV LHC.  The black curves show the effect of running the reclustering/declustering on all constituents of the original anti-$k_T$ (R=0.5) jet.  The red curves show the effect of restricting the C/A reclustering radius to $R = 600$~GeV/$p_T$(jet).
}
\label{fig:tail}
\end{figure}

Since the JHU tagger is constructed to identify localized high-$p_T$ subjets within a jet, it is not capable of removing QCD emissions above the $\delta_p$ scale.  Consequently, the declustered mass distributions still exhibit a broad radiative tail.  While this represents only a $O(10\%)$ fraction of top-jets in the $p_T > 1.6$ TeV sample, the tail grows as the top-jet $p_T$ increases.  This can be seen in Figure~\ref{fig:tail}, which includes a more extreme case of top-jets with $p_T > 3$~TeV at the 33~TeV LHC.  Removal of this hard contamination might be possible with a modified declustering procedure that folds in more information about the top mass.\footnote{The {\sc HEPTopTagger} is of this type.  However, its performance in the very high-$p_T$ range is not ideal, as it tends to significantly overgroom the top-jets for a broad choice of parameters.}  A simple workaround is to just reduce the active size of the top-jet.  The decay particles of the top are confined within a cone of $R \sim 4m_t/p_T$, whereas the FSR emissions of the top mainly occur outside of this region (the ``dead cone'' phenomenon).  Shrinking the active area also reduces the chance of picking up hard ISR.  As an illustration of such a procedure, we apply a maximum radius to the C/A reclustering step of $R = 600\;{\rm GeV}/p_T({\rm jet})$.  Figure~\ref{fig:tail} shows how the modest tail of the 1.6~TeV sample and the more significant tail of the 3~TeV sample are both suppressed, while the central peak is enhanced.  Nonetheless, some radiative broadening remains, and attempting to further ameliorate the effect by using even smaller radii causes the distribution to instead develop a {\it low}-mass tail and a larger declustering failure rate due to incomplete containment.  Similar behavior is observed at particle-level.  Whether this represents a fundamental limitation to the quality of boosted top kinematic reconstruction is a question that would be interesting to explore.

\begin{figure}[tbp]
\begin{center}
 \includegraphics[scale=0.4, angle=0]{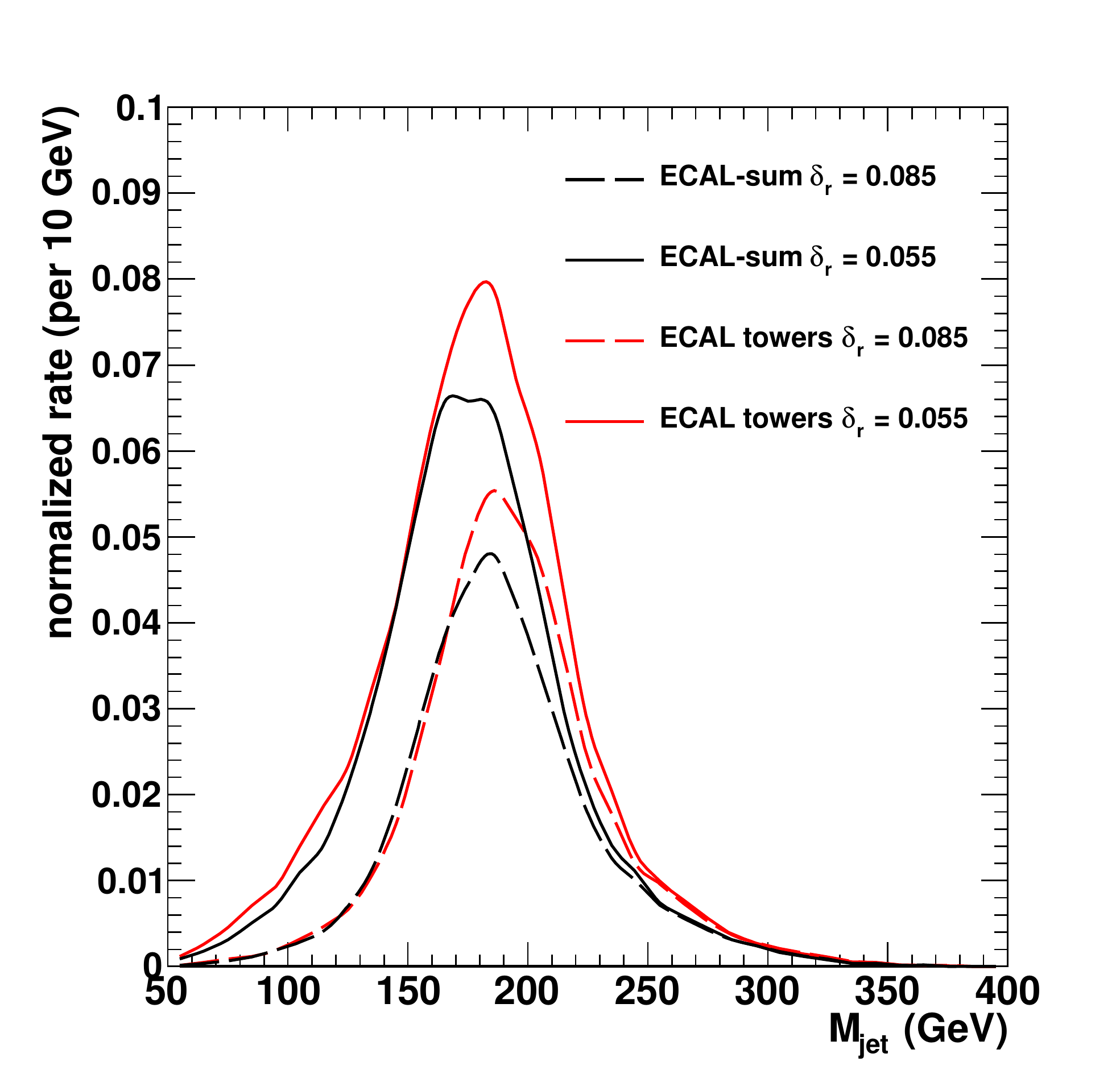}
\end{center}
\caption{
Top-jet mass distributions for $p_T$(jet)$ > 3$~TeV, using different ECAL+HCAL models and different $\delta_r$ settings.  All have been reclustered with $R = 600$~GeV/$p_T$(jet).  Black curves show nominal results where the ECAL towers have been summed to determine the vector of each ECAL+HCAL unit.  Red curves show results where the ECAL towers are instead left at their original spatial resolution, and uniformly rescaled to the full ECAL+HCAL energy.  Dashed lines are the nominal $\delta_r = 0.085$, and solid lines are the more refined $\delta_r = 0.055$.  The highest declustering efficiency (solid red) is 70\%, and the lowest efficiency (dashed black) is 38\%.}
\label{fig:JHU_shrinkingCone_differentResolutions_pt3000}
\end{figure}

One other feature of Figure~\ref{fig:tail} that is easy to see is that the overall declustering rate falls significantly at these extremely high $p_T$'s, from roughly 75\% at 1.6~TeV to 38\% at 3~TeV (both with the shrinking-$R$).  Even though some of the radiative tail events are recaptured, most of these fail as well.  This residual inefficiency is partially due to the limitations of the ECAL+HCAL detector model, but mostly because the $\Delta R$ scales are shrinking while $\delta_r$ is held fixed at 0.085.  Figure~\ref{fig:JHU_shrinkingCone_differentResolutions_pt3000} shows what happens when $\delta_r$ is reduced to 0.055, and also if we further refine the detector model.  For the latter, the four ECAL towers associated to each HCAL tower are uniformly rescaled to the full ECAL+HCAL energy (as suggested in~\cite{Katz:2010mr,Son:2012mb}), rather than first taking the ECAL towers' vector-sum.  This treatment allows approximate resolution of subjets smaller than an HCAL tower.  The smaller $\delta_r$ and finer-grained detector each allow more jets to pass the declustering, up to 70\% for the latter, though the mass resolution is clearly suffering.  For the two modifications, the Gaussian width increases from 30~GeV to 35~GeV.


In summary, while a simple treatment of the detector and top-tagging algorithms can lead to badly degraded mass resolution and failures of declustering at $p_T$ much larger than 1~TeV, much of the discriminating power can in fact be recovered.  Different subdetectors with different spatial resolutions can be combined to allow small-scale substructures to survive, even if tracker-based particle flow methods are not applied or break down.  In the context of the JHU tagger, reducing $\delta_r$ appears to be crucial to take advantage of these smaller structures.  Contaminating radiation from the hard event is a persistent nuisance, and grows with momentum unless the top-jet radius is proportionally reduced.  There is a remaining tension between eliminating this radiation and eliminating the top decay products, a problem that may benefit from a more sophisticated substructure strategy.

\subsection{Summary}

Studies of high-$\ptjet$ physics require boosted techniques to identify top decays.
Below we summarize the main conclusions:

\begin{itemize}

\item
Future LHC runs will have large enough statistics to reach 
the highly-boosted regime for top quark reconstruction. 
The transverse 
momenta of top quarks can be high enough ($\ptjet \gsim 700$~GeV) that all top decay products can be contained
within a standard jet radius of 0.4--0.6, and resolved approaches become ineffective.
Even larger statistics will be accumulated within the semi-boosted regime ($p_T \simeq 350$--700~GeV), where resolved
approaches still often fail.  In this intermediate regime, it is also possible to collect
the decay products within a single jet, provided it is a ``fat jet'' with $R \simeq 1.0$. 

\item  Pileup in the highly-boosted regime is manageable .  Standard jet area subtraction methods can preserve much of the structure of the hadronic top-jet mass distributions, though there is a tendency to over-correct.  Other jet grooming methods such as trimming, pruning, and filtering should be exploited by appropriately tuning their parameters.  These methods can also be easily interfaced with top-taggers that probe the jet's detailed substructure, rather than just its total four-vector.  Semi-boosted tops are expected to have pileup sensitivity intermediate between the highly-boosted and threshold tops, and would be worth a dedicated study.

\item  Hadronic top-jet mass resolution degrades with increasing $p_T$ due to a combination of physics and instrumental effects.  Jet substructure approaches to top-tagging can partially compensate by cleaning out radiation from FSR and ISR.  A remaining hard radiative component can be diminished by using a jet radius that shrinks with $p_T$.  The instrumental limitations will require detailed study under full ATLAS and CMS detector simulation, but simple studies with a spatially-refined {\sc Delphes} calorimeter suggest that combining information from detector subsystems can substantially improve the mass resolution.  Top-tagging with jet substructure is still viable even at multi-TeV transverse momenta.



\end{itemize}

\clearpage
\section{Top quarks at Linear Colliders}
\label{sec:sec_lc}
Generally speaking, a $e^+e^-$ linear collider (LC) is the ideal machine for high precision top quark physics~\cite{bib:ilc-tdr-dbd,bib:clic-cdr}. 
In contrast to the situation at hadron colliders, the leading-order pair production process $\ee\to t \bar t$ goes directly through the $t\bar{t} Z^0$ and $t\bar{t} \gamma$ vertices.  There is no concurrent QCD production of $t$~quark pairs, which increases greatly the potential for a clean measurement. This section will present experimental details on physics potential, which has been published
elsewhere.

\subsection{Experimental environment}
A future linear collider will operate at centre-of-mass energies between 250\,GeV up to 3\,TeV thus covering the kinematical regime for top pair production, which starts at around 350\,GeV. Being the most advanced proposal, the ILC is designed to produce an integrated luminosity of $500\,\invfb$ at a centre-of-mass energy of $500\,\gev$ within four years of running. A major asset is that the ILC allows for running with polarized beams. At $500\,\gev$ the expected beam polarization is $\pem=\pm80\%$ for the electron beam and $\pep=\mp30\%$ for the positron beam.The production cross section varies with the polarization according to

\beq
\sigma_{\pem,\pep} = \frac{1}{4}\left[(1- \pem \pep)(\sigma_{-,+}+\sigma_{+,-})+(\pem-\pep)(\sigma_{+,-}-\sigma_{-,+})\right] 
\label{eq:tot-cross}
\eeqn

At $500\,\gev$ about 300k top pair events are expected in the case of non-polarized beams for an integrated luminosity of $500\,\invfb$.
At energies above the $\ttbar$ threshold different polarization modes permit to enrich the samples with $t$~quarks of a given helicity. Due to the electroweak production all processes occur at roughly the same rate and the total number of interactions is such that no online trigger is needed. The event selection and classification happens exclusively offline. In the energy regime below $1\,\tev$ background effects like $\gamma\gamma$ with low $p_t$ hadrons do not compromise the precision measurements, see also later in the text. 
Unlike LHC, pile-up and underlying event will not affect top reconstruction at the LC.

Details of linear collider detectors can be found elsewhere~\cite{bib:ilc-tdr-dbd,bib:clic-cdr}. Their design assures an optimal jet energy resolution based on particle flow. The aim is to measure every particle of the final state. The momentum of charged  final state particles  are measured with a high precision tracking system and the energy of neutral particles in the calorimeters. A separation of charged and neutral final state particles require highly granular calorimeters~\cite{Brient:2002gh}. Jet energy resolutions of about 4\% for jet  energies between $250\,\gev$ and $1\,\tev$ seem to be achievable.

A most relevant component for top quark physics is the vertex detector. The first layer of the vertex detector approaches the interaction point as close as 16\,mm.  The small material budget and the comparatively low occupancy allow for a precise determination of primary and secondary (and tertiary) interaction vertices. This potential is exploited in tagging of the $b$~quarks.
jets and the measurement of the $b$~quark charge. 

\subsubsection{Event generation and analysis framework}

All the studies presented in this section are based on a full detector simulation. For ILC studies events typically are generated with version 1.95 of the {\sc WHIZARD} event generator~\cite{Kilian:2007gr,Moretti:2001zz} in the form of six fermion final states of which $\ttbar$ events form a subsample. The generated events are then passed to the {\sc PYTHIA} simulation program to generate parton shower and subsequent hadronization. For the study of the $t$~quark mass, as carried our for the CLIC CDR,  {\sc PYTHIA} was also used for the generation of the signal for parts of the background.  Events are subject to a full simulation of the LC detectors and subsequent event reconstruction. For this the software frameworks {\sc Marlin} and the SiD reconstruction are available. 

\subsection{Top quark reconstruction}

\subsubsection{Semi-leptonic channel} 

The charged lepton in semi-leptonic decays (see the definition in Sect.~\ref{sec:intro}) allows for the determination of the  $t$~quark charge. The $t$~quark mass is reconstructed from the hadronically decaying $\Wboson$~boson, which is combined with one of the $\bottom$~quark jets. All jets are reconstructed using the Durham jet algorithm. Leptons are identified using typical selection criteria. The authors of~\cite{ild:bib:benchmark:doublet,Amjad:2013tlv} propose a method in which the lepton from the $W$~boson decay is either the most energetic particle in a jet or has a sizable transverse momentum w.r.t. neighbored jets. More specific the following criteria are applied
\beq
x_T=p_{T,lepton}/M_{jet} > 0.25\,\,\,\,\,\mathrm{and}\,\,\,\, z=E_{lepton}/E_{jet} >0.6,
\eeqn
where $E_{lepton}$ is the energy and $p_{T,lepton}$ the transverse momentum of the lepton within a jet with energy $E_{jet}$ and mass $M_{jet}$.
The efficiency to identify the decay lepton is about 70\% where the selection has a tendency to reject low momentum leptons. This number includes $e$, $\mu$ and $\tau$~leptons.
Alternatives to this method isolate the lepton within a cone. The selection efficiencies in that case are somewhat smaller and are not applied for $\tau$ leptons.

\begin{figure}
\begin{center}
\includegraphics[width=0.7\textwidth]{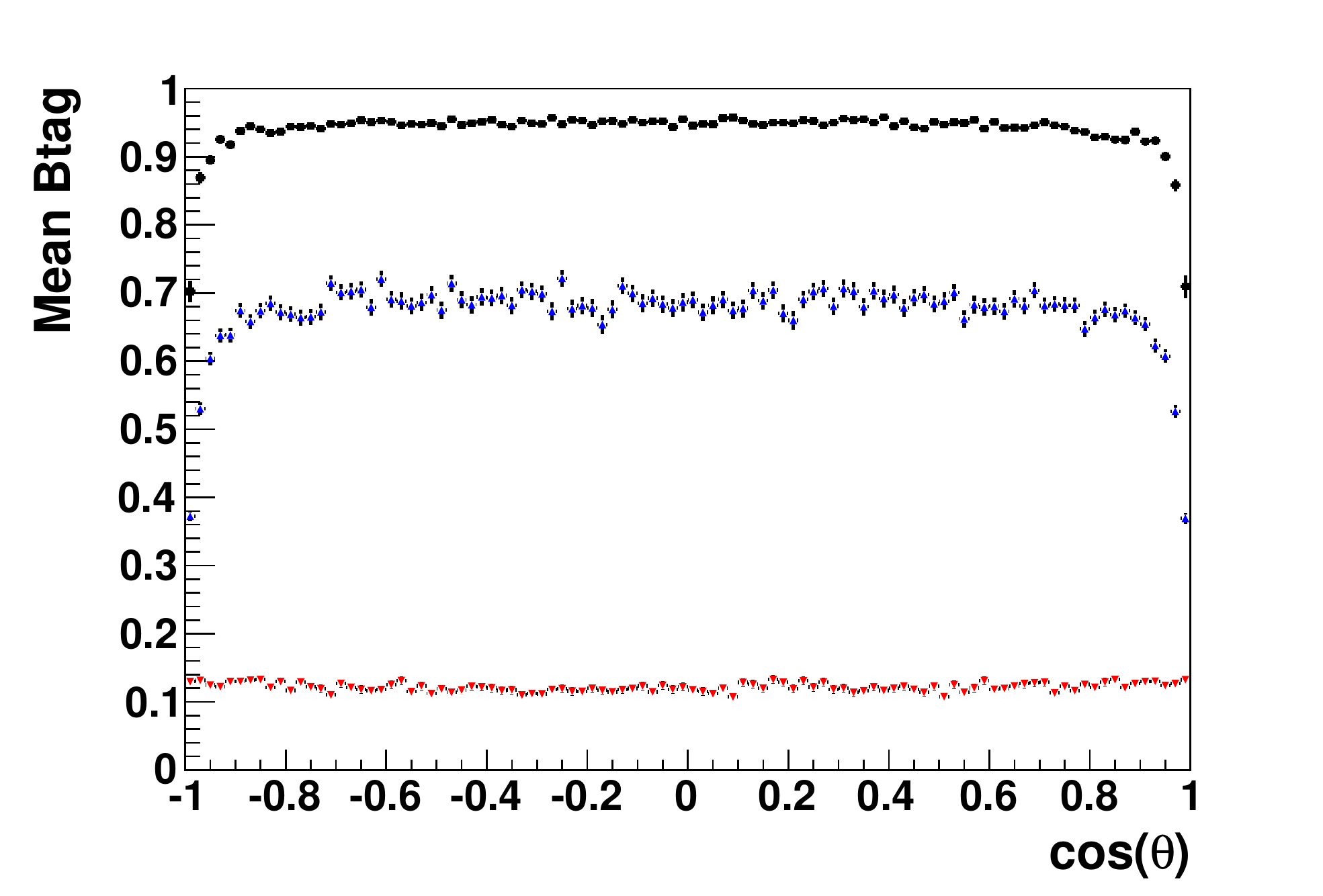}
\caption{\sl The $b$-tag values as a function of the polar angle of the jets with the highest $b$-tag value (black dots) and of that with the second highest $b$-tag value (blue dots). The third line (red dots) shows the $b$-tag values for light quark jets. }
\label{fig:bt}
\end{center}
\end{figure}

Two of the remaining four jets must be identified as being produced by the $\bottom$~quarks of the $\tpq$~quark decay. 
The $b$-likeness or {\em b-tag} is determined with the {\tt LCFIPlus} package. Secondary vertices in the event are analyzed by means of the jet mass, the decay length and the particle multiplicity. The jets with the highest $b$-tag values are selected. As shown in Fig.~\ref{fig:bt} the higher $b$-tag value is typically 0.92 while the smaller one is still around 0.65. Jets produced by light quarks lead to a considerable smaller $b$-tag value.

These values are nearly independent of the polar angle of the $b$~quark jet but drop towards the acceptance limits of the detector. Finally, the two remaining jets are associated with the decay products of the $W$~boson. 

The final reconstruction is achieved by the correct pairing of $b$~quark jets and $W$~bosons. The method for the pairing depends largely on the purpose of the analysis. In case of a determination of the $t$ quark mass the selection targets events close to the $t$ mass peak, while in the analysis of $t$ quark couplings events from the Breit-Wigner tails of the top mass spectrum are allowed. 

In case of top mass measurements the final pairing is realized by a kinematic fit using constraints based on the assumption of a $\ttbar$ event to improve the precision on the parameters of interest. The kinematic fit as used in~\cite{Seidel:2013sqa} retains about 50\% of events classified as semi-leptonic top decays.  
An alternative is to assign these constraints to a test variable and select events with a minimum value of the test variable. The selection efficiency of this variable yields about twice as much events as the kinematic fit. 

The total selection efficiency depends also on the lepton identification. The authors of~\cite{Amjad:2013tlv} give a total selection efficiency of about 56\% for semi-leptonic $\ttbar$ events including events with a $\tau$~lepton in the final state. 

\subsubsection{Fully hadronic channel}
When the $\ttbar$ pair decays fully hadronically the final state comprises at least six jets (see the definition in Sect.~\ref{sec:intro}). The $b$~jets are identified in the same way as in the semi-leptonic case. As the number of combinations for the pairing into top quarks is higher a correct $b$-tagging is of even bigger importance than in the semi-leptonic case. Typical selection efficiencies of fully hadronic $\ttbar$ events are 20-30\%.

\subsection{Background rejection}

Table~\ref{tab:channel-production} taken from~\cite{Seidel:2013sqa} gives an overview of relevant backgrounds to $\ttbar$ production at the $\ttbar$ threshold and at a centre-of-mass energy of $500\,\gev$. 

\begin{table}
\centering
\begin{tabular}{c|c|c|c}
\hline
type & final & $\sigma$ & $\sigma$\\
& state & 500 GeV & 352 GeV\\
\hline
\hline
Signal ($m_{\rm{top}}$ = 174 GeV)& $t \bar t$   & 530 fb& 450 fb\\
\hline
Background& $W W$       & 7.1 pb & 11.5 pb      \\
Background& $Z Z$       &  410 fb & 865 fb \\
Background& $q \bar q$  &  2.6 pb & 25.2 pb\\
Background& $W W Z$     &  40 fb & 10 fb\\
\end{tabular} 
\caption{Signal and considered physics background processes, with their approximate cross section calculated for centre-of-mass energies of 500\,GeV 352\,GeV.}
\label{tab:channel-production}
\end{table}

\begin{figure}
\centering
  \includegraphics[width=0.49\textwidth]{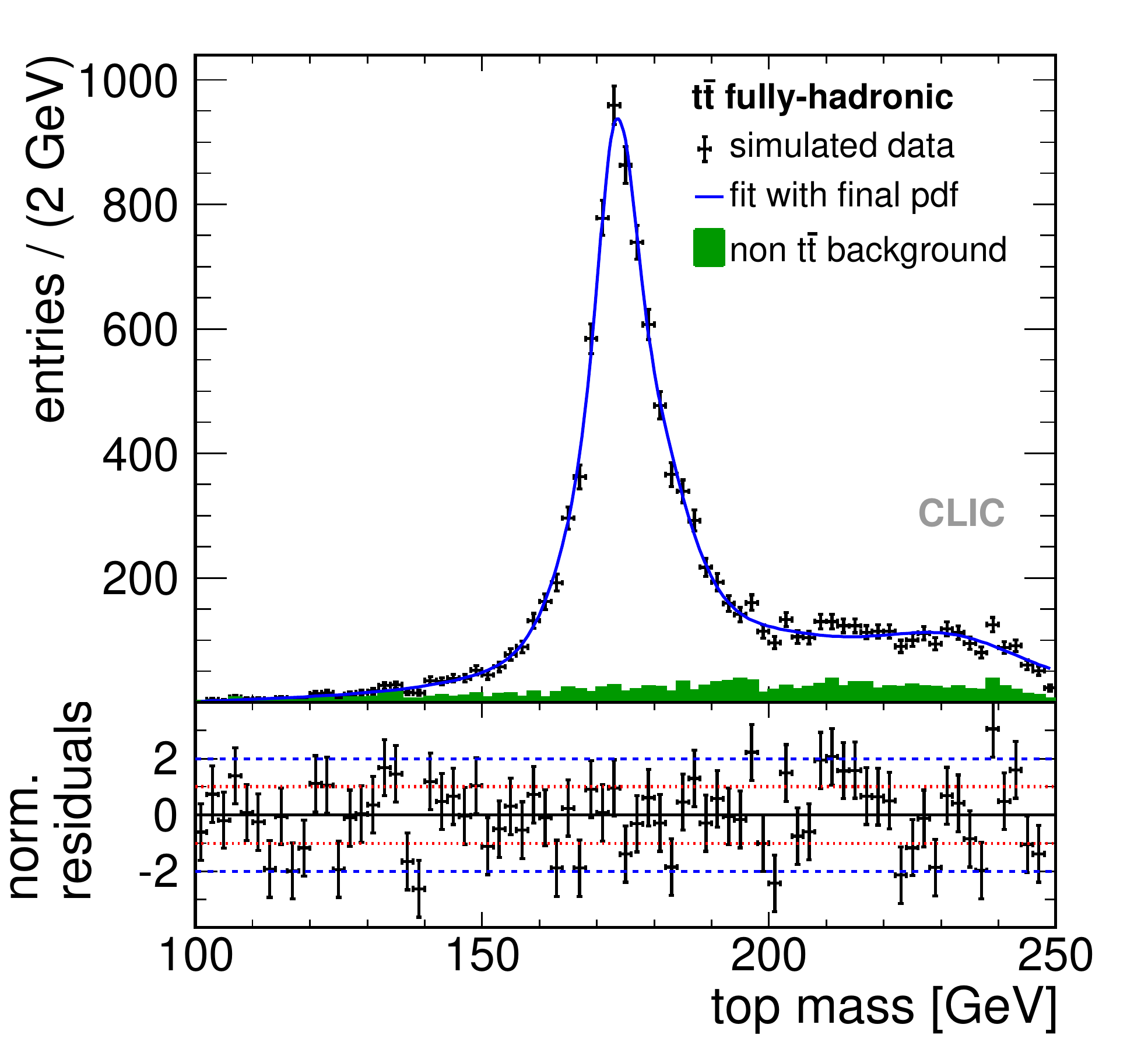}
   \includegraphics[width=0.49\textwidth]{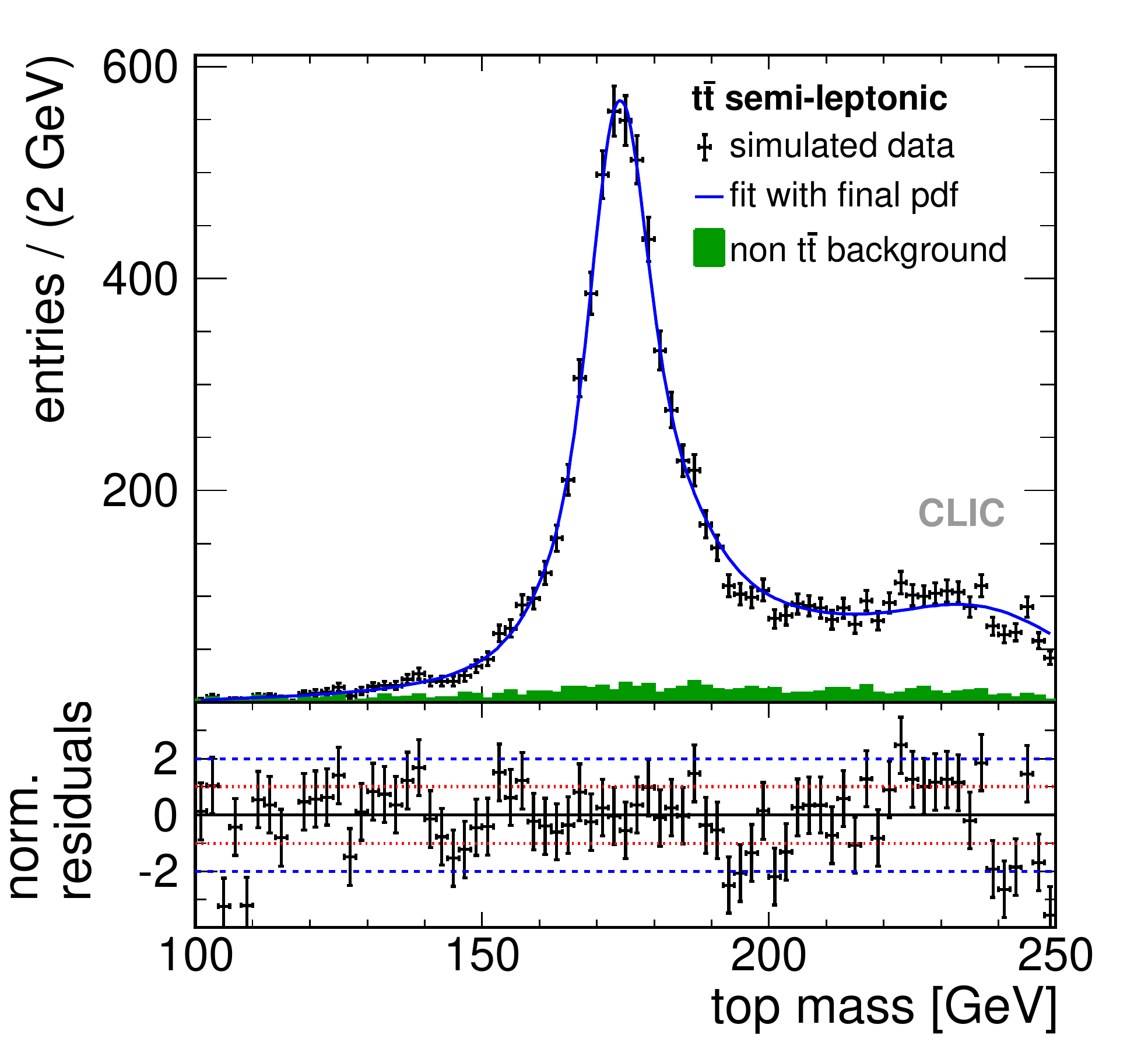}
        \caption{Distribution of reconstructed top mass for events classified as fully-hadronic (left) and semi-leptonic (right). The data points include signal and background for an integrated luminosity of 100\,fb$^{-1}$. The pure background contribution contained in the global distribution is shown by the green solid histogram. The top mass is determined with an unbinned likelihood fit of this distribution, which is shown by the solid line.}
   \label{fig:TopMass}
 \end{figure}

These background processes can be very efficiently suppressed. A tool common to all analysis is the $b$~tag value. This suppresses about 97\% of the dominant $W W$ background. The authors of~\cite{ild:bib:benchmark:doublet} exploit also the event thrust which rejects in addition 80\% of the remaining events. In~\cite{Devetak:2010na} and~\cite{Seidel:2013sqa} it is shown that the kinematic fit is also a powerful tool to suppress background. The detailed set of cuts to suppress background depends also on the actual analysis.  Fully hadronic $t$~quark decays are a background to semi-leptonic $t$~quark decays and vice versa. Additional cuts comprise cuts on the $t$~quark and $W$ mass and of the invariant mass of the hadronic total hadronic final state. In~\cite{Seidel:2013sqa} a binned likelihood technique is applied to suppress residual background. All analyses manage to suppress the background to a negligible level as is shown in Fig.~\ref{fig:TopMass} published in~\cite{Seidel:2013sqa}. This figure shows the resulting top mass spectrum at a centre-of-mass energy of $500\,\gev$.    

\begin{figure}[h]
  {\centering 
 \subfigure[Durham]{\label{fig:jetrecowhad_a} \includegraphics[width=0.45\textwidth]{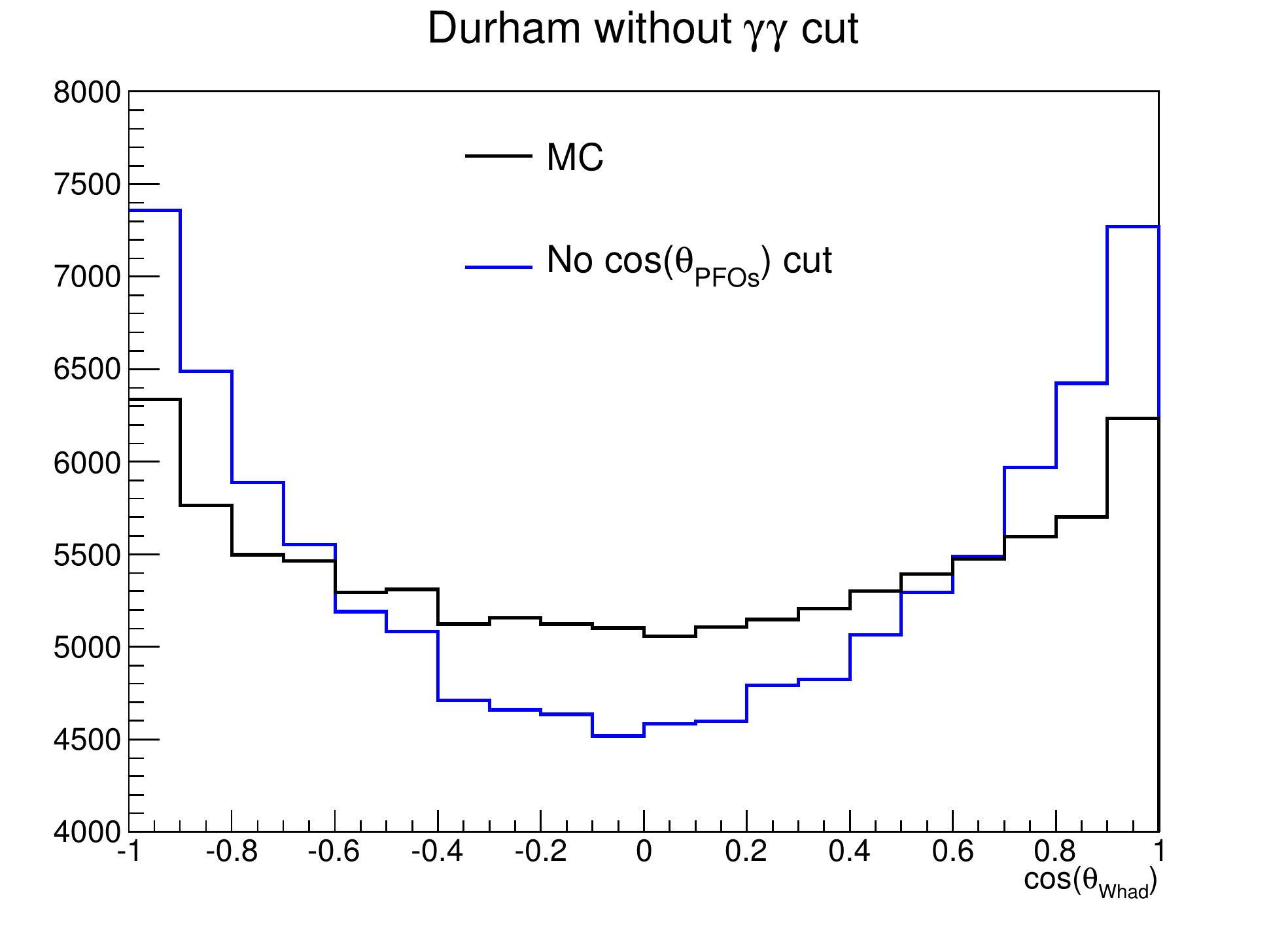}}
\subfigure[$k_t$, $R=1.5$]{\label{fig:jetrecowhad_d} \includegraphics[width=0.45\textwidth]{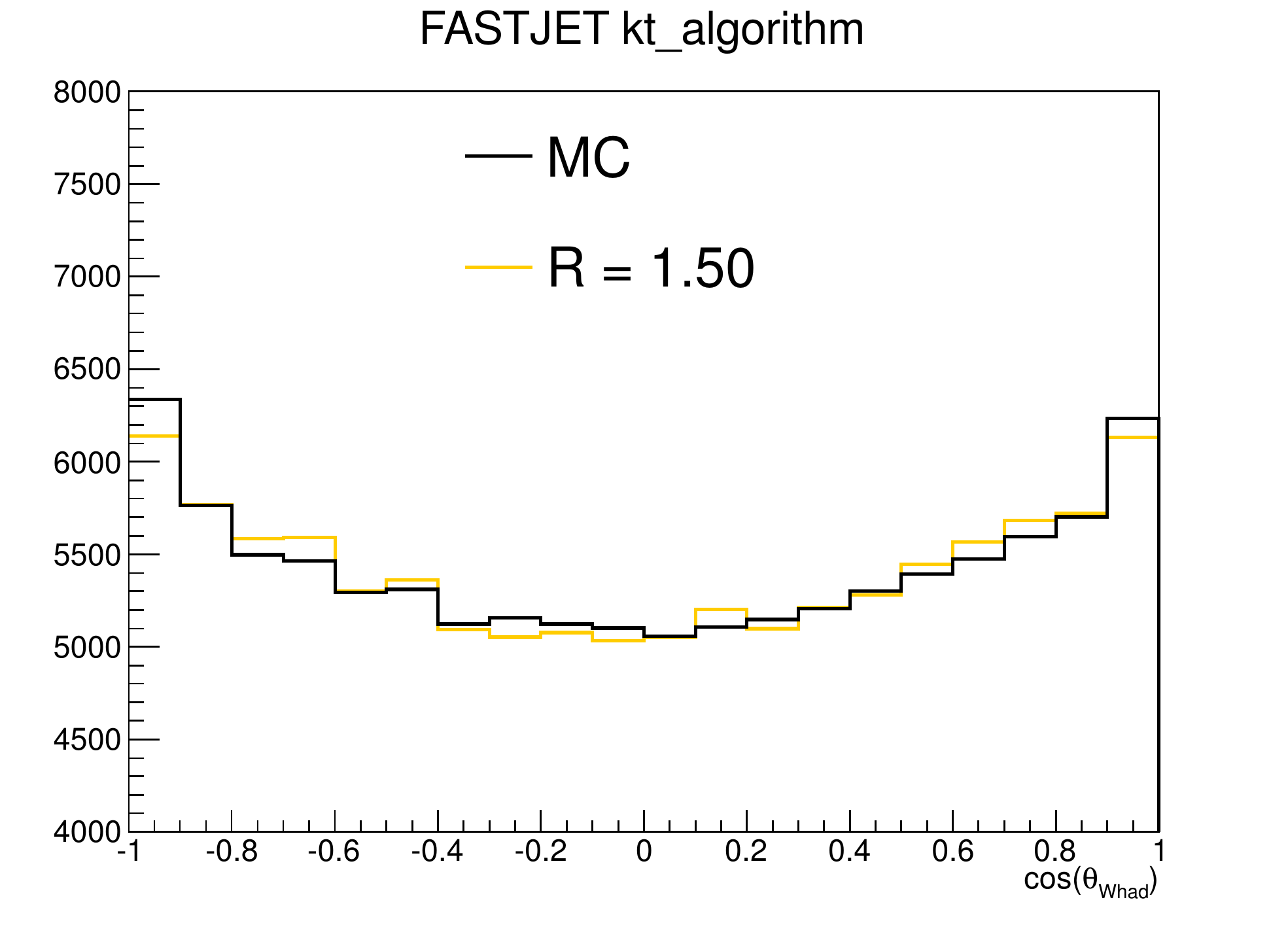}} \caption{The polar angle distribution of the hadronically decaying $W$ for four different jet algorithms.}
        \label{fig:jetrecowhad}
        }
        \end{figure}

\paragraph{Rejection of $\gamma\gamma$ background}
Superposed on each hard $e^+e^-$ scatter a number of background sources lead to additional particles.  
Two of these have been studied in some detail. Photons radiated off the incoming electron and positron beams in
the intense field of the opposing beam (beamstrahlung) produce $e^+e^-$ pairs (coherent and incoherent pair production~\cite{schulte:1996}). 
The electrons and positrons produced this way are very soft and their effect is limited to the innermost and most forward detector elements. A number of processes known collectively as multi-peripheral $\gamma \gamma \rightarrow ${\it hadrons} production yield a small number of additional particles (typically 1.7 low-multiplicity events per bunch crossing for the ILC).  The polar angle distribution of these particles is markedly forward.
Particles from $\gamma \gamma \rightarrow ${\it hadrons} tend to be harder and can reach the outer layers of the detector and affect the overall detector performance, in particular jet reconstruction. 
An excellent rejection of this $\gamma\gamma$ background is achieved by the application of the longitudinally invariant $k_t$ algorithm~\cite{Catani:1993hr,Ellis:1993tq}. This is demonstrated in Fig.~\ref{fig:jetrecowhad}. 
The figure shows the reconstructed polar angle distribution of the hadronically decaying $W$ boson compared with the generated distribution. The result is shown for the ''traditional'' Durham algorithm and for the longitudinally invariant $k_t$ algorithm. The improvement achieved by the longitudinally invariant $k_t$ algorithm is obvious. The study has been carried out as well for CLIC energies of up to $3\,\tev$ and it was shown that also in this more hostile environment this $\gamma \gamma$ background can be controlled.

\subsection{Top mass determination}
The determination of the mass of the $t$~quarks is carried out differently depending of the centre-of-mass energy of the $e^+e^-$ collisions~\cite{Seidel:2013sqa}. 

{\em In the continuum}, i.e. at energies well above the $\ttbar$~threshold, the spectrum shown in Fig.~\ref{fig:TopMass} is fitted by a function that includes a Breit-Wigner convoluted with a detector resolution function. The analysis results for a sample of $100\,\invfb$ in a combined value of $m_t=174.133\pm0.080\,\gev$ for the top mass and $\sigma_t=1.55\pm0.22$ for the top quark width.

{\em At the $\ttbar$ threshold} the $t$~quark mass is determined by a threshold scan. The considerations above on background and selection efficiency suggest that the $\ttbar$ cross section can be measured with very high precision. For the scan data samples of $10\,\invfb$ each have been simulated for ten centre-of-mass energies around the threshold. The $t$~quark mass and of the strong coupling $\alpha_s$ is extracted from a template fit between the reconstructed simulated data and corresponding distributions for different values of  $m_t$ and $\alpha_s$ in a finely grained grid. The threshold scan allows for determining the $t$~quark mass to a statistical precision of around 30\,MeV. Due to the different luminosity spectrum the precision is expected to be somewhat better at the ILC than at CLIC. Residual systematic errors from experiment negligible. Another error source is the normalization of the theory prediction. The simultaneous determination of $m_t$ and $\alpha_s$  yields a statistical error for $\alpha_s$ of 0.0008 and theory errors between 0.0009 and 0.0022 depending on the assumption on the normalisation.  It should be noted that as of today the determination of the $t$~quark mass via a threshold scan is the method that is controlled best in terms of theoretical uncertainties.  However, it is expected that the conversion to a \mbox{$\overline{\rm MS}$} mass will drive the total error to around 100\,MeV. 

\subsection{Forward backward asymmetry}

A key observable to study the chiral structure at the $\ttbar X$ vertex is the forward backward asymmetry $\afbt$. 
For the determination of the forward-backward asymmetry $\afbt$, the number of events in the hemispheres of the detector w.r.t. the polar angle $\theta$ of the $t$~quark is counted, i.e.
\beq
\afbt = \frac{N(\mathrm{cos}\theta>0)-N(\mathrm{cos}\theta<0)}{N(\mathrm{cos}\theta>0)+N(\mathrm{cos}\theta<0)}.
\eeqn

This observable can be and has been studied for fully hadronic and semi-leptonic final states. The elements of these analysis will be outlined in the following.
\subsubsection{Semi leptonic final states}

Here, the polar angle of the $t$~quark is calculated from the decay products in the hadronic decay branch.
The direction measurement depends on the correct association of the $b$~quarks to the jets of the hadronic $b$~quark decays. 
The analysis is carried out separately for a left-handed polarized electron beam and for a right handed polarized beam.
Therefore, two different situations have to be distinguished. 
\begin{itemize}
\item In case of a {\em right}-handed electron beam the sample is expected to be enriched with $\tpq$-quarks with {\em right}
-handed helicity~\cite{Parke:1996pr}. Due to the $V-A$ structure of the standard model an energetic $\Wboson$~boson is emitted into the flight direction of the $\tpq$-quark. The $W$~boson decays into two energetic jets.
The $\bottom$~quark from the decay of the $\tpq$~quark are comparatively soft. Therefore, the direction of the  $t$~quark is essentially reconstructed from the direction of the energetic jets from the $W$~boson decay. This scenario is thus insensitive towards a wrong association of the jet from the $b$~quark decay to the jets from the $W$~boson decay. 

\item In case of a {\em left}-handed electron beam the sample is enriched with $\tpq$~quarks with {\em left}-handed helicity.
 In this case the $\Wboson$~boson is emitted opposite to the flight-direction of the $\tpq$~quark and gains therefore only little kinetic energy. In fact for a centre-of-mass energy of 500\,GeV the $\Wboson$~boson is nearly at rest. On the other hand the $b$~quarks are very energetic and will therefore dominate the reconstruction of the polar angle of the $t$~quark.
In this case a wrong association of the jets with that from the $\bottom$~quark can flip the reconstructed polar angle by $\pi$ giving rise to migrations in the polar angle distribution of the $\tpq$~quark. 
\end{itemize}

The explanations above apply correspondingly to polarized positron beams and $\bar{\tpq}$-quarks.

\begin{figure}[tbp]
\begin{center}
\includegraphics[width=0.7\textwidth]{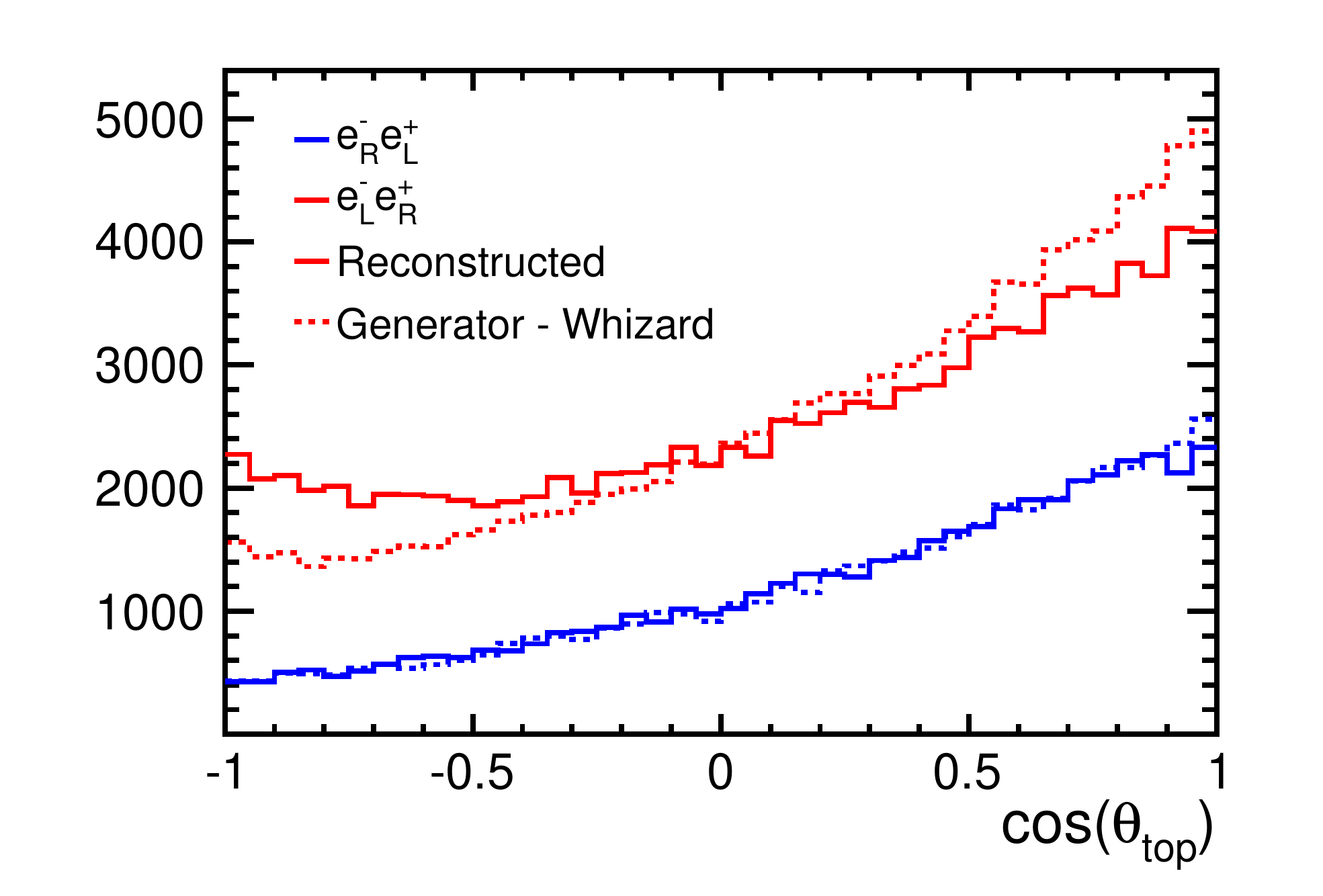}
\caption{Reconstructed forward-backward asymmetry compared with the prediction by the event generator {\sc WHIZARD} {\protect \cite{Kilian:2007gr}} for two configurations of the beam polarizations.
}
\label{fig:ambig_rec}
\end{center}
\end{figure}

\begin{figure}[tbp]
\begin{center}
\includegraphics[width=0.7\textwidth]{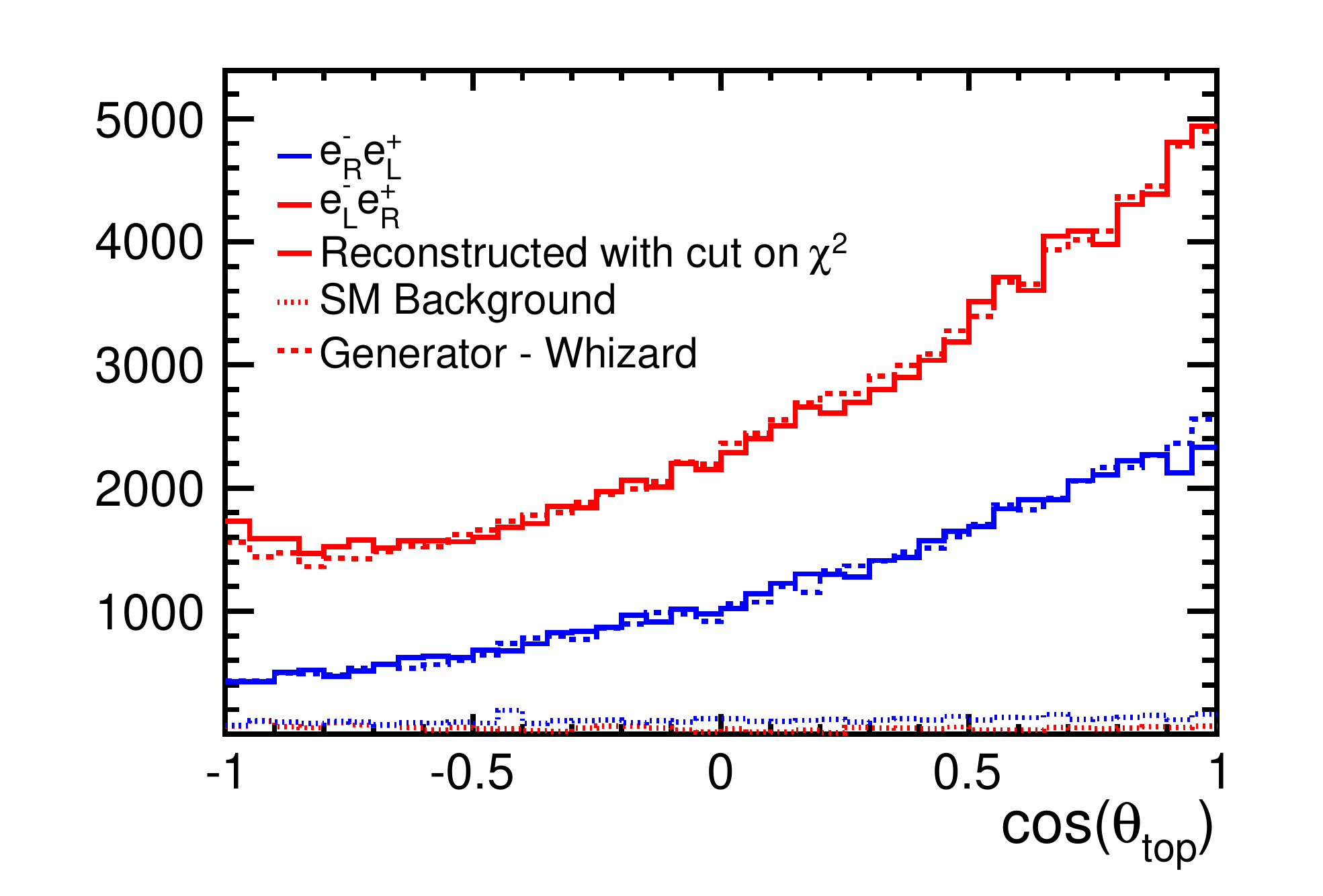}
\caption{\sl 
Reconstructed forward backward asymmetry compared with the prediction by the event generator WHIZARD after the application of a on $\chi^2<15$ for the beam polarizations
$P,P' =-1,+1$ as explained in the text. Note that no correction is applied for the beam polarizations $\pem,\pep =+1,-1$. This figure shows also the residual Standard Model background.
}
\label{fig:ambig_reccut}
\end{center}
\end{figure}

The described scenarios are encountered as shown in Fig.~\ref{fig:ambig_rec}. 
First, the reconstructed spectrum of polar angles of the $t$~quark in the case of right handed electron beams is in reasonable agreement with the generated one. On the other hand the reconstruction of $\cos{\mathrm{\theta_{\tpq}}}$ in case of left-handed $\tpq$~quarks suffers from considerable migrations. 
As discussed, the migrations are caused by a wrong association of jets stemming from $b$~quarks to jets stemming from $W$~decays. This implies that the reconstruction of observables will get deteriorated. This implication motivates to restrict the determination of $\afbt$ in case of $\pem,\pep=-1,+1$ to cleanly reconstructed events. For this the  
authors of~\cite{Amjad:2013tlv} define a test variable $\chi^2$. The reconstructed polar angle distribution of the $t$~quark is compared with the generated one for different cuts on $\chi^2$. The preliminary results shown in Fig.~\ref{fig:ambig_reccut} are given for a value of $\chi^2<15$ and demonstrate the excellent agreement between the generated and reconstructed polar angle distributions. With this the forward backward asymmetry can be determined to a precision of about 2\%.
A review of the procedure to handle the ambiguities will however be made in future studies. In the ideal case the ambiguities can be eliminated by a proper measurement of the charge of the $b$~quark from the $t$~quark decay. The measurement of the $b$~quark charge will be outlined in the following section.

\subsubsection{Fully hadronic final states}

The measurement of $\afbt$ in the fully hadronic final states requires the determination of the charge of the $b$~quark from the $t$ or $\bar{t}$ quark decay. The $b$~quark charge is obscured by fragmentation and hadronisation and in 50\% of the cases the$b$~quark fragment into a neutral $B$~meson. Charged $B$~mesons allow in principle for an unambiguous measurement of the 
$b$~quark charge. Such a measurement is supported by the small distance between the interaction point and
the first vertex layer that is the fist part of a high precision vertex and tracking system. 

For the reconstruction of the vertex charge the authors of~\cite{Devetak:2010na} propose to combine two estimators. 
The first is the momentum weighted vertex charge defined as:  

\begin{equation}
  \label{eq:momweighted}
  Q_{VTX}= \frac{\sum_{j} p_{j}^k Q_{j}}{\sum_{j} p_{j}^k}.
\end{equation}

\begin{figure}[ht]
\begin{center}
\subfigure{ \includegraphics[width=0.45\textwidth]{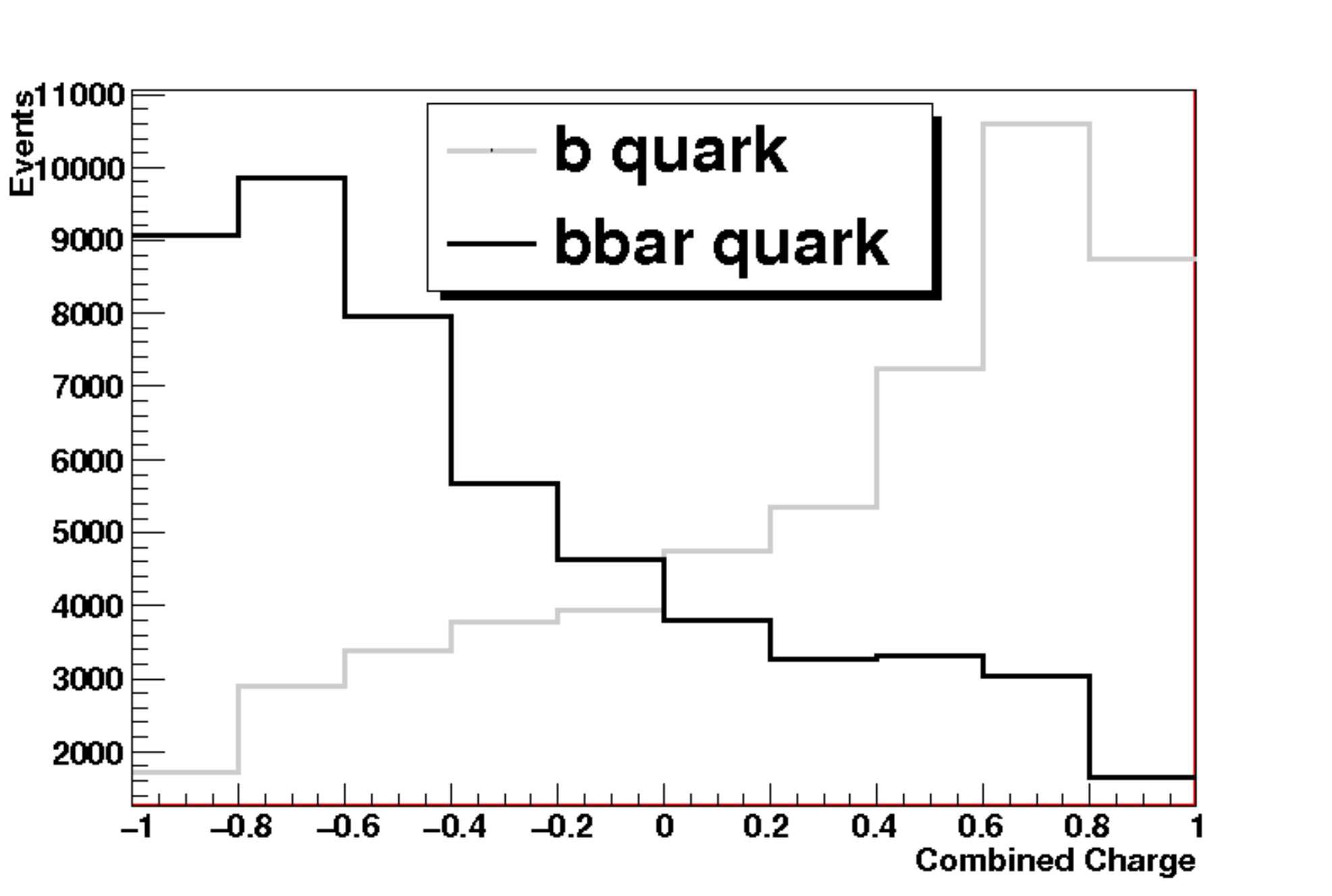} \label{f:combinedcharge}}
\subfigure{ \includegraphics[width=0.45\textwidth]{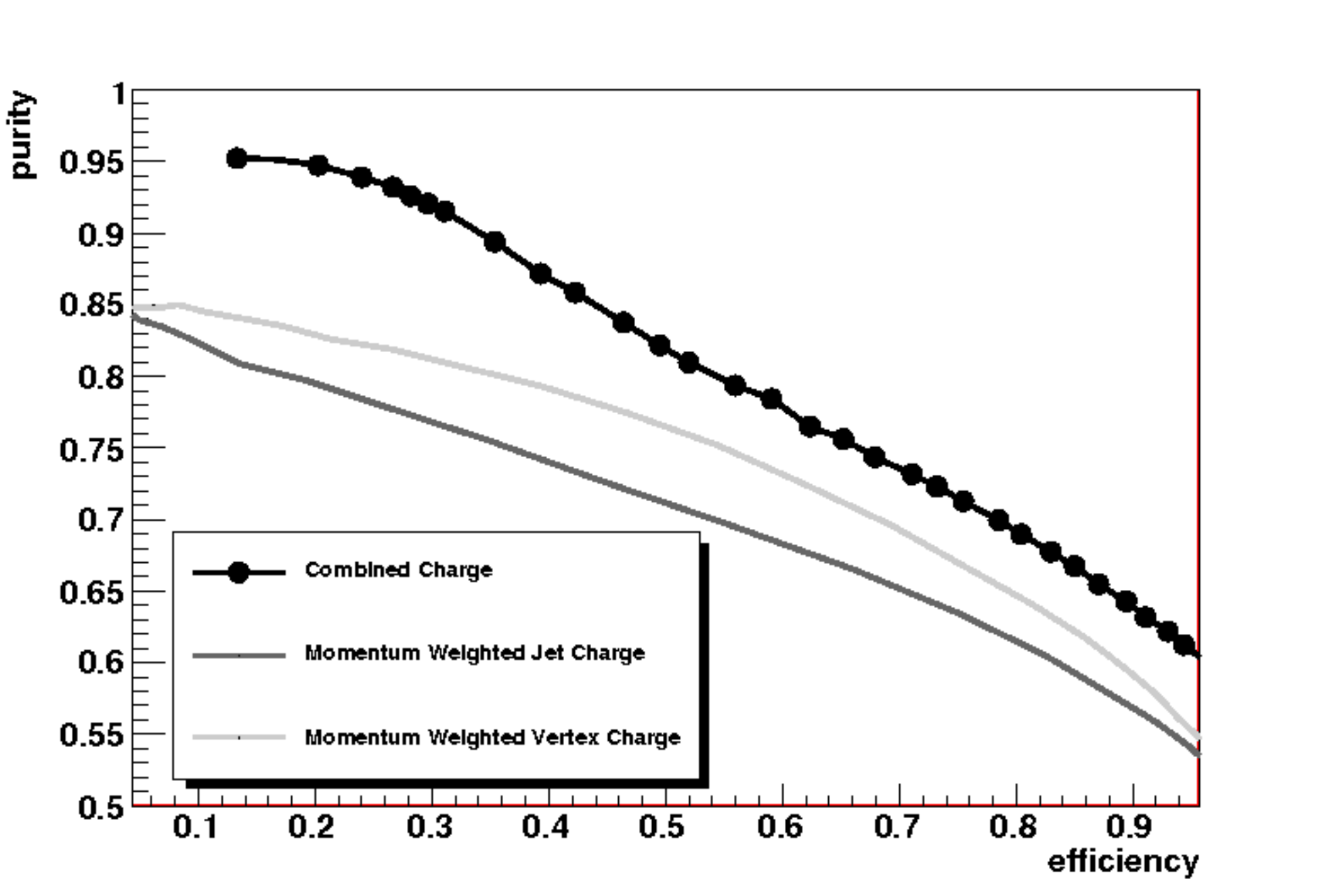}\label{f:combinedchargepe}}
\caption[Combined Charge]{Combined charge a) distributions for $b$ quark and $\bar{b}$ quark jets b) purity versus 
efficiency curves for $b$ quark and $\bar{b}$~quark jets for combined charge, momentum weighted vertex charge and 
momentum weighted jet charge. Shown for a sample using a $t$~quark mass of $174\,\gev$ after all event selections have been applied. The figures are taken from~\cite{Devetak:2010na}.}
\end{center}
\end{figure}

The second one is a generalization of the Eq.~\ref{eq:momweighted} but now all tracks charged tracks within a $b$~quark jet are taken into account.  Both estimators are combined into a test variable $C$ for which $C>0$ for $b$~quarks and $C<0$ for $\bar{b}$~quarks. The Fig.~\ref{f:combinedcharge} shows the performance of the test variable for $b$ and $\bar{b}$~quarks. The purity of the sample can be enhanced by tightening the cuts in the event selection, however with a penalty on the efficiency. An example for the dependence of the purity on the selection efficiency is shown in Fig.~\ref{f:combinedchargepe}. In the study the combination of several turned out to be beneficial for the correct reconstruction of the $b$~quark charge. 

The typical efficiency in fully hadronic $\ttbar$ decays is between 20\% and 25\%. In that case the $\afbt$ can be measured to a precision of about 3\% for an integrated luminosity of $250\,\invfb$~\cite{bib:ilc-tdr-dbd}(NDLR: SiD part). This result is compatible with the one given in~\cite{bib:ttbar-afbfh}, which only uses the charge of tracks pointing to a secondary $b$~quark vertex.

\subsection{Helicity angle}
\begin{figure}
\begin{center}
\includegraphics[width=0.7\textwidth]{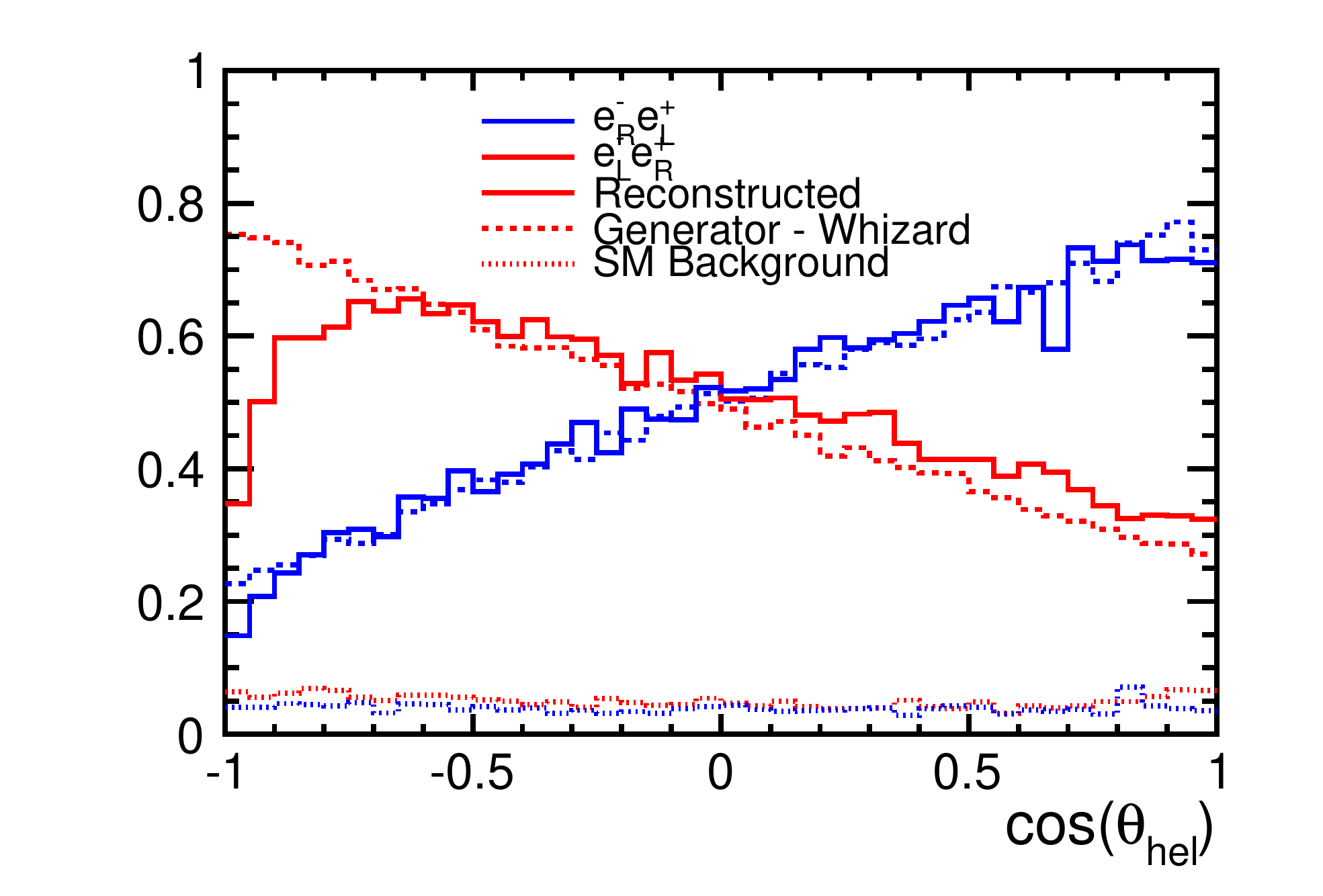}
\caption{\sl  Polar angle of the decay lepton in the rest frame of the $t$~quark.}
\label{fig:hel_dist}
\end{center}
\end{figure}

The helicity approach has been suggested for top studies at Tevatron~\cite{Berger:2012nw}. In the rest system of the $t$~quark, the angle of the decay products from the $W$~boson $\theta_{hel}$  is distributed like:

\beq
\frac{1}{\Gamma} \frac{d\Gamma}{d\cthel} = \frac{1+ \kappa \lambda_t \cthel}{2}  
\eeq{costhel}
Where $\lambda_t$ varies between $+1$ and $-1$ depending on the fraction of right-handed ($t_R$) and left-handed top quarks ($t_L$) in the sample. In addition it is $\kappa=1$ for semi-leptonic $W$~boson decays for which this studies is carried out. 
In case of $\mu$ and $e$, the measurement of the decay lepton is particular simple at a LC. The Fig.~\ref{fig:hel_dist} taken from~\cite{Amjad:2013tlv} shows the helicity angle distribution for two configurations of the initial beam polarization. In both cases the reconstructed distribution reproduces very well the generated one. over a broad range in $\cthel$. The dip at small values of $\cthel$ can be explained by residual inefficiencies in the measurement. of low momentum leptons. Please note that for the extraction of e.g. $t$~quark to couplings to $\gamma$ and $Z^0$ the slope of the distribution is used. For the determination of the slope the available lever arm is largely sufficient.

\clearpage
\section{Conclusions}
\label{sec:sec_opportunities}
The main conclusions of the studies described in this paper are given below:

\begin{itemize}

\item
High-luminosity runs at the LHC will be unfavorable for many high-precision SM measurements
that require reconstruction of jets with transverse momenta below 100~GeV.  
Uncertainties on jet measurements will be dominated by our understanding of pileup corrections, rather than by 
the measurement uncertainties of signal jets as for the 2010--2012 LHC measurements. 
We expect that the jet-energy uncertainty will increase by a factor two or more for $\mmu>100$.
This conclusion was obtained using
model-independent considerations, and is still on the optimistic side given several idealized assumptions.

\item
The above conclusion has broad implications for top measurements near threshold (cross sections, top masses etc.),
new physics searches, and other SM searches (such as $Ht\bar t$)
that require the reconstruction of low-$p_T$ jets. 
Currently, jet-related uncertainties are the dominant uncertainties for top reconstruction.
For the $\mmu > 100$ pileup scenario, the uncertainty on the $t\bar{t}$ cross section and top masses 
will be a factor of two or more larger than
for the previously published measurements using 2010--2012 data sets.  It is unlikely that any single or
combined measurement will reach a precision better than the expected theoretical uncertainties, unless
methods are employed that give a larger role to leptonic and/or tracker-based observables.



\item
Future LHC runs (at 13--14~TeV and possibly 33~TeV) should have large enough statistics to reach
the highly-boosted regime for top-quark reconstruction where resolved
approaches becomes ineffective. The transverse
momenta of the top quarks will be high enough 
that all of the decay products can be contained within standard jets
of radius 0.5--0.6. The most favorable kinematic regime for highly-boosted top reconstruction is $\ptjet \ge 0.8$~TeV  assuming  $R=0.5$.
An inclusive observation of highly-boosted hadronic top quarks is possible using jet mass distributions after $b$-tagging, assuming that
the $b$-tagging reduces the contribution from light jets by a factor of two compared to $b$-jets, 
and that the shape of jet mass distribution near 170~GeV is understood within $10\%$.

\item
Pileup effects can be less severe for highly-boosted top quarks than for near-threshold top quarks due to the higher energy scales
and smaller detector areas involved.  However, the pileup still must be carefully removed to preserve internal kinematic
features such as the top-jet mass.  Dedicated jet grooming techniques
such as trimming, pruning, and filtering can yield more robust pileup subtraction than the area-based four-vector subtraction method, and
allow easy interface to other jet substructure methods for top-tagging.

\item
Hadronic top-jet mass resolution and the efficiency of jet substructure algorithms for top-tagging quickly degrade as $\ptjet$ exceeds 1~TeV,
for example suffering a factor of two broadening of the mass peak between $\ptjet \simeq 0.8$~TeV and $\ptjet \simeq 1.6$~TeV with the 
EFlow {\sc Delphes} detector.  This effect is 
due to the spatial granularity of the detector and contamination from initial- and final-state radiation.
The effect can be highly ameliorated by incorporating
more spatially-segmented information from the ECAL, by applying jet substructure methods to reduce contaminating semi-hard radiation, 
and by restricting the active area of the top-jet to particles within $\Delta R \sim 4 m_t / \ptjet$ of the jet center.  
Similarly, any minimum-$\Delta R$ parameters in a top-tagger should be scaled as $1/\ptjet$.

\item A future linear collider $e^+e^-$  will allow for studying electroweak production of $\ttbar$ pairs. There is no QCD background. Due to the electroweak production all interesting processes occur at roughly the same rate and the total number of interactions is such that no online trigger is needed. 

\item Unlike at the LHC there is no or little event pile up. This is in particular true for the ILC. There will be no multiple $e^+e^-$ interactions. Overlay events may come from $\gamma \gamma$ interactions.  Ongoing studies show that this residual  pile-up can be controlled with adequate jet algorithms.

\item Being the most advanced proposal, the ILC is designed to produce an integrated luminosity of $500\,\invfb$ at a centre-of-mass energy of $500\,\gev$ within four years of running.
Physics backgrounds to $\ttbar$ production can be reduced to a negligible level. The needed selection cuts will always leave a signal sample of about 100~k events.

\item The $t$~quark mass can be measured to a precision of better than 40\,MeV where systematic studies have shown that the statistical error is the dominating error. 
It is expected, however, that the conversion to the theoretical well defined \mbox{$\overline{\rm MS}$} mass will drive this uncertainty to around 100\,MeV. The precise $t$~quark mass will e.g. lead to a precise picture on the vacuum stability of the universe. 

\item A major asset of the linear collider are the polarized beams. The increased number of observables permit to disentangle $\ttbar$ couplings to 
the photon and the $Z^0$~boson. The beam polarization will allow for enriching the samples with $t$~quarks of left or right handed helicities. New physics scenarios predict
measurable differences in the production rate of right-handed $t$~quarks w.r.t. to the Standard Model. 

\item For the first time in energy frontier machines the charge of the $b$~quark will be measured at a purity of 60\% and better. This is in particular beneficial for observables like
$\afbt$ which is sensitive to modifications of the chiral structure at the $\ttbar X$ vertex. Note, that the chiral structure will also be tested with observables as the helicity angle.

\item Electroweak couplings can be determined at the percent level. It is important that experimental and theoretical errors are kept at the same level. This requires a precise measurement of e.g. the luminosity and the beam polarization. Currently, both parameters are expected to be controlled to better than 0.5\%. As shown elsewhere this allows for reaching sensitivity to new physics at scales beyond 10\,TeV. Keywords here are compositeness and/or extra dimensions. 

\item  In general the realization of machine and detectors must not compromise the precision physics. This is can be the biggest challenge in the coming years.

\end{itemize}

\section*{Acknowledgements}
We would like to thank many collegues for the discussion of these
results, especially T.~LeCompte, J.~Proudfoot,  A.~Schwartzman and S.~Padhi.           
The submitted manuscript has been created by UChicago Argonne, LLC,
Operator of Argonne National Laboratory (``Argonne'').
Argonne, a U.S. Department of Energy Office of Science laboratory,
is operated under Contract No. DE-AC02-06CH11357.

\bibliographystyle{l4z_pl}
\pagestyle{plain}
\bibliography{topdet}

\end{document}